\documentclass{jfm}

\usepackage{graphicx}
\usepackage{newtxtext}
\usepackage{newtxmath}
\usepackage{natbib}
\usepackage{hyperref}
\hypersetup{
    colorlinks = true,
    urlcolor   = blue,
    citecolor  = black,
}

\newcommand{\RomanNumeralCaps}[1]
\linenumbers

\usepackage{xcolor}

\newcommand{\reva}[1]{\textcolor{black}{#1}}
\newcommand{\revb}[1]{\textcolor{black}{#1}}
\newcommand{\revc}[1]{\textcolor{black}{#1}}
\newcommand{\revaa}[1]{\textcolor{black}{#1}}
\newcommand{\revbb}[1]{\textcolor{black}{#1}}
\newcommand{\revcc}[1]{\textcolor{black}{#1}}

\title{Coherent structures in Newtonian and viscoelastic turbulent planar jets}

\author{Christian Amor\aff{1},
  Adri{á}n Corrochano\aff{2},
  Giovanni Soligo\aff{1,3},
  Soledad Le Clainche\aff{2}
  \and Marco Edoardo Rosti\aff{1}\corresp{\email{marco.rosti@oist.jp}}}

\affiliation{\aff{1}Complex Fluids and Flows Unit, Okinawa Institute of Science and Technology Graduate University, 1919-1 Tancha, Onna-son, Kunigami-gun, 904-0497, Okinawa, Japan
\aff{2}School of Aerospace Engineering, Universidad Polit{é}cnica de Madrid, Plaza del Cardenal Cisneros 3, 28040, Madrid, Spain
\aff{3}Currently at: Flow Matters Consultancy BV, Groene Loper 5, 5612 AE Eindhoven, The Netherlands}

\begin{document}
\maketitle

\begin{abstract} 
The addition of a small amount of long-chain polymers confers viscoelastic properties to Newtonian flows. The resulting non-Newtonian solution now exhibits different dynamics, such as enhanced mixing at low Reynolds, where elastic instabilities can trigger elastic turbulence even though inertial turbulence is absent. Here, we study this phenomenon in viscoelastic planar jets and, in particular, we do it from the perspective of coherent structures to understand how elastic turbulence is triggered and sustained, which remain barely explored in this setup. We introduce the spatio-temporal Koopman decomposition for extracting the dominant flow patterns, and we compare them with those from Newtonian planar jets at high Reynolds number. Global flow structures are similar between jets, with low-frequency streaks and high-frequency wave packets dominating the turbulent dynamics. However, structures are strikingly different in the near field, where elasticity-driven streaks affect the dynamics in the potential core of the viscoelastic planar jet, modifying the bulk flow and interacting with the flow instability. The analysis of the polymer field reveals stretched polymer filaments and centre-mode structures, which support the implication of the near-field streaks on sustaining elastic turbulence in three-dimensional viscoelastic planar jets.
\end{abstract}


\section{Introduction}\label{sec:intro}
Turbulent flows are ubiquitous in nature and industrial applications, although their interpretation is challenging due to their inherent complexity. A common strategy to tackle their study relies on coherent structures---organised regions in space with similar features that persist in time---which provide a low-dimensional representation of otherwise high-dimensional turbulent flows. This approach has been exploited for more than half a century, with some early works such as those by \cite{Kline1967StreaksWBF} who revealed streaky patterns in turbulent boundary layers, or \cite{Crow1971JFMWP} and \cite{Brown1974JFMSL} who identified organized structures in turbulent jets and mixing layers, respectively. More recently, the dynamics of coherent structures have been linked to relevant engineering problems, such as drag reduction in wall-bounded turbulent flows \citep{Jimenez2018JFMWBTCohStr} and noise generation in high-speed turbulent jets \citep{Jordan2013ARFMWPNoise}.

Here, we focus on large-scale coherent structures, a few of which are sufficient to picture the dominant flow dynamics. There are different ways of computing these large-scale coherent structures, with mathematical tools such as modal decompositions being a common data-driven approach for their identification \citep{Taira2017AAIAModalDecomp}. These algorithms aim to describe the flow as a superposition of spatial modes, each accompanied by characteristic values that represent either their energy content or dynamic traits. Modal decompositions generally require large amounts of data, and can be used to compress high-dimensional data (hundreds thousands of grid points in a numerical simulation) into more interpretable low-dimensional forms. This simpler description of turbulent flows is also suitable for reduced-order models, where modal decompositions can be combined with Galerkin methods \citep{Rowley2017ARFMPODGalerkin}, run in parallel with numerical simulations \citep{Amor2023JCPHODMDonthefly}, or integrated into deep learning models \citep{LeClainche2022PorousWall, AbadiaHeredia2022ESWAPODDL} to accelerate computations.

In this work, we employ modal decompositions to identify coherent structures in turbulent planar jets. In particular, we use the higher order dynamic mode decomposition (HODMD) \citep{leclainche2017hodmd}. Similarly to standard dynamic mode decomposition (DMD) \citep{schmid2010dmd}, HODMD extracts the dominant nonlinear dynamics directly from the data. HODMD (and DMD) represents the flow as a linear superposition of so-called modes, each corresponding to coherent structures that oscillate either in time, space, or in time and space. The key innovation of HODMD lies in its use of delay embeddings, which augment the original data by incorporating time-delayed snapshots \citep{takens1981delay}. This improves the spectral resolution of the data and filters small amplitude frequencies \citep{LeClainche2017hodmdZNMF}, thus making the method more suitable for analysing highly complex, nonlinear flows \citep{leclainche2020evp, LeClainche2022PorousWall}. Here, we extend the application of HODMD to compute the coherent structures in Newtonian and viscoelastic turbulent planar jets. Specifically, we validate the method for high-Reynolds Newtonian planar jets, and we extend the work by \cite{Amor2024JPCS5MTW} for low-Reynolds viscoelastic planar jets.

Coherent structures in Newtonian turbulent jets are characterised by three distinct mechanisms. At first, the Kelvin-Helmholtz instability causes wave packets---coherent axisymmetric structures---that grow linearly near the inlet \citep{Cringhton1976JFMStabCE, Cohen1987JFMInstJet, Suzuki2006JFMWP, Gudmundsson2011JFMWP, Cavalieri2013JFMWP}. A second mechanism, the Orr mechanism, occurs downstream the potential core, where the flow response is rather nonlinear \citep{Garnaud2013JFMOrr, Tissot2017PRFOrr, schmidt2018KHOrr}. In this region, Orr wave packets have greater gain than those associated to Kelvin-Helmholtz, specifically at low frequency and zero wavenumber \citep{schmidt2018KHOrr}. The third response mechanism is the lift-up \citep{Boronin2013JFMLiftUp, JimenezGonzalez2017PoFLiftUp, nogueira2019streaks}. The lift-up and the associated coherent structures, often called streaks, are an important mechanism in wall-bounded \citep{Brandt2014EJMLiftUp} and planar free shear flows \citep{Pierrehumbertt1982JFMShearLayer, Bernal1986PlaneShear, Metcalfe1987JFMShearFlow}. In turbulent jets, streamwise vortices have a fundamental effect on the dynamics and statistical properties of the flow \citep{Liepmann1992JFMJetVort}, since these streamwise vortices (and streaks) are the most energetic for non-zero wavenumbers \citep{Citriniti2000JFMStreakVort}, scaling inversely to the distance from the inlet \citep{Jung2004JFMStreakVort, Tinney2008JFMStreakVort}.

A comparable characterisation of coherent structures in viscoelastic jets is still lacking. Viscoelasticity, for instance introduced by adding flexible polymers to a Newtonian solvent, significantly changes the flow dynamics and, as a consequence, it is expected to also modify the coherent structures. In viscoelastic turbulent flows, polymers interact with the energy cascade by extracting energy from the large-scale eddies, that they either return to the flow at smaller scales or dissipate through polymer diffusion \citep{valente2014effect}. Even though their influence is more evident at small scales, their range of influence, identified by changes in the scaling exponent of the energy spectrum, depends on the polymer elasticity \citep{rosti2023large} and inertia \citep{Singh2025JFMITET}. Polymers can also destabilise laminar flows through purely elastic instabilities \citep{Larson1992RheoActaElInst, Shaqfeh1996ARFMElInst}, leading to elastic turbulence \citep{Groisman2000NatureET, Singh2024NatCommET}. These have been extensively studied in curvilinear flow configurations, such as the flow between concentric cylinders \citep{larson1990purely, groisman1998mechanism}, counter-rotating parallel disks \citep{McKinley1991JNNFMParPlate, Oztekin1993JFMParPlate}, and in curved channels \citep{groisman2001efficient, groisman2004elastic}. However, curvature is not a prerequisite for their occurrence, and elastic instabilities can occur in straight shear flows too \citep{Pan2013PRLElaInstPlaneChan, Qin2017PRLElaInstPlaneChan, Jha2021PNASKHPlaneChan, Lellep2024PNASvChanET, Rota2024PRFvChan}. Recently, elastic turbulence has been found also hiding in the smallest scales of polymeric turbulence at large Reynolds number \citep{Garg2025PRLETSmallScales}.

In jet flows, elastic instabilities appear at markedly lower Reynolds number compared to classic Newtonian jets. \cite{yamani2021spectral} observed in round jets the transition to elasto-inertial turbulence---when inertial and elastic effects are of the same order of magnitude \citep{samanta2013elasto}. The elastic turbulent state can be sustained at much lower Reynolds number if elasticity is sufficiently large \citep{soligo2023non}. In the latter case, the destabilising role of elasticity remains unclear. While elasticity stabilises the sinuous (antisymmetric) mode and partly stabilises the varicose (symmetric) mode at high Reynolds number \citep{rallison1995instability}, elasticity is rather destabilising at moderate Reynolds numbers, though this effect has a non-monotonic trend: increasing elasticity first destabilises the flow, though it stabilises if elasticity is sufficiently large \citep{ray2015absolute}, as also confirmed experimentally \citep{yamani2021spectral, yamani2023spatiotemporal}. This effect may be induced by the competing influence of elasticity at separated regimes: in the linear regime, elasticity enhances the instability of the jet \citep{ray2015absolute}, but it inhibits the roll-up process in the non-linear regime, that yields to overall flow stabilisation \citep{Kumar1999JNNFMInstShearLayer, Guimaraes2023PRFJetInst}. In the low-Reynolds number limit of elastic turbulence, polymers alone sustain the turbulence, when inertial effects are negligible \citep{Groisman2000NatureET}.

To address this gap and clarify the mechanism, we apply nonlinear dynamic mode decomposition (HODMD) \textit{(i)} to compare the global coherent structures in Newtonian and viscoelastic turbulent planar jets and \textit{(ii)} to examine the influence of polymer elasticity in the near field region. Data-driven analysis is well-suited for investigating global instabilities in highly nonlinear systems, where traditional stability analysis is often intractable. For instance, the spatial decomposition based on HODMD (or DMD) can represent the flow as growing and/or decaying time-dependent perturbations that propagate upstream or downstream from their point of origin. This approach was successfully applied by \cite{yamani2023spatiotemporal}, who employed local DMD analysis to characterise elasto-inertial instabilities in viscoelastic planar jets. In our study, we decompose the nonlinear dynamics through a spatio-temporal analysis, in which the sequential application of HODMD yields a representation of the flow as a superposition of travelling waves. This methodology, namely spatio-temporal Koopman decomposition \citep{LeClainche2018STKDJNonlinSci}, has previously proven effective in capturing the unstable modes of the flow past a circular cylinder at low Reynolds number \citep{LeClainche2018STKDCyl}, as well as the modes associated with streak breakdown in the turbulent channel flow of an elastoviscoplastic fluid \citep{leclainche2020evp}.

The article is organized as follows. Section \ref{sec:meth} introduces the direct numerical simulations and the data-driven analysis employed in this work. Section \ref{sec:res} compares the coherent structures in Newtonian and viscoelastic turbulent planar jets, with a detailed analysis of the near-field structures in the viscoelastic jet in \S\ref{sec:inj}. Section \ref{sec:trc} focuses on the polymer field in the viscoelastic jet, and finally \S\ref{sec:concl} summarises the main conclusions.

\section{Methods}\label{sec:meth} 
\subsection{Direct numerical simulations}\label{subsec:dns}
Data are obtained by means of direct numerical simulations. We employ our in-house solver \textit{Fujin} \citep{rosti_2026a} (\url{https://www.oist.jp/research/research-units/cffu/fujin}) for simulating the Newtonian and viscoelastic turbulent planar jets. Most of the data considered here are taken from previously published results \citep{soligo2023non, Soligo2025JFMNewtonJet}.

The problem is governed by the incompressible Navier-Stokes equations:
\begin{equation}
    \nabla \cdot \boldsymbol{u} = 0,
    \label{eq:a11}
\end{equation}
\begin{equation}
    \rho \left(\partial_t \boldsymbol{u} + \nabla \boldsymbol{u} \boldsymbol{u} \right) = - \nabla p + \mu_s \nabla^2 \boldsymbol{u} + \nabla \cdot \tau,
    \label{eq:a12}
\end{equation}
where $\boldsymbol{u}$ and $p$ are the velocity and pressure fields, $\rho$ is the density, $\mu_s$ is the dynamic viscosity of the solvent, and $\tau$ is the non-Newtonian stress tensor. In the Newtonian jet, $\tau = 0$ and $\mu_s$ is equivalent to the viscosity of the fluid ($\mu_0 = \mu_s$). On the other hand, the viscoelastic jet requires an extra equation to model $\tau$. Here, we use the Oldroyd-B model:
\begin{equation}
    \lambda \overset{\nabla}{\tau} + \tau = \mu_p \left( \nabla \boldsymbol{u} + \nabla \boldsymbol{u}^{\top} \right),
    \label{eq:a21}
\end{equation}
where $\lambda$ is the relaxation time of the polymer, i.e., the time required by the polymer to reach equilibrium if perturbed by an external forcing, and $\mu_p$ is the dynamic viscosity of the polymer, with $\mu_0 = \mu_s + \mu_p$ being the total viscosity. Here, $\overset{\nabla}{\tau}$ is the upper-convective derivative, defined as:
\begin{equation}
    \overset{\nabla}{\tau} = \partial_t \tau + \boldsymbol{u} \cdot \nabla \tau - \left( \nabla \boldsymbol{u}^{\top} \cdot \tau + \tau \cdot \nabla \boldsymbol{u} \right).
    \label{eq:a22}
\end{equation}
Note that the model describes a purely viscoelastic fluid without shear-thinning viscosity. The non-Newtonian stress $\tau$ can be written in terms of the conformation tensor $\mathbf{C}$, a second-order, positive-definite tensor that represents the average value of the end-to-end distance of the polymers, i.e., $\tau \left( \mathbf{C} \right) = \mu_p \left( \mathbf{C} - \mathbf{I} \right)/ \lambda$, with $\mathbf{I}$ the tensorial identity.

The equations are discretized on a staggered, uniform, Cartesian grid, with $x$ being the streamwise, $y$ the jet-normal, and $z$ the spanwise direction. Scalar variables, namely pressure, density, viscosity and extra stresses, are stored at the centre of the cells, and velocity at the faces. The spatial derivatives are approximated using second-order, central finite differences, while time integration is performed using a second-order explicit Adams-Bashforth scheme, coupled with a fractional step method \citep{kim1985application} to solve the pressure coupling. We adopt a matrix-logarithm formulation \citep{fattal2004constitutive, hulsen2005flow} to overcome the numerical instability at high polymer elasticity \citep{keunings1986high}, and we use a \revb{fifth-order weighted essentially non-oscillatory scheme} \citep{shu2009high, sugiyama2011full} to solve the advection term in the upper-convective derivative. 

The fluid is injected through a plane slit with height $2h$ and length $L_z$ from the left side of the domain box, \revc{that is filled with the same fluid as that injected through the plane slit}. \revb{We impose on the left boundary ($x = 0$) the no-slip and no-penetration boundary conditions}, \revc{except for the inlet slit (centred at $y = 0$), where a top-hat profile is imposed to the streamwise velocity \citep{Silva2002JFMInflow, Stanley2002JFMInflow}: 
\begin{equation}
   \frac{u}{U} = \frac{1 + \alpha}{2} + \frac{1 - \alpha}{2} {\rm tanh} \left[ \frac{h}{4 \vartheta} \left( \frac{2 |y|}{h} - 1 \right)  \right],
   \label{eq:a1}
\end{equation}
where $\vartheta$ is the shear layer momentum thickness, and $\alpha = 0.1$ in the Newtonian case and $\alpha = 1$ in the viscoelastic case; a small co-flow of $0.1 U$ is added in the Newtonian jet.} \revcc{This choice is meant for minimising the effect of the upper and lower boundaries in the Newtonian jet, where, as we describe below, the vertical length of the computational box is significantly smaller compared to the corresponding for the viscoelastic jet. Simulating the Newtonian jet in a similar domain than the viscoelastic jet would be computationally unfeasible, thus the different choice of co-flow.} 
\revb{Free-slip and no-penetration boundary conditions are imposed in the lower and upper boundaries ($y = \pm L_y / 2$), periodic boundary conditions are used in the spanwise direction ($z = 0$, $z = L_z$), and a non-reflective outflow boundary condition \citep{orlanski1976simple} is enforced at the outlet ($x = L_x$).}

\reva{The Reynolds number is based on the half-slit height $h$ and the inlet velocity $U$, and is defined as $Re = U h / \nu_0$, where $\nu_0$ is the total kinematic viscosity of the fluid, equal to $\mu_s/\rho$ in the Newtonian case and to $\left( \mu_s +\mu_p \right)/\rho$ in the viscoelastic one. We consider a Newtonian jet at a high Reynolds number $Re = 6750$ and a viscoelastic jet at a low Reynolds number $Re = 20$. We choose a large enough Reynolds number to achieve fully developed turbulence in the Newtonian jet \citep{Soligo2025JFMNewtonJet}; contrarily, Newtonian jets at $Re = 20$ are laminar, so any observed unstable behaviour in the viscoelastic jet is attributed to the non-Newtonian property of the fluid \citep{soligo2023non}.} The viscoelastic jet requires two additional non-dimensional parameters. The first one is the Deborah number, $De = \lambda U  / h$, that is set to $100$, \revb{such that the elasticity number, measured as $E = De / Re$, is greater than unity; in other words, turbulence in this regime is sustained solely by elasticity, with inertial effects being negligible. We denote this regime elastic turbulence even though $Re$ is not smaller than unity, since $Re \lesssim De$ and since the flow would be laminar in the absence of polymers \citep{Steinberg2001PRLET}. At the same $Re$ but one order of magnitude smaller $De$ ($De = 10$), thus $E \lesssim 1$, the resulting viscoelastic jet has properties similar to those from the elastic turbulent planar jet studied here, but with the flow relaminarizing downstream after a certain distance from the inlet \citep{soligo2023non}. Once fully-developed, elastic turbulence shows similar qualitative properties notwithstanding the value of $De$ \citep{Singh2024NatCommET}. Finally, the last parameter is the viscosity ratio $\beta = \mu_s / \mu_0 = \mu_s / \left( \mu_s + \mu_p \right)$, that is set to $0.98$, thus indicating a dilute polymer solution.} 

The computational domain extends for $140h \times 140h \times 35h$ in the streamwise, jet-normal and spanwise directions for the Newtonian case, and $320h \times 480h \times 26.7h$ in the viscoelastic case, and each domain is discretized using $5760 \times 5760 \times 1440$ and $1440 \times 2340 \times 128$ grid points, respectively. The size of the two domains are different to ensure a minimal effect of the bounding box on the flow behaviour. Indeed, due to the difference of the Reynolds numbers, the computational box in the low Reynolds number viscoelastic jet must be longer and taller to contain a fully developed region of the flow, which develops further away from the inlet, compared to the Newtonian turbulent jet. \revbb{The level of turbulence in the viscoelastic jet remains, however, much less intense than in the Newtonian jet: the Taylor Reynolds number in the Newtonian jet reaches $Re_{\lambda} \approx 120$ in the developed region beyond $x/h \gtrsim 80$, while for the viscoelastic one it is limited to $\approx 4$ at the same distance.} The chosen setup is the results of various tests done with different box sizes, performed by \cite{soligo2023non} and \cite{Soligo2025JFMNewtonJet}, respectively.

\subsection{Higher order dynamic mode decomposition}\label{subsec:hodmd}
Higher order dynamic mode decomposition \citep{leclainche2017hodmd} (HODMD) is an extension of the standard dynamic mode decomposition (DMD) \citep{schmid2010dmd}. Similarly to DMD, HODMD decomposes spatio-temporal data $\boldsymbol{v} \left( \boldsymbol{x}, t \right)$ into a finite set of $M$ DMD modes, each associated with a frequency $\omega_m$ and a growth rate $\delta_m$. In addition, HODMD computes an amplitude coefficient $a_m$ that weights each spatial mode $\boldsymbol{u}_m(\boldsymbol{x})$ in the reconstruction, quantifying the contribution of each mode to the overall flow dynamics. More precisely, HODMD (and DMD) reconstructs the data as follows: 
\begin{equation}
    \boldsymbol{v} \left( \boldsymbol{x}, t_k \right) \simeq \sum_{m = 1}^{M} a_m \boldsymbol{u}_m \left( \boldsymbol{x} \right) e^{\left( \omega_m i + \delta_m \right) t_k}, \quad k = 1, \ldots, N_t.
    \label{eq:b31}
\end{equation}

To estimate eq.~(\ref{eq:b31}), the data are arranged in matrix form $\mathbf{V}_1^{N_t}$, with each column containing a snapshot $\boldsymbol{v}_k$---an instantaneous field from an experiment or numerical simulation. This matrix has dimension $I \times J \times N_t$, where $I$ is the number of flow features, $J = N_x \times N_y \times N_z$ the number of spatial grid points, and $N_t$ the total number of snapshots---they should be spaced equally in time. 

There are two fundamental differences between HODMD and the standard DMD method. First, HODMD introduces a pre-processing dimensionality reduction step. The snapshot matrix $\mathbf{V}_1^{N_t}$ is projected onto a low-dimensional basis, yielding a compact representation that exploits redundancies and mitigates noise. Second, HODMD augments the projected data using delay embeddings \citep{takens1981delay}. The combination of DMD with time-delayed observables, e.g., in the form of a Hankel matrix, has proven effective for approximating the spectral properties of the Koopman operator in nonlinear dynamical systems \citep{Arbabi2017HankelDMD, Brunton2017HAVOK, Kamb2020TimeDelayKoopman}. 
This implementation makes HODMD suitable for extracting spectral information from temporally broadband data, which enables carrying out the standard DMD method in highly complex nonlinear flows.

In what follows, we outline the steps comprising HODMD. For further details on DMD, the reader is referred to \cite{schmid2010dmd}. 

\subsubsection{Step 1: Dimensionality reduction via singular value decomposition}\label{subsubsec:svd}
Truncated singular value decomposition (SVD) is performed to the snapshot matrix: 
\begin{equation}
    \mathbf{V}_1^{N_t} \simeq \mathbf{U} \mathbf{\Sigma} \mathbf{T}^{\top},
    \label{eq:b32}
\end{equation}
where the diagonal matrix $\mathbf{\Sigma}$ contains the singular values $\sigma_1 > \sigma_2 > \ldots > \sigma_N$, and $\mathbf{U}^{\top} \mathbf{U} = \mathbf{T}^{\top} \mathbf{T}$ are $N \times N$ unit matrices. Based on a threshold $\varepsilon_1$, the spatial dimension $I \times J$ of the original data is reduced to a set of linearly independent vectors of dimension $N$, with $N \ll I \times J$ the spatial complexity of the reduced data. 
The reduced snapshot matrix is defined as: 
\begin{equation}
    \hat{\mathbf{V}}_1^{N_t} = \mathbf{\Sigma} \mathbf{T}^{\top},
    \label{eq:b33}
\end{equation}
that has dimension $N \times N_t$. 
The reduced snapshot matrix is computationally more tractable than the original one, thus making possible the implementation of the delay embeddings, in particular if data are very high-dimensional, which is the usual case in large-scale problems. 

In this work, we implement high-order SVD (HOSVD) \citep{tucker1966hosvd} instead of standard SVD. HOSVD is better suited for tensorial data, as it captures the relationships among different dimensions. Truncation can be performed separately for each dimension, allowing the complexity to be adjusted independently. This is advantageous for our case, where planar jets exhibit greater complexity in the streamwise and jet-normal directions than in the spanwise one. However, this flexibility comes at the cost of increased parametrisation: the number of thresholds increases from one (uniform truncation in space and time) to four (three in space and one in time). To simplify this, we adopt either a single threshold $\varepsilon_{1}$ applied to all dimensions, or two thresholds, $\varepsilon_{1s}$ in space, and $\varepsilon_{1t}$ in time. 
Although HOSVD increases the computational cost, its use provides a more optimal implementation of HODMD for multi-dimensional data \citep{LeClainche2017hodmdZNMF}.

\subsubsection{Step 2: Time-delay embedding}
\begin{figure}
    \centering
    \includegraphics[scale=1.1, keepaspectratio]{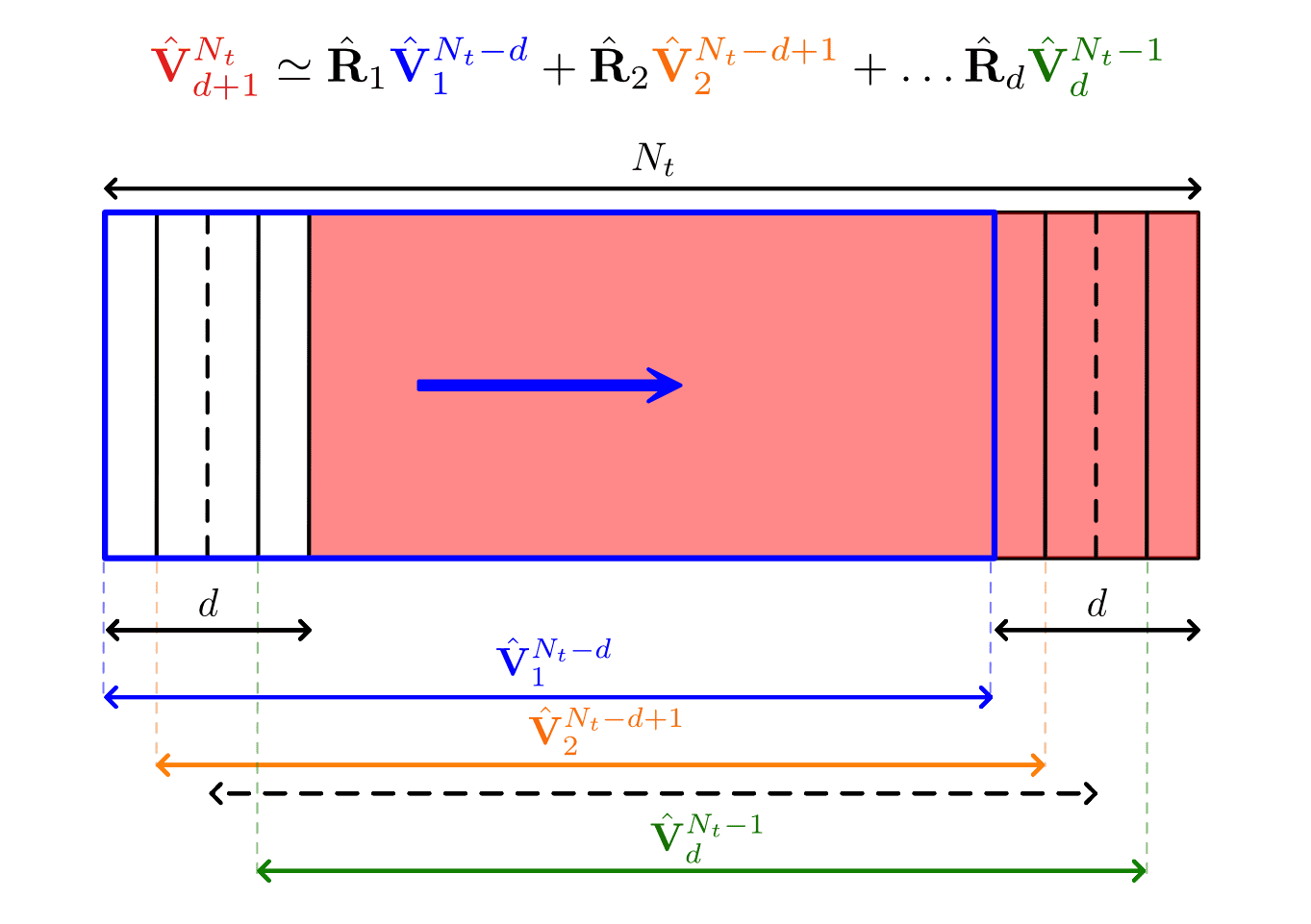}
    \caption{Sketch of the delay embedding. A window of length $N_t - d$ (blue outline) moves through the reduced snapshot matrix, producing the delayed matrix $\hat{\mathbf{V}}_{d+1}^{N_t}$ (red box), which is expressed as a linear superposition of the $d$ preceding reduced snapshot matrices.} 
    \label{fig:sketch-hodmd}
\end{figure}

Next, HODMD introduces the following assumption:
\begin{equation}
    \hat{\mathbf{V}}_{d+1}^{N_t} \simeq \hat{\mathbf{R}}_1 \hat{\mathbf{V}}_1^{N_t - d} + \hat{\mathbf{R}}_2 \hat{\mathbf{V}}_2^{N_t - d + 1} + \ldots + \hat{\mathbf{R}}_d \hat{\mathbf{V}}_d^{N_t - 1},
    \label{eq:b35}
\end{equation}
The operator $\hat{\mathbf{R}}$ maps the temporal evolution of the reduced observables onto an infinite-dimensional linear space \citep{Rowley2009Koopman, Mezic2013ARFM}. Standard DMD exploits this concept and HODMD refines it by incorporating time-lagged snapshots that generalise the original Koopman assumption from DMD to the high-order expression in eq.~(\ref{eq:b35}). 
Figure~\ref{fig:sketch-hodmd} illustrates this process, where a window of length $N_t - d$ slides through $\hat{\mathbf{V}}$. Note that HODMD reduces to standard DMD when $d = 1$ (see also in eq.~(\ref{eq:b35})).

Equation~(\ref{eq:b35}) can be recasted as:
\begin{equation}
    \tilde{\mathbf{V}}_2^{N_t - d + 1} = \tilde{\mathbf{R}} \tilde{\mathbf{V}}_1^{N_t - d}.
    \label{eq:b36}
\end{equation}
The matrices $\tilde{\mathbf{V}}$ have dimension $d N  \times \left( N_t - d\right)$, which can be large. However, the much larger spatial dimension underscores the importance of the dimensionality reduction step, which makes the implementation of the time-lagged snapshots feasible. 

Then, another step of SVD removes potential redundancies introduced in eq.~(\ref{eq:b35}):
\begin{equation}
    \tilde{\mathbf{V}}_1^{N_t - d + 1} \simeq \tilde{\mathbf{U}} \tilde{\mathbf{\Sigma}} \tilde{\mathbf{T}}^{\top} = \tilde{\mathbf{U}} \bar{\mathbf{T}}_1^{N_t-d+1}.
    \label{eq:b37}
\end{equation}
The number of modes is again truncated following $\varepsilon_1$, which determines the new spatial complexity $N^{\prime} \leq N$---if two thresholds are employed, $\varepsilon_{1t}$ is used instead. Lastly, pre-multiplying eq.~(\ref{eq:b36}) by $\tilde{\mathbf{U}}$ and invoking eq.~(\ref{eq:b37}) yields to:
\begin{equation}
    \bar{\mathbf{T}}_2^{N_t-d+1} = \bar{\mathbf{R}} \bar{\mathbf{T}}_1^{N_t-d},
    \label{eq:b38}
\end{equation}
where $\bar{\mathbf{R}} \simeq \tilde{\mathbf{U}}^{\top} \tilde{\mathbf{R}} \tilde{\mathbf{U}}$ has dimension $N^{\prime} \times N^{\prime}$.

\subsubsection{Step 3: Computation of the DMD modes}
The DMD routine is applied to eq.~(\ref{eq:b38}). The matrix $\bar{\mathbf{T}}_1^{N_t-d}$ is decomposed using SVD (no truncation needed):
\begin{equation}
    \bar{\mathbf{T}}_1^{N_t-d} = \mathbf{U} \Sigma \mathbf{V}^{\top}.
    \label{eq:b39}
\end{equation}
Next, $\bar{\mathbf{R}}$ is approximated such as:
\begin{equation}
    \bar{\mathbf{R}} \simeq \bar{\mathbf{T}}_2^{N_t-d+1} \mathbf{V} \mathbf{\Sigma} \mathbf{U}^{\top}.
    \label{eq:b40}
\end{equation}
The spectral properties of $\bar{\mathbf{R}}$, characterised by its $N^{\prime}$ eigenvalues $\mu_m$ and eigenvectors $\bar{\boldsymbol{q}}_m$, provide information about the dynamics of the reduced snapshots, 
that can be reconstructed as:
\begin{equation}
    \hat{\boldsymbol{v}}_k \simeq \sum_{m=1}^M a_m \boldsymbol{q_m} \mu_m^{k-1}, \quad k = 1, \dots, N_t,
    \label{eq:b41}
\end{equation}
with the frequencies $\omega_m$ and growth rates $\delta_m$ in eq.~(\ref{eq:b31}) given by:
\begin{equation}
    \delta_m + i \omega_m = {\rm log}\left( \mu_m \right) / \Delta t,
    \label{eq:b42}
\end{equation}
with $\Delta t$ the (constant) sampling rate between temporal snapshots. The amplitudes of the DMD modes are computed using least-squares fitting in eq.~(\ref{eq:b41}) \citep{chen2012optimdmd}. Finally, eq.~(\ref{eq:b31}) is recovered from multiplying eq.~(\ref{eq:b41}) by $\mathbf{U}$ and invoking eq.~(\ref{eq:b33}).

A second threshold $\varepsilon_2$ defines the number of modes $M$, or spectral complexity, in eq.~(\ref{eq:b31}), such that:
\begin{equation}
    a_{M+1} / a_1 \leq \varepsilon_2. 
    \label{eq:b43}
\end{equation}
By discarding low-amplitude modes, we obtain a sparse representation of the flow dynamics that only contains the most relevant structures. This indeed reduces the accuracy by neglecting small-scale features, but also prevents overfitting noise and spurious artefacts, thereby improving the generalisability of the reconstruction.

\subsection{Spatio-temporal Koopman decomposition}\label{subsec:stkd} 
The interpretability of the modes computed with HODMD can be improved by further decomposing them in space. This is the basis for the spatio-temporal Koopman decomposition (STKD) \citep{LeClainche2018STKDJNonlinSci}, that applies HODMD sequentially in time and space for approximating nonlinear data as a linear superposition of periodic or quasi-periodic standing and/or travelling waves. 

To do so, HODMD is applied individually to each DMD mode in eq.~\ref{eq:b31}. For example, applying it in the spanwise direction yields:
\begin{equation}
    \boldsymbol{u}_m \left( x, y, z_r \right) \simeq \sum_{n = 1}^{N} a_n \hat{\boldsymbol{u}}_{mn} \left( x, y \right) e^{\left( \nu_{mn} i + \alpha_{mn} \right) z_r}, \quad r = 1, \ldots, N_z,
    \label{eq:b44}
\end{equation}
with $\kappa_{mn}$ and $\alpha_{mn}$ the spanwise wavenumber and the growth rate, respectively, associated to each spanwise-temporal mode. 

Finally, combining eq.~(\ref{eq:b44}) with eq.~(\ref{eq:b31}) gives: 
\begin{equation}
    \boldsymbol{v} \left( x, y, z_r, t_k \right) \simeq \sum_{m, n = 1}^{M, N} a_{mn} \hat{\boldsymbol{u}}_{mn} \left( x, y \right) e^{\left( \delta_m + \omega_m i \right) t_k + \left( \nu_{mn} + \alpha_{mn} i \right) z_r}, k = 1, \ldots, N_t,\ r = 1, \ldots, N_z.
    \label{eq:b45}
\end{equation}
As a result, the data are now described as a superposition of travelling waves with spanwise-temporal amplitude $a_{mn} = a_m a_n$, whose phase velocity is defined as $c_{mn} = \omega_m / \alpha_{mn}$.

\section{Global analysis}\label{sec:res}
\subsection{Data visualization}\label{subsec:datvis}

\begin{figure}
    \centering
     \includegraphics[width=\textwidth, keepaspectratio]{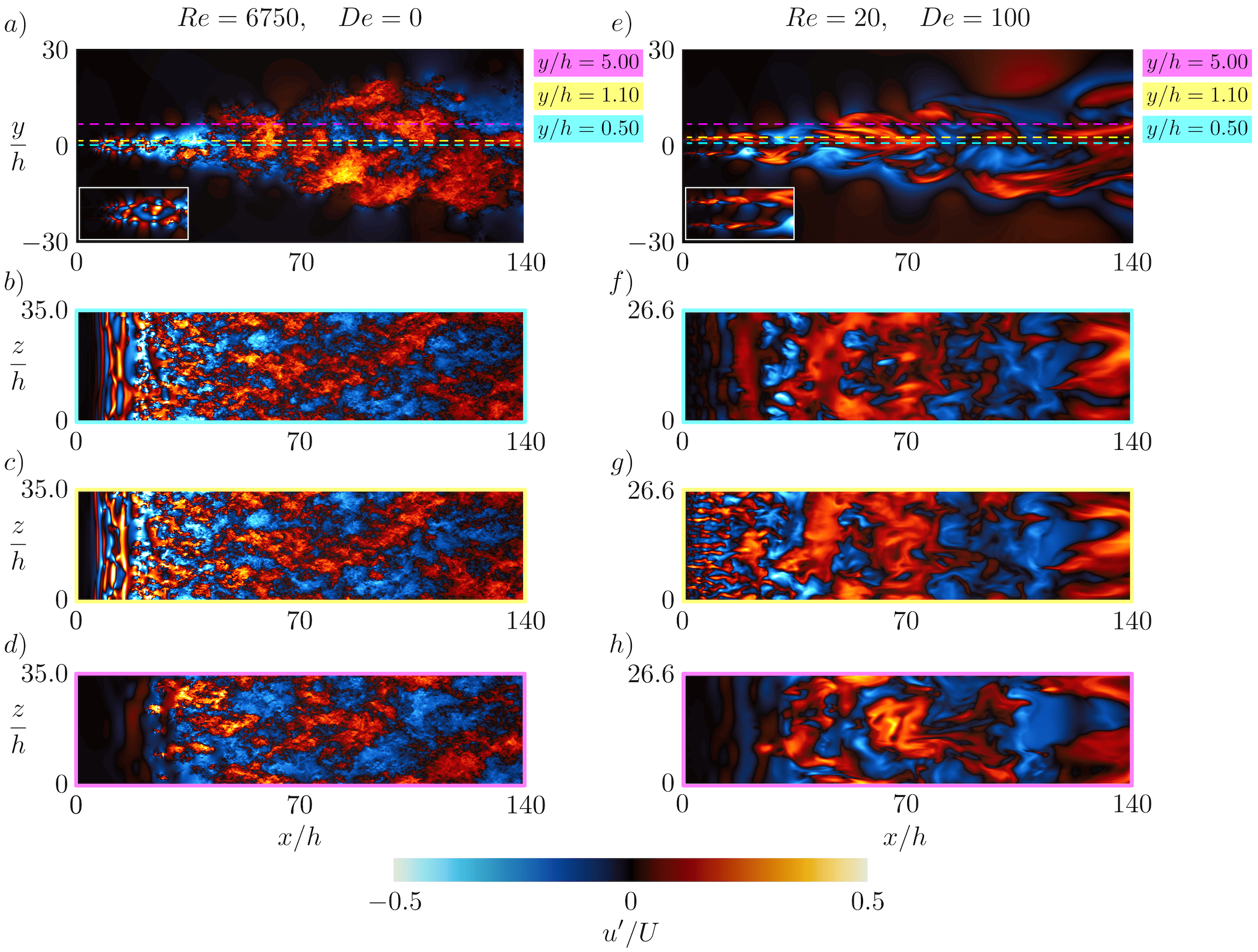}
    \caption{Instantaneous fields of the streamwise fluctuating velocity $u^{\prime}$ for the Newtonian (left) and the viscoelastic (right) turbulent planar jets. Panels \textit{a} and \textit{e} show an $xy$-plane extracted at $z = 0$. Insets provide a closer view of the near-field up to \reva{$x/h \approx 20$}. The dashed coloured lines indicate the $xz$-planes shown in panels \textit{b, c} and \textit{d} for the Newtonian and \textit{f, g} and \textit{h} for the viscoelastic jets.}
    \label{fig:new_vjet}
\end{figure}
Let us start by visually comparing the flow of the Newtonian and viscoelastic jets. Figure~\ref{fig:new_vjet} shows an instantaneous field of the streamwise velocity fluctuation $u^{\prime}$ with values comparable in the two cases despite the large difference in Reynolds number. Panels \textit{a} and \textit{e} are side views of the jets at $z = 0$, and we can observe that the far-field region is dominated by large-size structures in both jets. Here, while small-scale turbulence is also visible in the Newtonian jet, structures are smoother in the viscoelastic jet, with little evidence of small-scale fluctuations. \revb{This is consistent with the turbulent kinetic energy decaying significantly faster across scales in elastic turbulence than in Newtonian turbulence, with a signature energy spectrum decay in wavenumber of $k^{-4}$ compared to the Kolmogorov one $k^{-5/3}$ \citep{Singh2024NatCommET}. It is worth mentioning that, despite the exponent $-4$ is also reported in viscoelastic planar jets at low Reynolds numbers \citep{Rota2026IJMFTaylor}, the energy spectrum in frequency decays as $\omega^{-3}$ \citep{yamani2021spectral, yamani2023spatiotemporal, soligo2023non}; the difference between scaling exponents is a consequence of Taylor's hypothesis not holding in elastic turbulence \citep{Rota2026IJMFTaylor}}. The more dominant presence of large-size structures in the viscoelastic jet also influences the near-field. Since the disturbance generated by the stronger coherent fluctuations can propagate upstream along the jet edges without being disrupted by the small-scale features, they reach even the proximity of the inlet, strongly affecting the onset of turbulence. This is not visible in the Newtonian case, where the small scale fluctuations rapidly destroy their coherency and prevents large-scale noise to propagate upstream.

Studying the near-field structures in more detail reveals the different nature of the flow instability (see insets in panels \textit{a} and \textit{e}). In the Newtonian planar jet, the flow is laminar with zero fluctuations close to the inlet, which start amplifying from around \reva{$x/h \approx 5$}. Here, alternating positive-negative $u^{\prime}$ lobes align along the upper and lower shear layers, consistently with Kelvin-Helmholtz rollers whose symmetry about the centreline matches the varicose (symmetric) instability, that dominates at high Reynolds numbers \citep{Antonia1983OrganizedJet, Thomas1986DevelopJet, deo2008influence, Soligo2025JFMNewtonJet}. The viscoelastic jet shows a notably different scenario. Streamwise velocity fluctuations rapidly appear in the form of streamwise-elongated structures along the shear layers of the potential core. This feature differs from the classical Kelvin-Helmholtz rollers, and indicates an alternative amplification pathway for the fluctuations, in which elasticity triggers an earlier transition---at lower Reynolds number and closer to the inlet---than in the Newtonian case at much higher Reynolds. \revc{The flow in both cases becomes unstable at larger distances from the inlet compared with some other works found in literature (see for instance \cite{Silva2002JFMInflow, Stanley2002JFMInflow, guimaraes2020direct}), since we do not add any disturbance at the inlet to induce an early transition.}

Next, we focus on top-views of the jet, with streamwise-spanwise planes extracted at three heights from the centreline, corresponding in the near-field to the jet core (panels \textit{b} and \textit{f}), above the shear layer (panels \textit{c} and \textit{g}), and in the outer flow (panels \textit{d} and \textit{h}). In the near-field of the Newtonian jet, spanwise-homogeneous Kelvin-Helmholtz rollers populate the shear layer, which destabilise and break-up moving downstream. In the far-field, streamwise-elongated structures appear (see for instance panel \textit{b}), reminiscent of the streaky structures reported by \cite{nogueira2019streaks} in round jets. These structures grow in size downstream as smaller structures merge. In the viscoelastic jet, roller-like structures are also present within the jet core (panel \textit{f}), but their amplitude is much smaller than in the Newtonian case. In contrast, short and thin streaky structures populate the near-field region along the shear layer (panel \textit{g}). They remain coherent at the potential core before they get disrupted by the large-scale fluctuations propagating upstream. These near-field streaks point toward a different, elasticity-driven pathway to turbulence in viscoelastic turbulent planar jets at low Reynolds numbers. Further downstream, the field is dominated by large-size, spanwise-coherent structures, with the flow remaining predominantly homogeneous in the spanwise direction and lacking the streamwise-elongated structures observed in the core of the Newtonian jet.

In the following sections, we aim to characterise the different structures observed here qualitatively, and investigate their dynamics and interactions.

\subsection{Flow structures}\label{subsec:flowstr} 
\begin{figure}
    \centering
    \includegraphics[width=\textwidth, keepaspectratio, trim=0 30 0 0, clip]{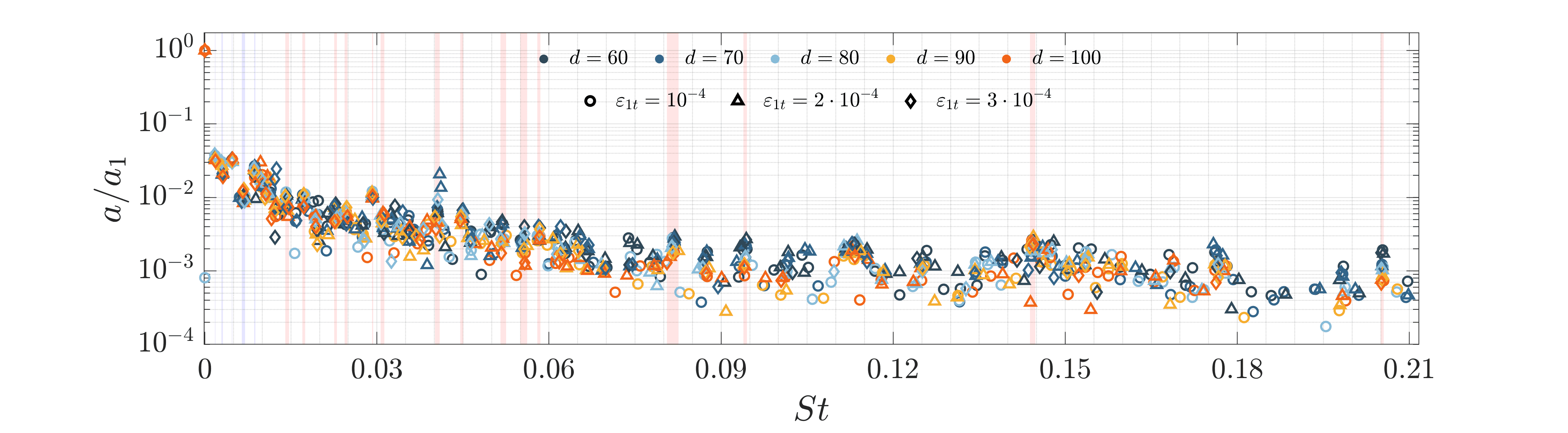}
    \put(-212,92){\revb{\textit{Newtonian}}}
    \vspace{+0.2cm}
    \includegraphics[width=\textwidth, keepaspectratio, trim=0 30 0 0, clip]{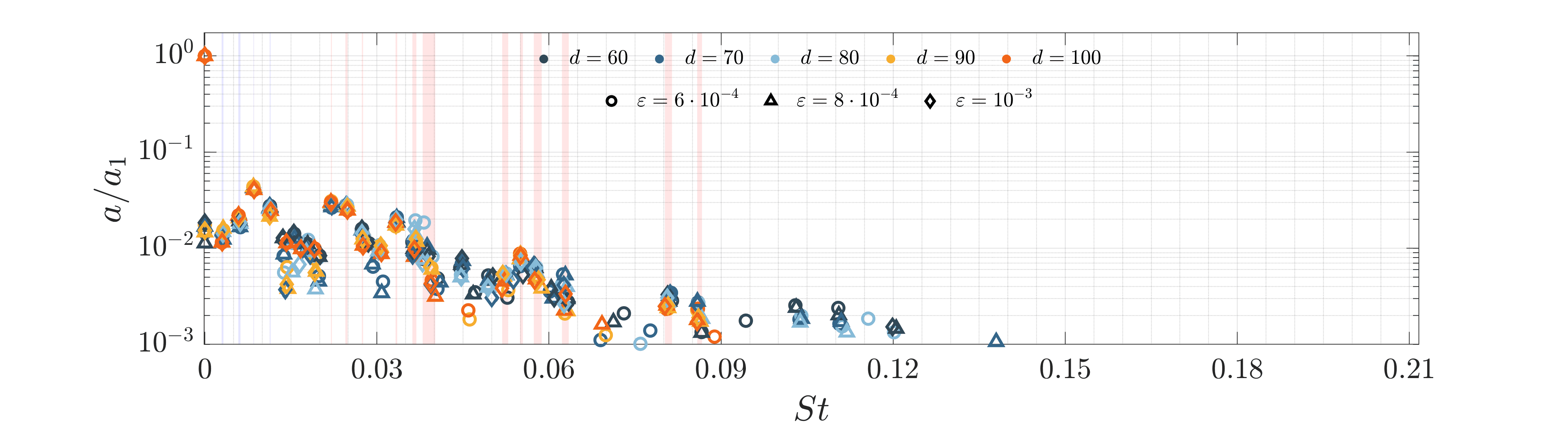}
    \put(-212,92){\revb{\textit{Viscoelastic}}}
    \vspace{+0.2cm}
    \includegraphics[width=\textwidth, keepaspectratio]{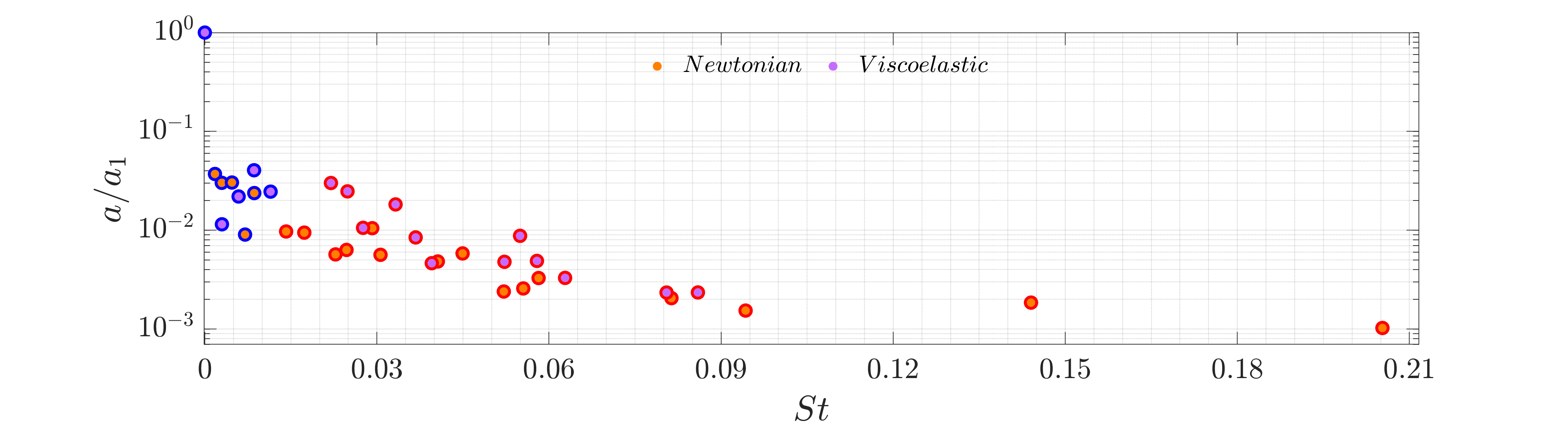}
    \put(-214,104){\revb{\textit{Robust modes}}}
    \caption{HODMD spectra overlapped for multiple calibrations for the Newtonian (upper panel) and viscoelastic (middle panel) jets. Non-dimensional frequency or Strouhal number, $St = f h / U$, are compared to the normalised amplitude, $a / a_1$, being $a_1$ the largest amplitude in each series. Shaded areas indicate robust modes: blue refers to low-frequency modes or streamwise-elongated structures, while red denotes high-frequency modes or spanwise-coherent structures. The thickness of each bar matches the maximum deviation of the frequency in each cluster. Robust modes from both jets are compared in the lower panel (marker outlines are colour-coded similar to the shaded areas).} 
    \label{fig:hodmd-spectra}
\end{figure}
We begin the characterisation of the global coherent structures in the Newtonian and viscoelastic jets using sequential HODMD or STKD. As a first step, data are decomposed in time to compute the temporal coherent structures. The wide range of temporal (and spatial) scales in both jets complicates the identification of flow patterns, so HODMD must be carefully calibrated to ensure physically meaningful results. Here, our interest is finding the large-amplitude modes associated with the dominant large-scale structures in the flow, which must be distinguished from the fairly large number of frequencies present in the data. To this end, we apply HODMD recursively with different combinations of $\varepsilon_1$, $\varepsilon_2$ and $d$. Modes that are robust across calibrations, i.e., their frequency appears consistently regardless of parameter choice, are considered physical. In contrast, spurious modes arising from the mathematical decomposition tend to be scattered throughout the spectrum without clustering at particular frequencies.

This step is computationally expensive due to the recursive application of HODMD, so data are pre-processed to reduce their size. First, data are cropped in the streamwise and jet-normal directions, since the original simulations are performed in a much larger domain to avoid confinement effects. The cropped sub-domains have dimensions $L_x = 140h$ and $L_y = 60h$; dimensions were chosen such that each box encloses a similar portion of the domain, which contains the jet and the fully-developed flow. Second, the data are uniformly downsampled in all spatial directions by a factor of two. After this reduction, the Newtonian jet data has dimensions $480 \times 207 \times 120$, and the viscoelastic one $337 \times 146 \times 64$. Therefore, \reva{the HODMD is performed to the flow within the reduced box having same lengths in the streamwise and jet-normal directions, and extending over the full span. The analysis is performed} using a statistically stationary data set consisting of $436$ snapshots for the Newtonian case and $300$ for the viscoelastic case, that are sampled at the interval of $\Delta t U / h = 2$ time units, that is sufficient for resolving the dynamics of the large-scale coherent structures in both jets. 

The HODMD calibration is summarised in fig.~\ref{fig:hodmd-spectra}. In the viscoelastic jet, we set $\varepsilon_1 = \varepsilon_2 = \varepsilon$, whereas in the Newtonian one we split the first threshold in space and time, $\varepsilon_{1s}$ and $\varepsilon_{1t}$, owing to the larger spatial complexity of the data. The calibration is then performed using $\varepsilon = 6 \cdot 10^{-4}, 8 \cdot 10^{-4}$ and $10^{-3}$ in the viscoelastic jet, and $\varepsilon_{1t} = 10^{-4}, 2 \cdot 10^{-4}$ and $3 \cdot 10^{-4}$, and $\varepsilon_{1s} = 2 \cdot 10^{-3}$ in the Newtonian one, with $d$ set to $60, 70, 80, 90$ and $100$ in both cases. We define two criteria for identifying robust modes: the first one over the frequency (the $\omega$-criterion) and the second one over the spatial shape of the DMD modes (the $u$-criterion). A mode with frequency $\omega_m$ fulfils the $\omega$-criterion if $|\omega_{m_i} - \omega_{m}| \leq \epsilon$, with $\epsilon = 5 \cdot 10^{-3}$. If this criterion is fulfilled in $75\%$ of the calibrations, modes are deemed common. Then, common modes are promoted to robust if their spatial shape $\boldsymbol{u}_m$ is similar across calibrations, i.e., they fulfil the $u$-criterion. This is evaluated computing the cosine similarity between DMD modes: ${\rm cos} ( \boldsymbol{u}_{m_i}, \boldsymbol{u}_{m_j} ) \geq \zeta$, with $\zeta = 0.8$. We acknowledge robustness if $\zeta$ is close to one, i.e., both modes are identical pairs; we request the $u$-criterion to be fulfilled in $75\%$ of the calibrations. In doing this, we find twenty-one modes in the Newtonian jet and sixteen modes in the viscoelastic jet that are robust (the same number of modes were deemed common). Moreover, the amplitude across calibrations within each robust cluster is comparable (the deviation is of the order of $10^{-2}$ at most) indicating that HODMD assigns a similar weight notwithstanding the choice of parameters. 

In the following, we show the results for the set of parameters $d = 80$, $\varepsilon_{1s} = 2 \cdot 10^{-3}$, and $\varepsilon_{1t} = \varepsilon_2 = 2 \cdot 10^{-4}$ in the temporal analysis of the Newtonian jet, and $d = 100$ and $\varepsilon_1 = \varepsilon_2 = 6 \cdot 10^{-4}$ in the viscoelastic one. We chose the calibration that retrieved the largest number of robust modes, though a simpler criterion could be choosing an intermediate $d$ and the lowest thresholds $\varepsilon_1$ and $\varepsilon_2$. The lower panel in fig.~\ref{fig:hodmd-spectra} shows that the robust modes are predominantly at low and intermediate frequencies, with only a few high-frequency modes present in the Newtonian jet. In both jets, the dominant part of the spectra is shifted toward lower frequencies (modes are computed globally, so the spectra reflect the dominance of the low-frequency structures due to energy distribution). The Newtonian spectrum, however, exhibits greater complexity. Smaller turbulent scales are hindered in the viscoelastic jet due to the low Reynolds number, while higher frequencies are more important for reconstructing the turbulent dynamics in the high-Reynolds number Newtonian jet. On the other hand, the spectrum of the viscoelastic jet is dominated by more pronounced peaks, which resemble those reported by \cite{suresh2008reynolds} in the power spectra of low-Reynolds, Newtonian jets. In their case, the peaks corresponded with sub-harmonics of the fundamental vortex formation frequency; here, these are roughly harmonics of the dominant frequency $St \simeq 0.008$ (see for instance $St \simeq 0.023$ and $0.056$).  

\begin{figure}
    \centering
    \includegraphics[scale=0.07, keepaspectratio]{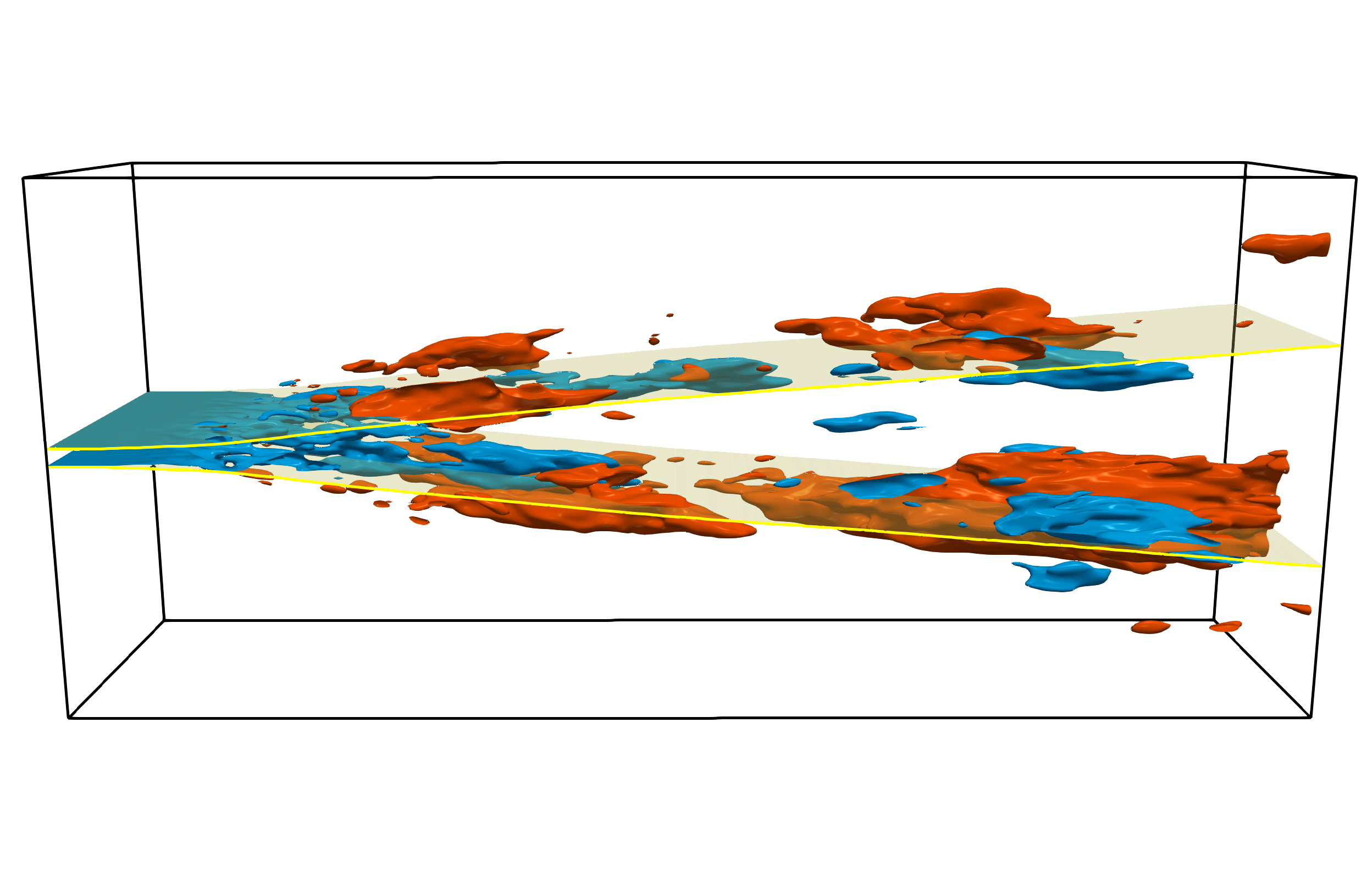}
    \put(-184,87){\small $30$}
    \put(-177,52){\small $0$}
    \put(-190,52){\large $\frac{y}{h}$}
    \put(-184,18){\small $-30$}
    \put(-169,12){\small $0$}
    \put(-94,12){\small $70$}
    \put(-15,12){\small $140$}
    \put(-185,98){\small $a)$}
    \put(-110,98){\textcolor{blue}{$St \simeq 0.002$}}
    \put(-115,115){\large \textit{Newtonian}}
    \hspace{0.5cm}
    \includegraphics[scale=0.07, keepaspectratio]{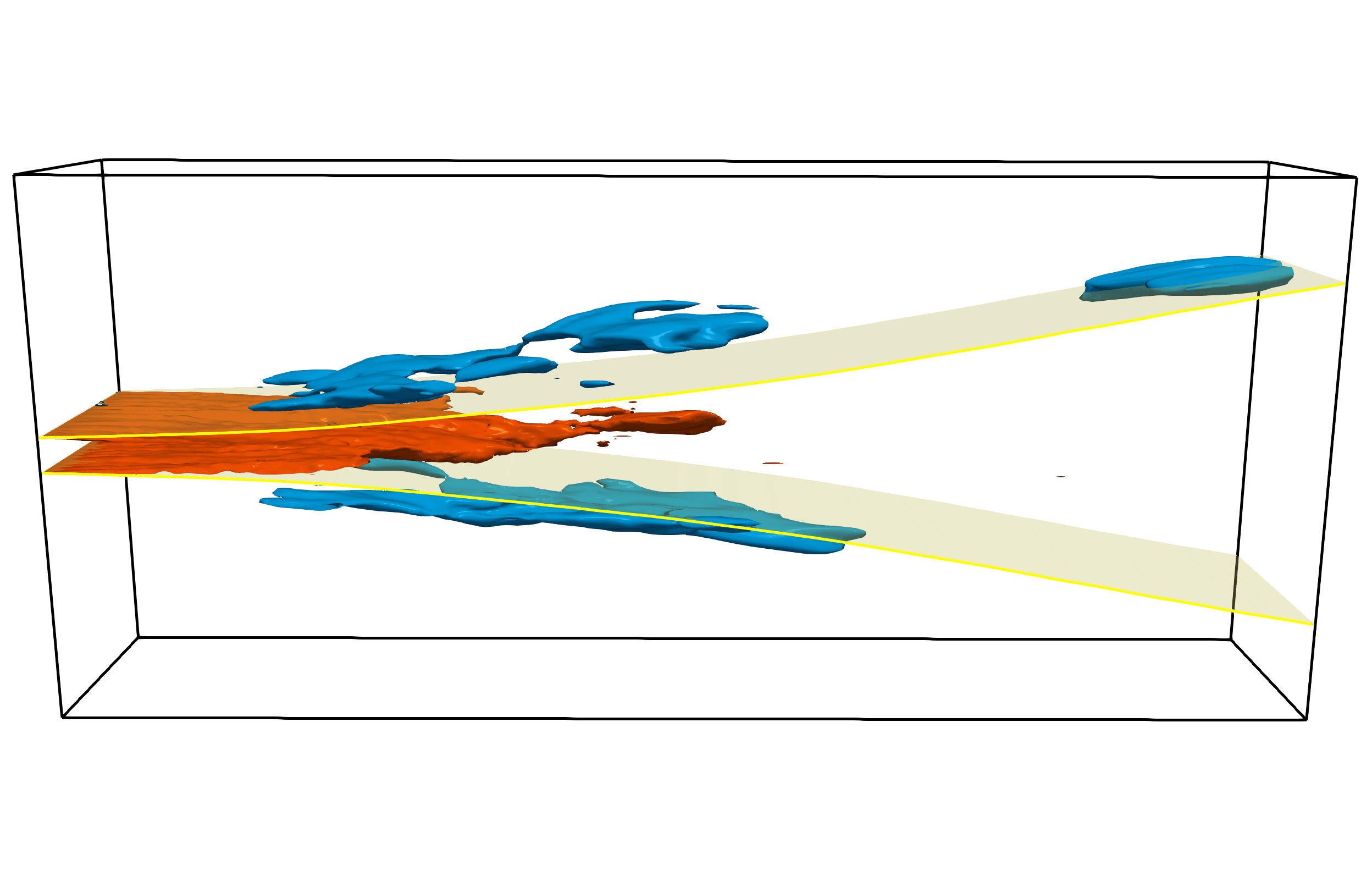}
    \put(-185,87){\small $30$}
    \put(-178,52){\small $0$}
    \put(-185,18){\small $-30$}
    \put(-169,12){\small $0$}
    \put(-94,12){\small $70$}
    \put(-20,12){\small $140$}
    \put(-185,98){\small $f)$}
    \put(-110,98){\textcolor{blue}{$St \simeq 0.003$}}
    \put(-120,115){\large \textit{Viscoelastic}}
    \vspace{-0.5cm}
    \includegraphics[scale=0.07, keepaspectratio]{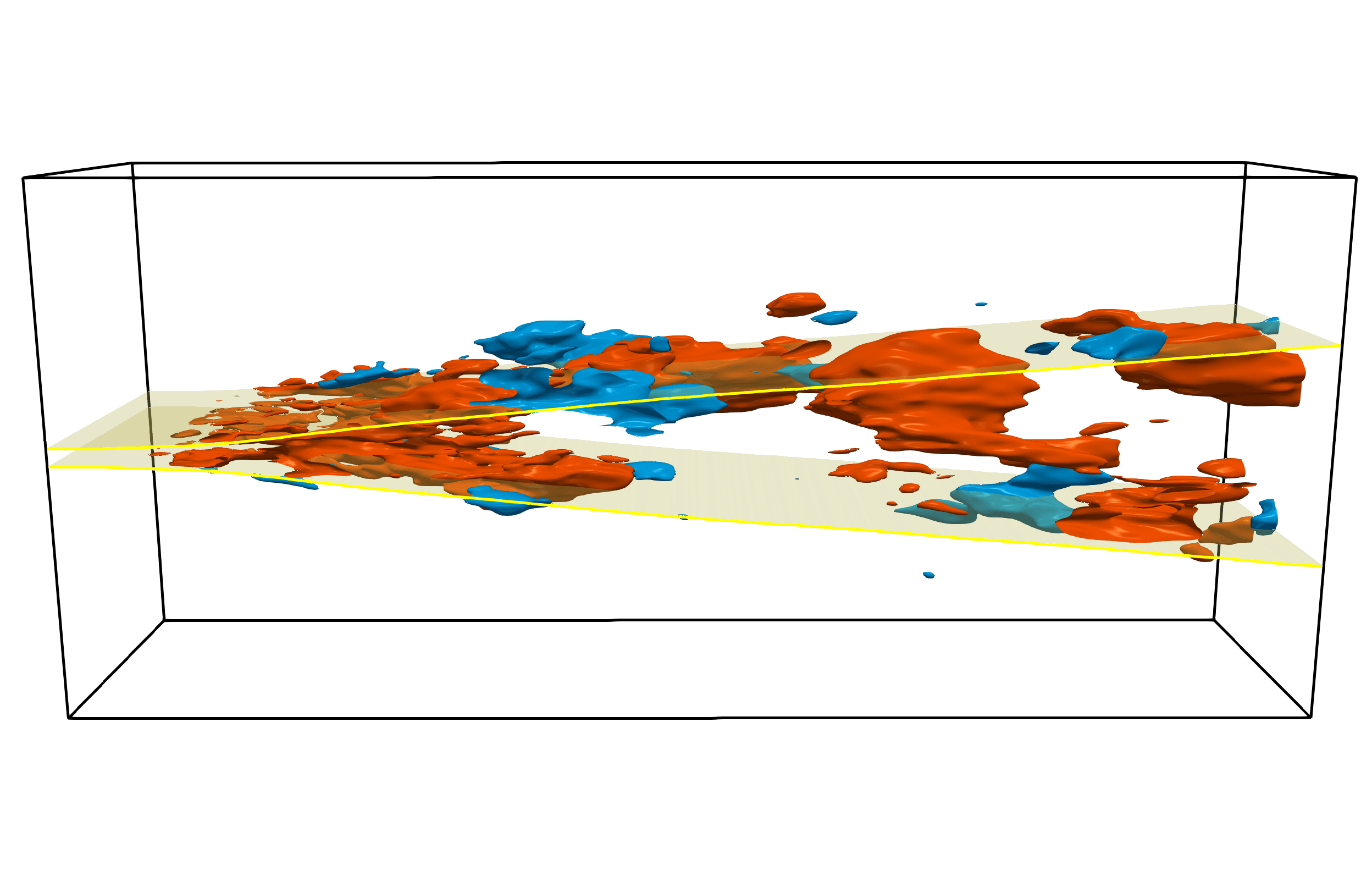}
    \put(-184,87){\small $30$}
    \put(-177,52){\small $0$}
    \put(-190,52){\large $\frac{y}{h}$}
    \put(-184,18){\small $-30$}
    \put(-169,12){\small $0$}
    \put(-94,12){\small $70$}
    \put(-15,12){\small $140$}
    \put(-185,98){\small $b)$}
    \put(-110,98){\textcolor{blue}{$St \simeq 0.005$}}
    \hspace{0.5cm}
    \includegraphics[scale=0.07, keepaspectratio]{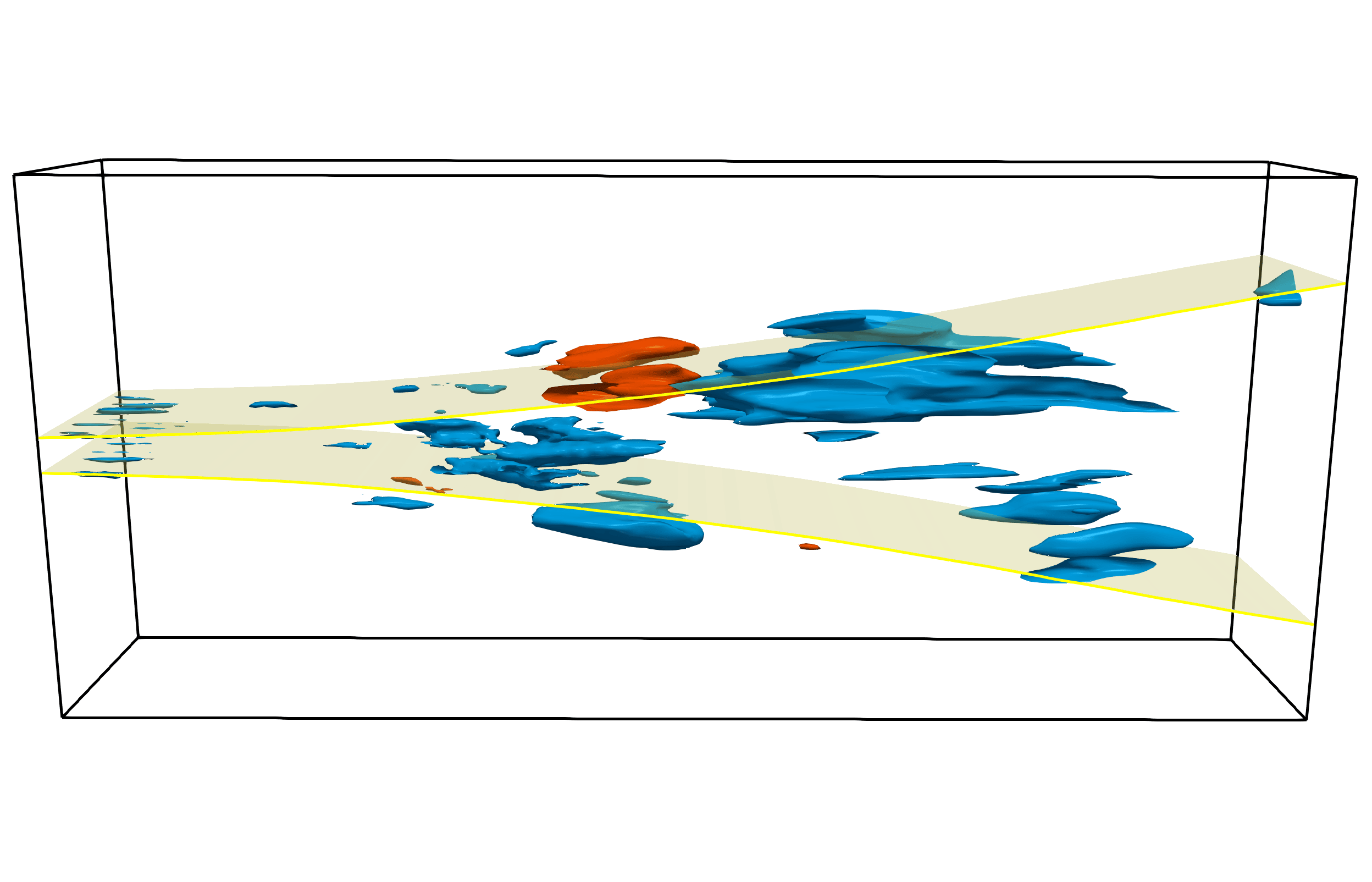}
    \put(-185,87){\small $30$}
    \put(-178,52){\small $0$}
    \put(-185,18){\small $-30$}
    \put(-169,12){\small $0$}
    \put(-94,12){\small $70$}
    \put(-20,12){\small $140$}
    \put(-185,98){\small $g)$}
    \put(-110,98){\textcolor{blue}{$St \simeq 0.008$}}
    \vspace{-0.5cm}
    \includegraphics[scale=0.07, keepaspectratio]{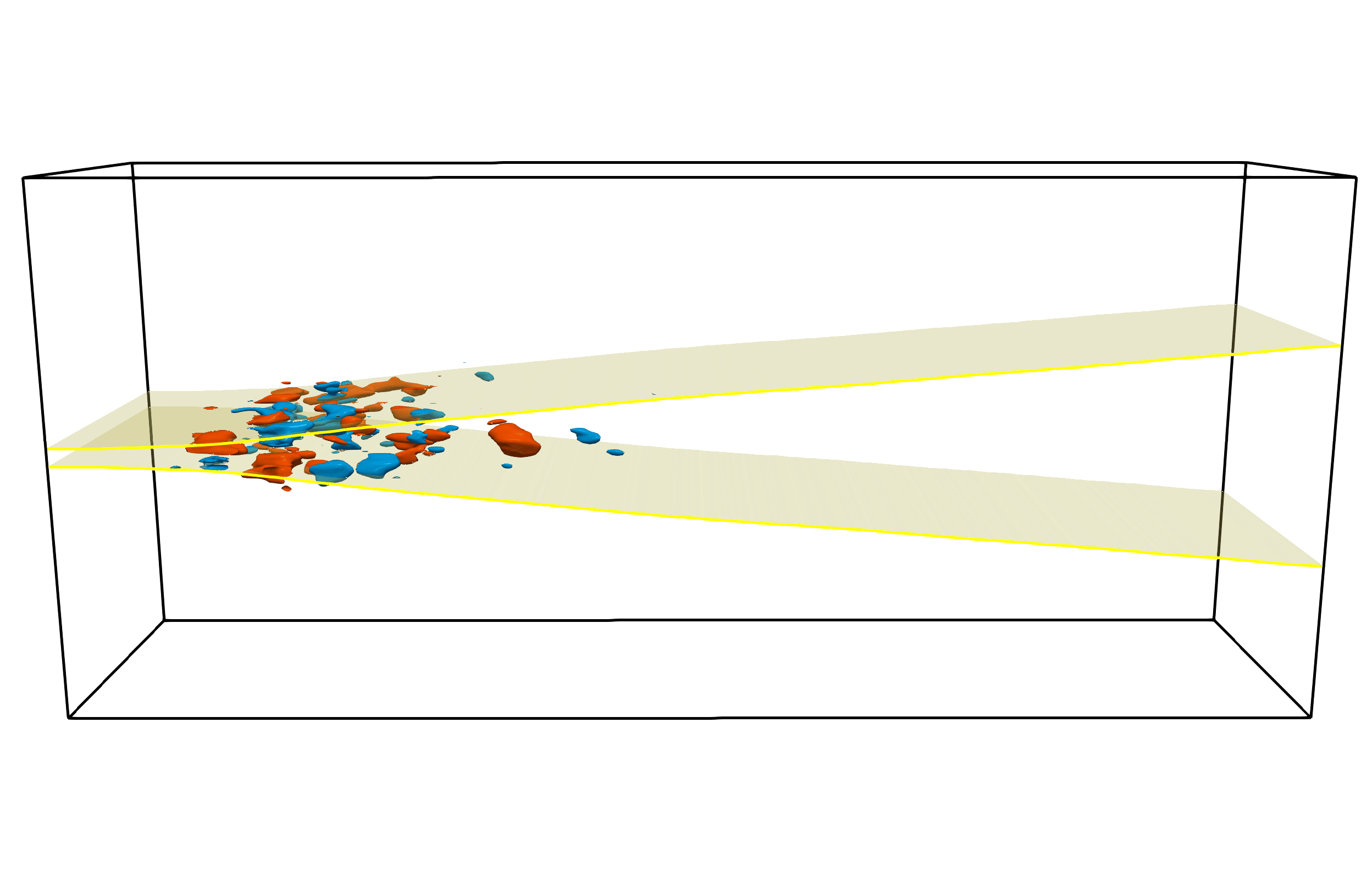}
    \put(-184,87){\small $30$}
    \put(-177,52){\small $0$}
    \put(-190,52){\large $\frac{y}{h}$}
    \put(-184,18){\small $-30$}
    \put(-169,12){\small $0$}
    \put(-94,12){\small $70$}
    \put(-15,12){\small $140$}
    \put(-185,98){\small $c)$}
    \put(-110,98){\textcolor{red}{$St \simeq 0.030$}}
    \hspace{0.5cm}
    \includegraphics[scale=0.07, keepaspectratio]{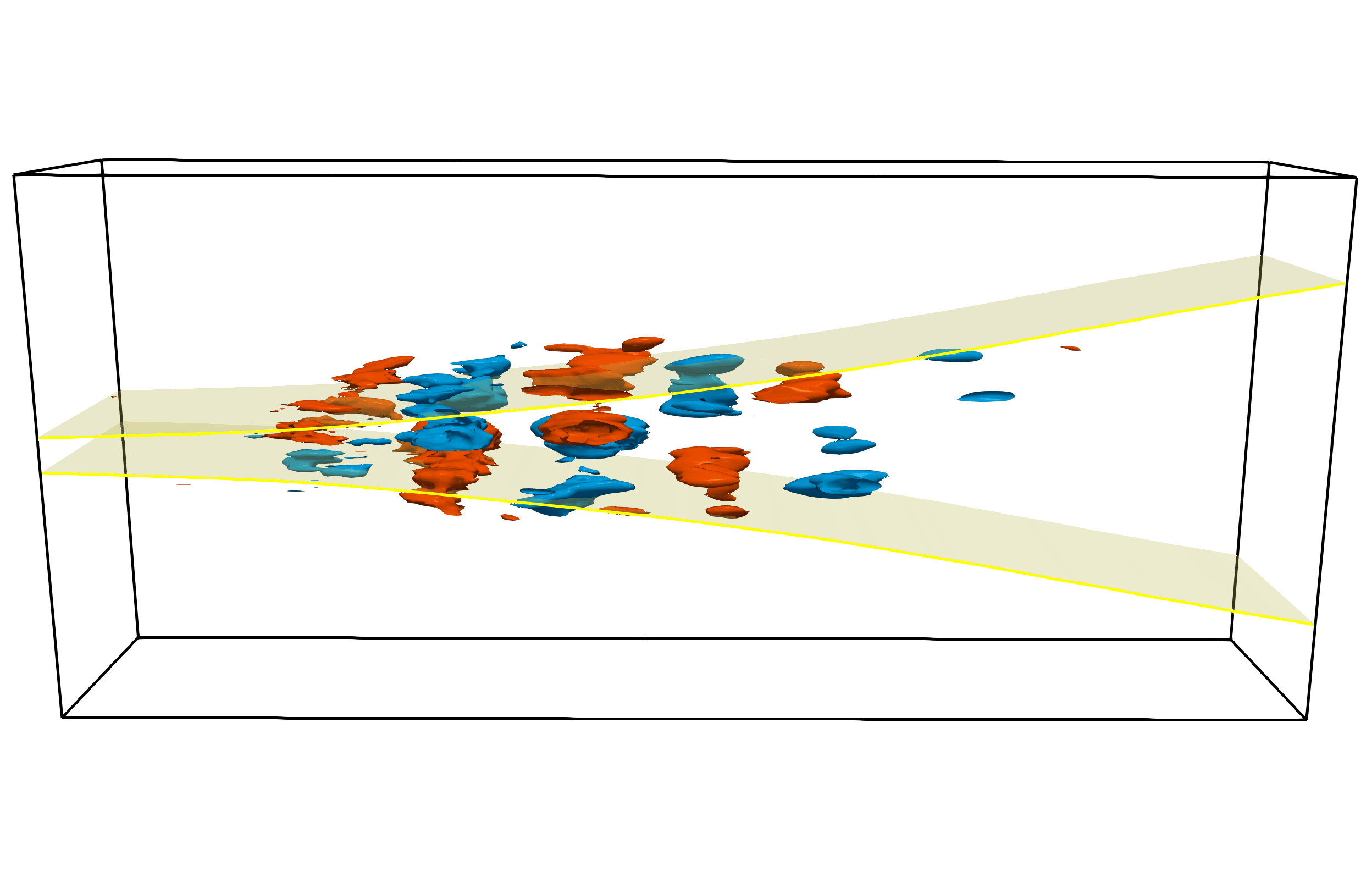}
    \put(-185,87){\small $30$}
    \put(-178,52){\small $0$}
    \put(-185,18){\small $-30$}
    \put(-169,12){\small $0$}
    \put(-94,12){\small $70$}
    \put(-20,12){\small $140$}
    \put(-185,98){\small $h)$}
    \put(-110,98){\textcolor{red}{$St \simeq 0.023$}}
    \vspace{-0.5cm}
    \includegraphics[scale=0.07, keepaspectratio]{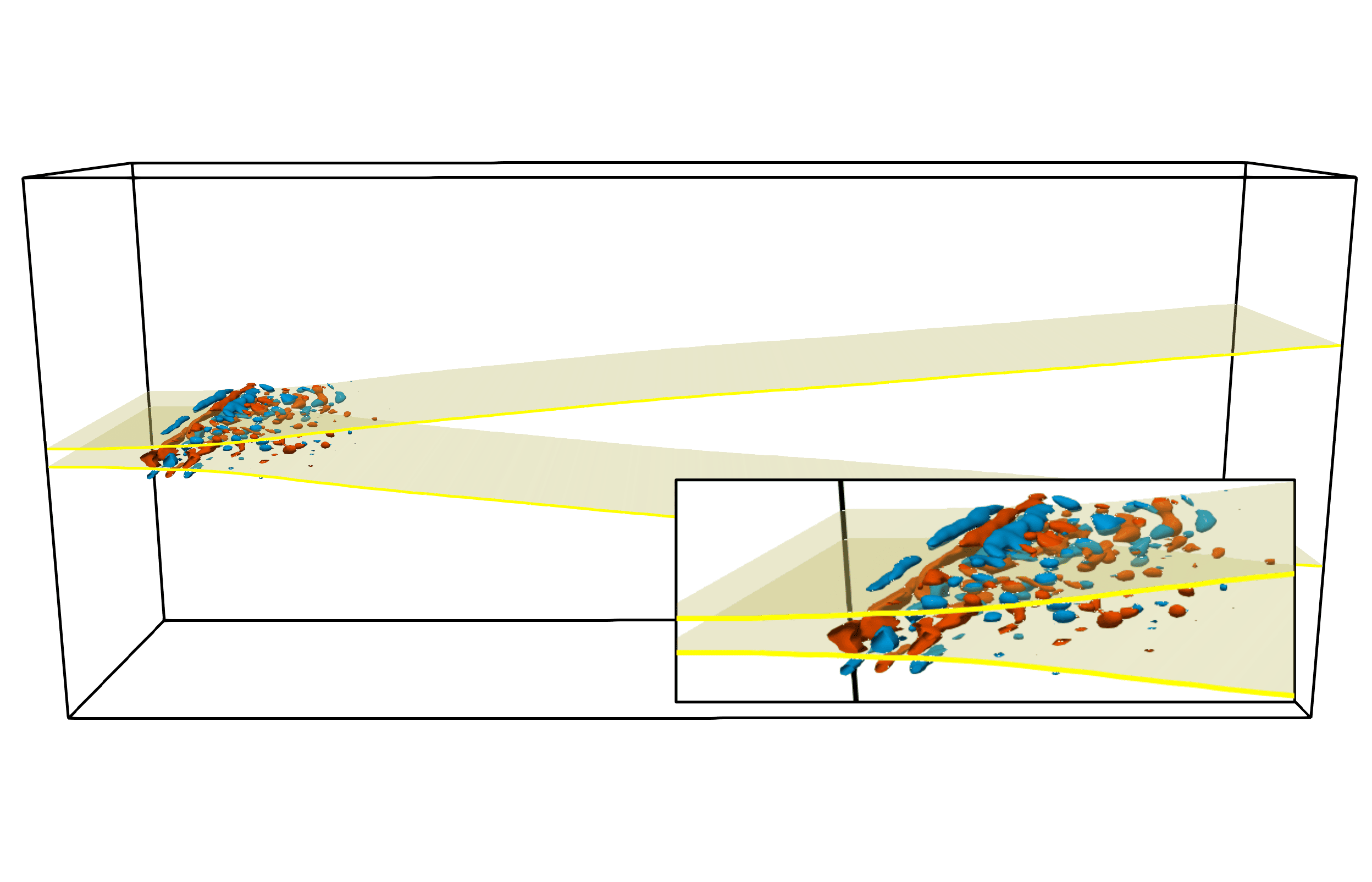}
    \put(-184,87){\small $30$}
    \put(-177,52){\small $0$}
    \put(-190,52){\large $\frac{y}{h}$}
    \put(-184,18){\small $-30$}
    \put(-169,12){\small $0$}
    \put(-94,12){\small $70$}
    \put(-15,12){\small $140$}
    \put(-185,98){\small $d)$}
    \put(-110,98){\textcolor{red}{$St \simeq 0.144$}}
    \hspace{0.5cm}
    \includegraphics[scale=0.07, keepaspectratio]{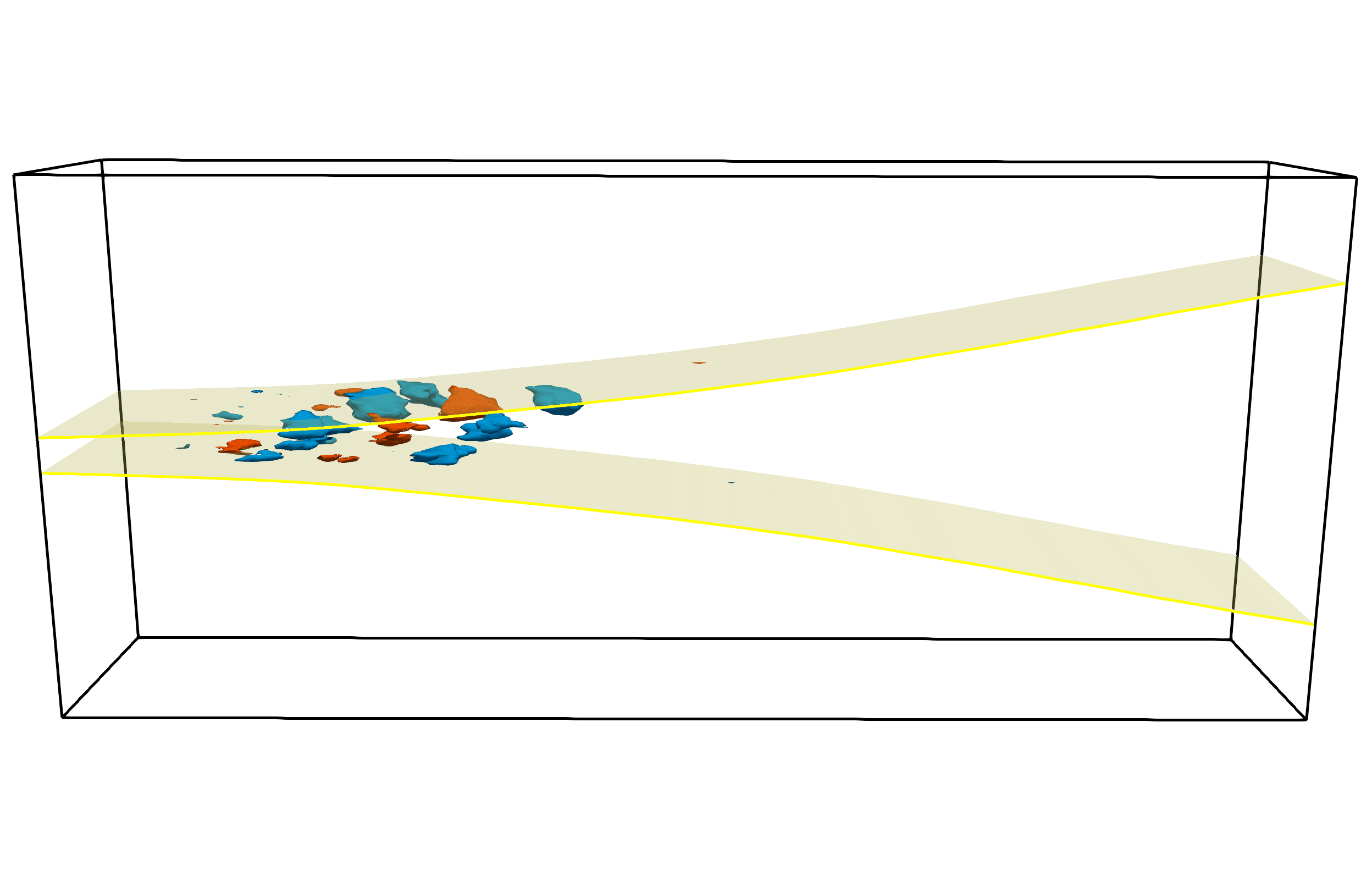}
    \put(-185,87){\small $30$}
    \put(-178,52){\small $0$}
    \put(-185,18){\small $-30$}
    \put(-169,12){\small $0$}
    \put(-94,12){\small $70$}
    \put(-20,12){\small $140$}
    \put(-185,98){\small $i)$}
    \put(-110,98){\textcolor{red}{$St \simeq 0.056$}}
    \vspace{-0.5cm}
    \includegraphics[scale=0.07, keepaspectratio]{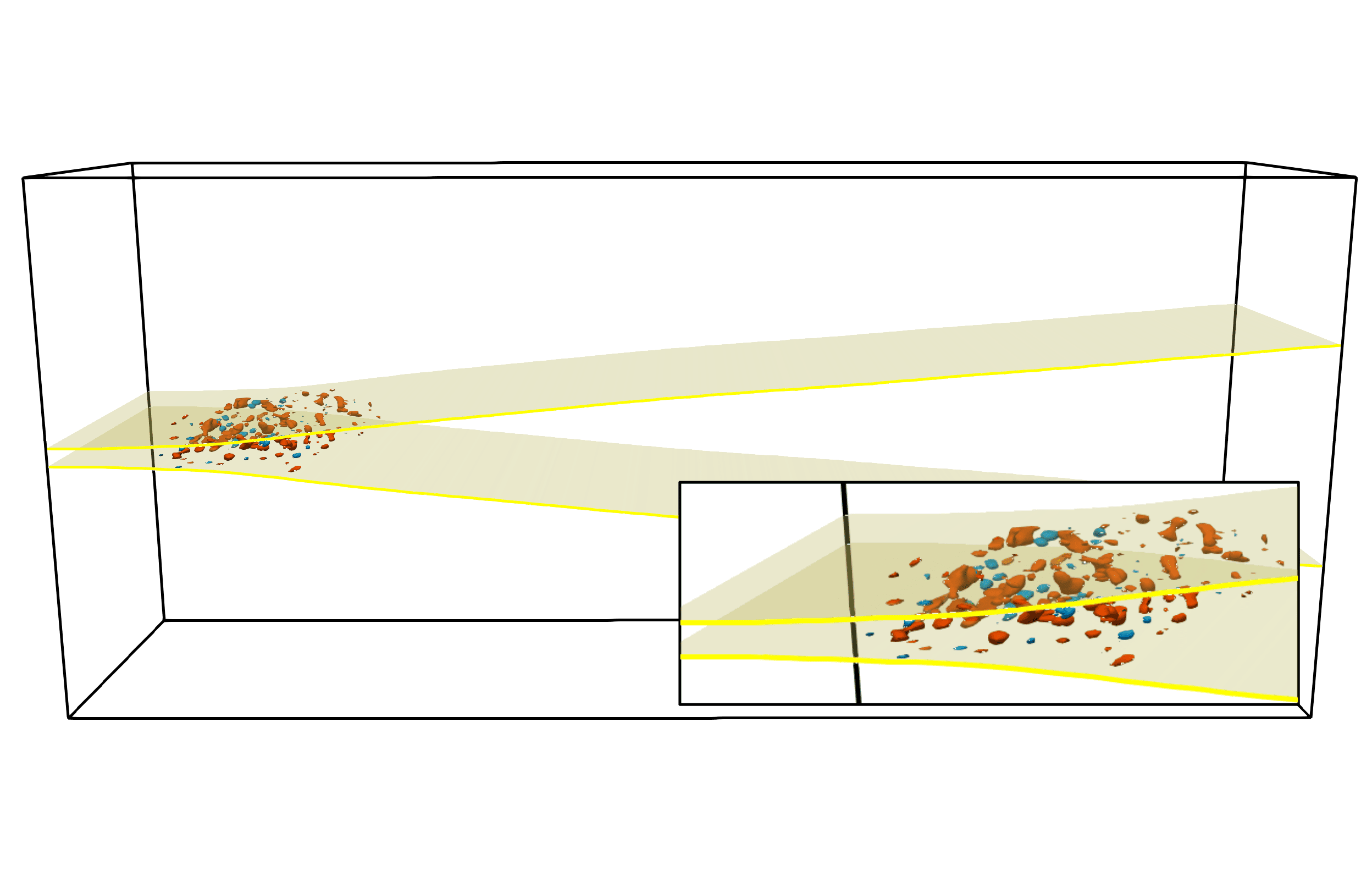}
    \put(-184,87){\small $30$}
    \put(-177,52){\small $0$}
    \put(-190,52){\large $\frac{y}{h}$}
    \put(-184,18){\small $-30$}
    \put(-169,12){\small $0$}
    \put(-94,12){\small $70$}
    \put(-15,12){\small $140$}
    \put(-96,0){$x/h$}
    \put(-185,98){\small $e)$}
    \put(-110,98){\textcolor{red}{$St \simeq 0.206$}}
    \hspace{0.5cm}
    \includegraphics[scale=0.07, keepaspectratio]{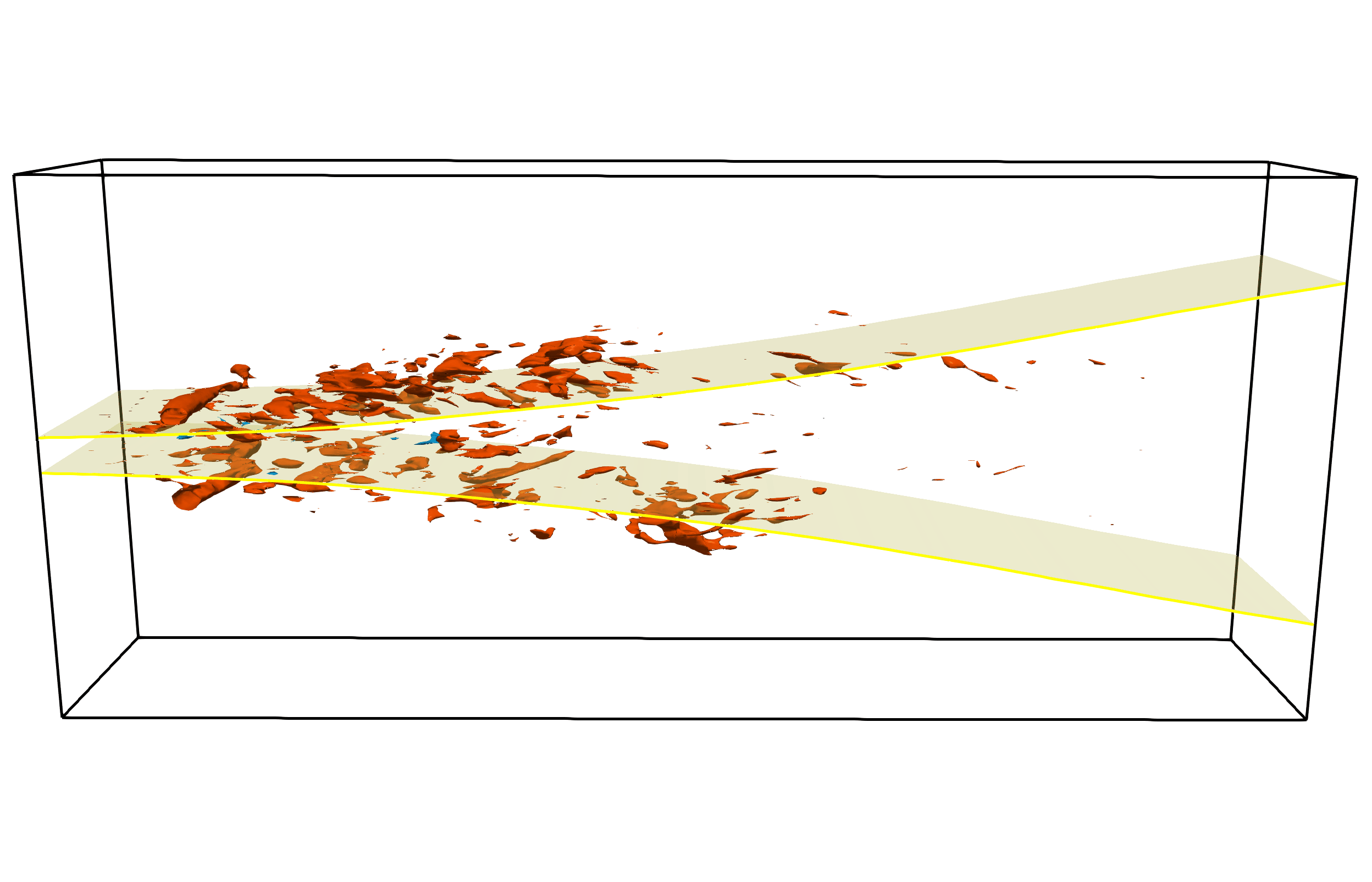}
    \put(-185,87){\small $30$}
    \put(-178,52){\small $0$}
    \put(-185,18){\small $-30$}
    \put(-169,12){\small $0$}
    \put(-94,12){\small $70$}
    \put(-20,12){\small $140$}
    \put(-96,0){$x/h$}
    \put(-185,98){\small $j)$}
    \put(-110,98){\textcolor{red}{$St \simeq 0.086$}}
    \caption{Spatial structure of robust DMD modes in the Newtonian and viscoelastic jets. Three-dimensional iso-surfaces represent the normalised streamwise velocity for magnitudes $+0.5$ (red) and $-0.5$ (blue). of the near-field up to \reva{$x/h \approx 30$}. The yellow semi-transparent surfaces mark the average jet thickness. Labels are colour-coded similar to the shaded areas in fig.~\ref{fig:hodmd-spectra} (blue for low-frequency or streamwise-elongated modes, and red for high-frequency or spanwise-coherent modes).}
    \label{fig:hodmd-modes}
\end{figure}
We next visualise the three-dimensional spatial shape $\boldsymbol{u}_m$ of a few relevant robust modes in the Newtonian and viscoelastic jets in fig.~\ref{fig:hodmd-modes}. Flow structures are broadly similar across low to intermediate frequencies between the cases: low-frequency modes (blue labels) represent streamwise-elongated structures, while higher-frequency modes show spanwise coherency (red labels). In the Newtonian jet, the lowest frequency (panel \textit{a}) indicates streamwise-elongated structures that emerge just downstream the potential core. These structures move along the edges of the jet, computed as the distance from the centreline where the local streamwise velocity equals half the value at the centreline, and they grow with downstream distance. In the viscoelastic jet, the slowest mode (panel \textit{f}) also shows streamwise-elongated structures, though they are more localised compared to the same structures in the Newtonian case, which extend globally and have a dominant presence in the far-field. Consistent with this picture, higher low-frequency modes (panels \textit{b} and \textit{g}) also capture the evolution of the streamwise-elongated structures. In the viscoelastic jet, this mode additionally highlights the short streaks in the potential core, similar than in fig.~\ref{fig:new_vjet}\textit{g}.

Higher frequencies show spanwise-coherent structures. In all cases, flow structures are confined rather than expanding globally like the low-frequency ones, and they are located symmetrically above and below the centreline of the jet. The shape and frequency of the modes at intermediate values (panels \textit{c} in the Newtonian jet and \textit{h} and \textit{i} in the viscoelastic one) match the description of Orr wave packets \citep{schmidt2018KHOrr}. At an even higher frequency, structures are displaced to the proximity of the inlet. In the Newtonian jet, these have enhanced spanwise-homogeneity, for instance at $St \simeq 0.144$ (panel \textit{d}) structures are located at the core and edges of the jet, the later resembling the Kelvin-Helmholtz rollers. The spatial arrangement of the mode, symmetric about the centreline, and their frequency, very close to that reported experimentally by \cite{deo2008influence} for high Reynolds number Newtonian jets, is consistent with the symmetric (varicose) mode, also highlighted by \cite{Soligo2025JFMNewtonJet} for the same database. It is noteworthy that, despite the method was applied globally, HODMD is also able to resolve the smaller-scale structures near the inlet. The method is also able to reconstruct very-high-frequency modes (panel \textit{e}), that arise from non-linear interactions between high-frequency modes at the near-field region. Similar structures are also highlighted at $St \simeq 0.086$ in the viscoelastic jet (panel \textit{j}). In this case, spanwise-homogeneous rollers appear along the jet edges at much smaller frequency compared to the Newtonian jet. \revb{Despite the roll-up of the shear layer is significantly different between the Newtonian and viscoelastic jets (see insets in panels $a$ and $e$ in fig. \ref{fig:new_vjet}), the shape and location of both structures at high frequency (spanwise-homogeneous and along the edges) point to Kelvin-Helmholtz rollers, though induced by a mechanism different than the one from Newtonian jets \citep{Guimaraes2023PRFJetInst}}. The same mode in the viscoelastic jet also represents small-scale structures along the jet edges downstream the potential core, indicating the breakdown of the upstream rollers.

\subsection{Spatio-temporal structures} \label{subsec:stkdglobal}
The temporal analysis revealed the presence of similar structures in both planar jets: streamwise-elongated at low-frequency, and spanwise-coherent at higher-frequency. To complete their description, we now decompose each temporal DMD mode in the spanwise direction. For this analysis, we employ $120$ snapshots, equidistant along the spanwise direction ($\Delta z / h \simeq 0.29$), in the Newtonian planar jet, and $64$ snapshots, also sampled uniformly in spanwise ($\Delta z / h \simeq 0.42$), in the viscoelastic one, with thresholds $\varepsilon_1 = \varepsilon_2 = 10^{-3}$ and $d = 1$ in both cases. The decomposition yields an expansion of spatial modes that depend on the streamwise and jet-normal directions, and their superposition reconstructs each temporal DMD mode in terms of steady and travelling waves. 

\begin{figure}
    \centering
    \includegraphics[scale=0.4, keepaspectratio]{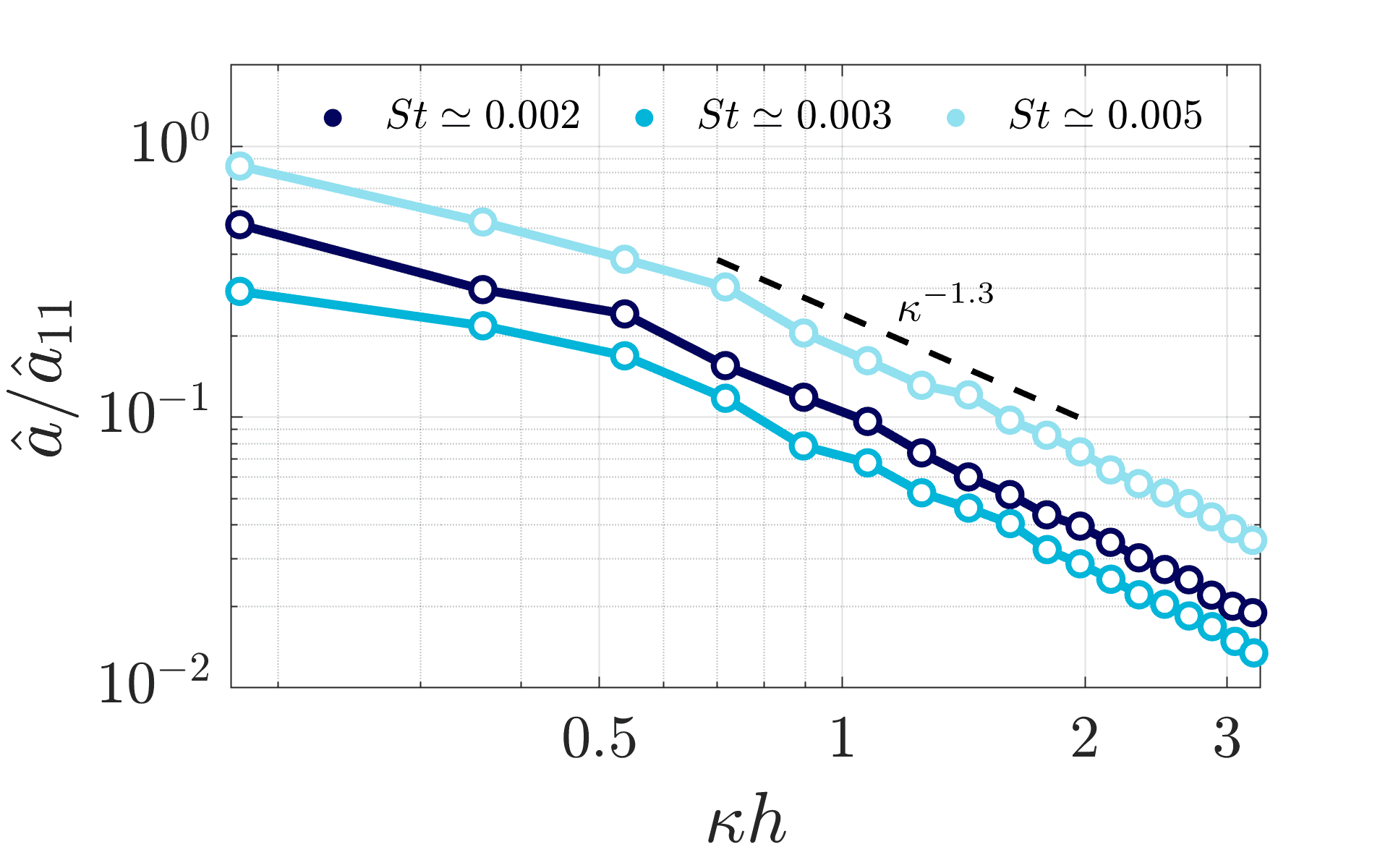}
    \put(-185,110){\revaa{\small $a)$}}
    \put(-110,115){\textit{Newtonian}}
    \includegraphics[scale=0.4, keepaspectratio]{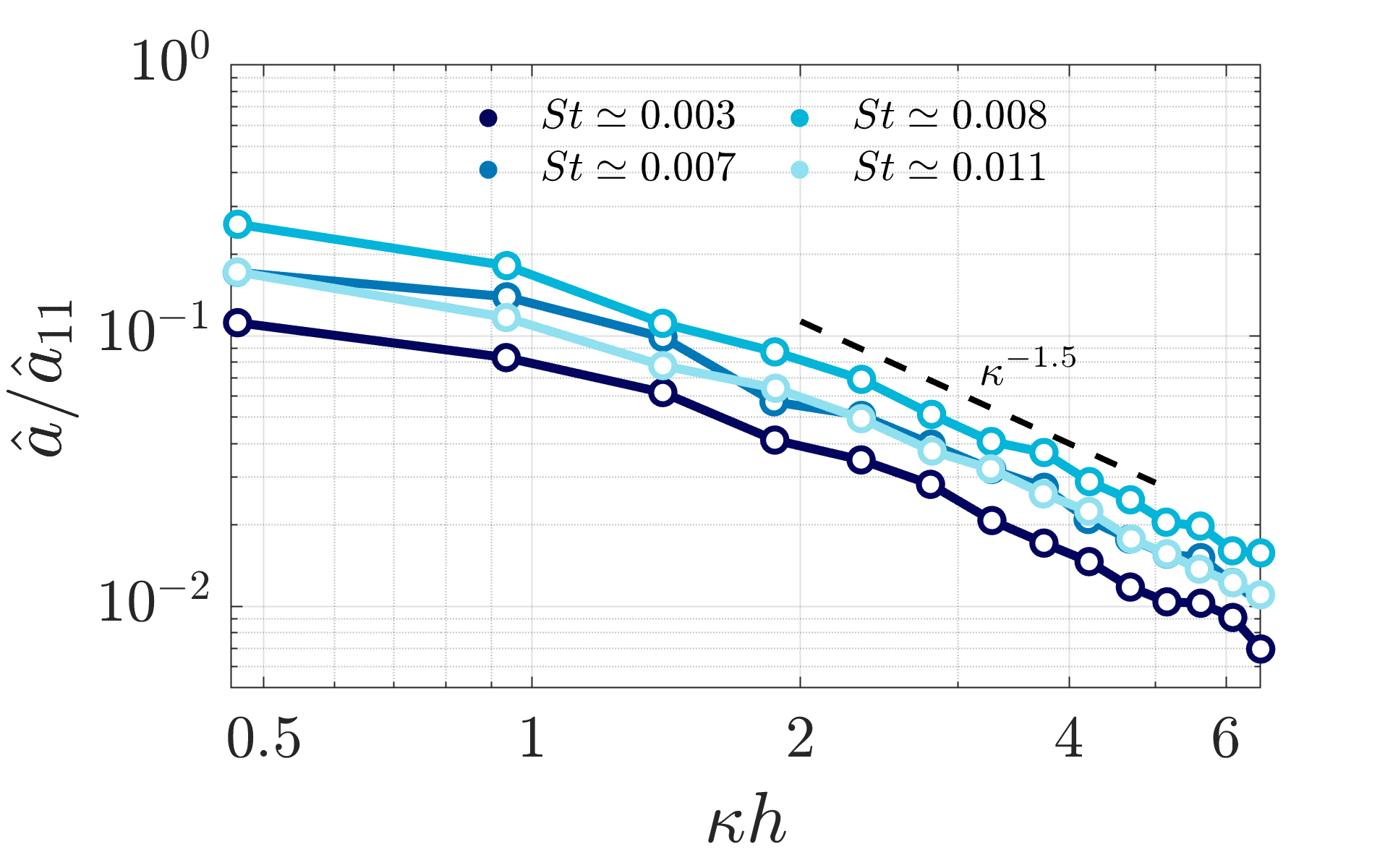}
    \put(-185,110){\revaa{\small $c)$}}
    \put(-110,115){\textit{Viscoelastic}}

    \includegraphics[scale=0.4, keepaspectratio]{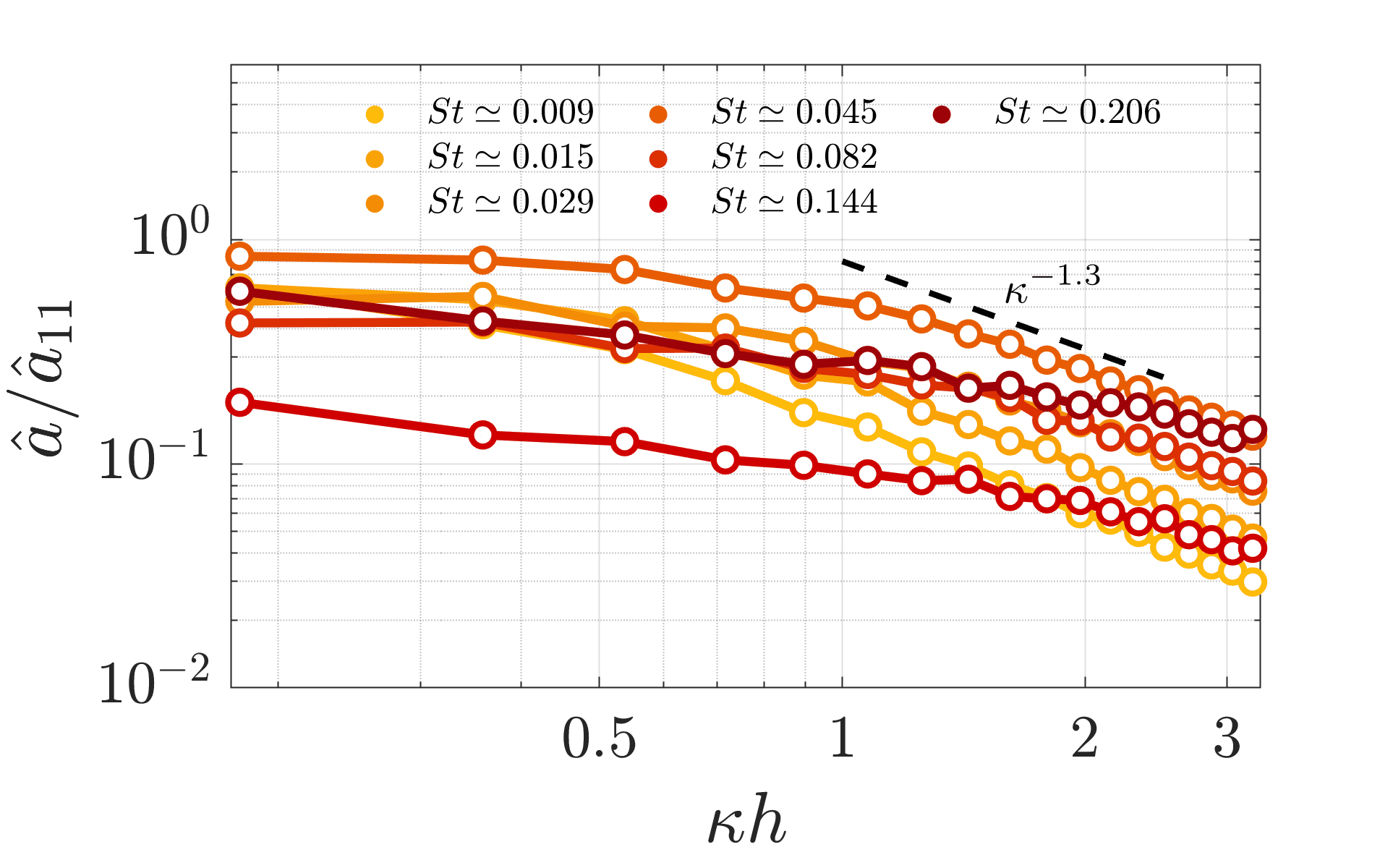}
    \put(-185,110){\revaa{\small $b)$}}
    \includegraphics[scale=0.4, keepaspectratio]{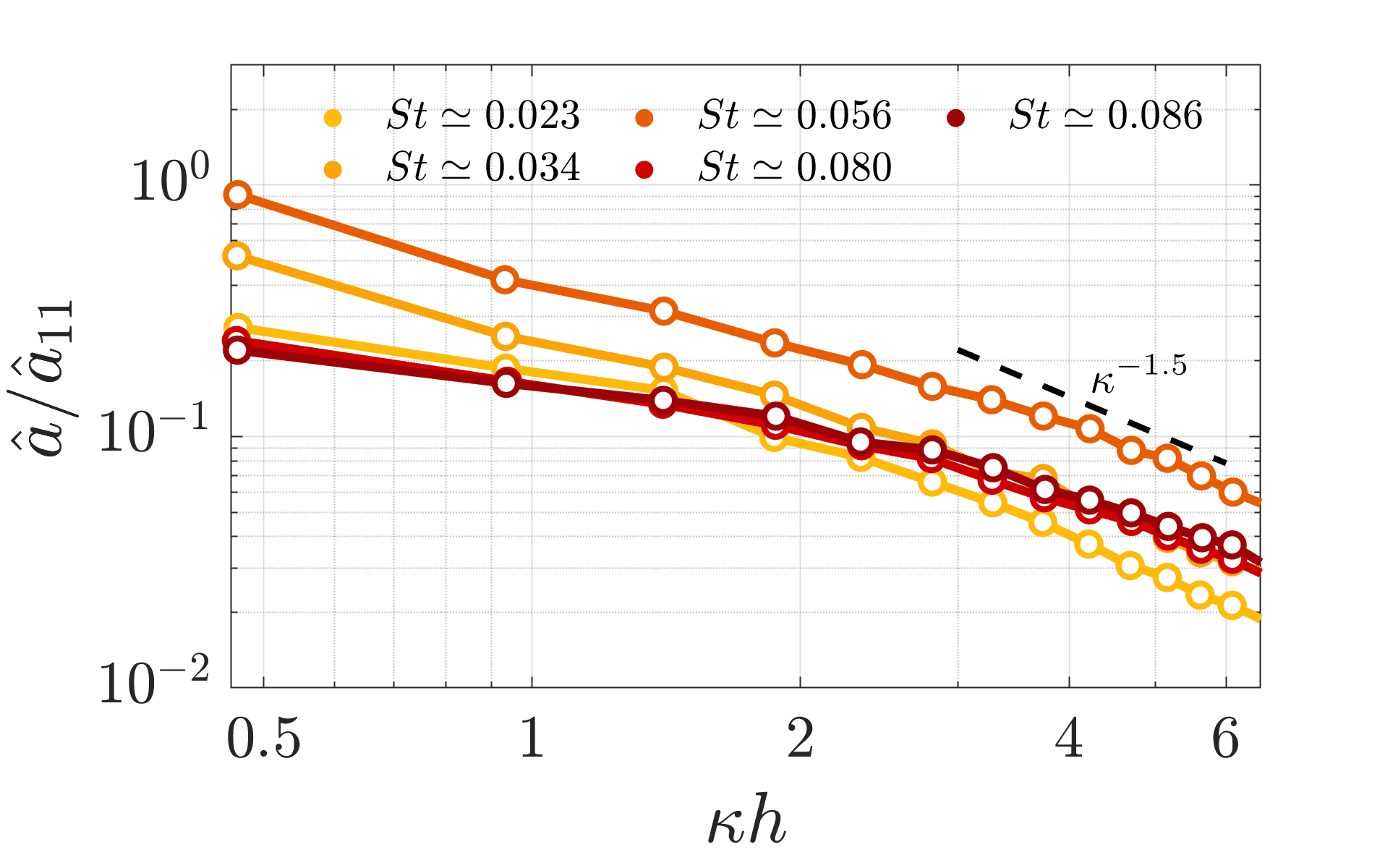}
    \put(-185,110){\revaa{\small $d)$}}
    \caption{\revc{Spatio-temporal spectra}. \revb{Normalised} spanwise wavenumber, $\kappa h$, compared to their normalised spatio-temporal amplitude, $\hat{a} / \hat{a}_{11}$, with $\hat{a}_{11}$ the largest amplitude in each series. Panels \textit{a} and \textit{c} correspond to low-frequency modes, while high-frequency modes are shown in \textit{b} and \textit{d}. \revc{A power-law decay of the normalized amplitude is suggested in each subpanel for high wavenumbers.}} \label{fig:stkd-spectrum}
\end{figure}
We show the spatial spectra for several robust frequencies in fig.~\ref{fig:stkd-spectrum}. Each amplitude quantifies the relative contribution of the different spanwise harmonics associated with each temporal mode. The smallest wavenumber is set by the spanwise domain length, $L_z$, such that $\kappa_{\min} = 2\pi/L_z$. In most cases, the spatio-temporal amplitudes decrease monotonically with the wavenumber, with the lowest non-zero wavenumber ($\kappa_1 = \kappa_{\min}$) carrying the largest contribution. The only exception are $St \simeq 0.029, 0.082$ in the Newtonian jet, where the second harmonic ($\kappa_2 = 2 \kappa_1 = 2 \kappa_{min}$) dominates. \reva{Such decay is similar notwithstanding the frequency, distinction made to the very high-frequency modes ($St \simeq 0.144$ and $0.206$ in the Newtonian jet, and $St \simeq 0.080$ and $0.086$ in the viscoelastic jet), whose normalized amplitude decays at a smaller rate. We also noticed two different decay rates, whether wavenumbers are above or below a threshold: $\kappa \approx 0.7$ at low frequency (\textit{a}) and $\kappa \approx 1$ at high frequency (\textit{b}) in the Newtonian jet, and $\kappa \approx 2$ at low frequency (\textit{c}) and $\kappa \approx 3$ at high frequency (\textit{d}) in the viscoelastic jet. The steeper decay of the normalized amplitudes in the viscoelastic jet highlights the greater complexity of the flow in the Newtonian jet.}

\revb{Another sign of complexity is the low-rank behaviour. A large separation between the amplitudes at $\kappa = 0$ and $\kappa > 0$ indicates that the spatial mode with the largest amplitude dominates at a given frequency. This separation is more pronounced in the viscoelastic jet by almost one order of magnitude. For instance, at low frequency (panel \textit{c}) the amplitude of the modes with wavenumber $\kappa_1$ is significantly lower compared to those from the Newtonian jet (panel \textit{a}), thus indicating that spanwise-homogeneous structures play a more dominant role in the slow dynamics of the viscoelastic jet, whereas spanwise-inhomogeneous (streamwise-elongated) structures are more relevant in the Newtonian jet (and subdominant in the viscoelastic one). At higher frequency, the amplitudes of the non-zero wavenumbers increase, and the low-rank behaviour reduces. However, at $St \simeq 0.144$ in the Newtonian jet (panel \textit{b}) and at $St \simeq 0.080$ and $0.086$ in the viscoelastic jet (panel \textit{d}), the low-rank behaviour is the most pronounced (the amplitude of $\kappa = 0$ is much larger than $\kappa_1$), hence the dominance of spanwise-homogeneous structures related to the Kelvin-Helmholtz instability at very high frequency \citep{schmidt2018KHOrr, pickering2020KHOrrLU}. The same low-rank behaviour is also observed at $St \simeq 0.023$ for the viscoelastic jet (panel \textit{d}), though related in this case with homogeneous Orr wave packets that dominate at low frequency \citep{schmidt2018KHOrr}.}

\begin{figure}
    \centering
    \includegraphics[scale = 0.28, keepaspectratio]{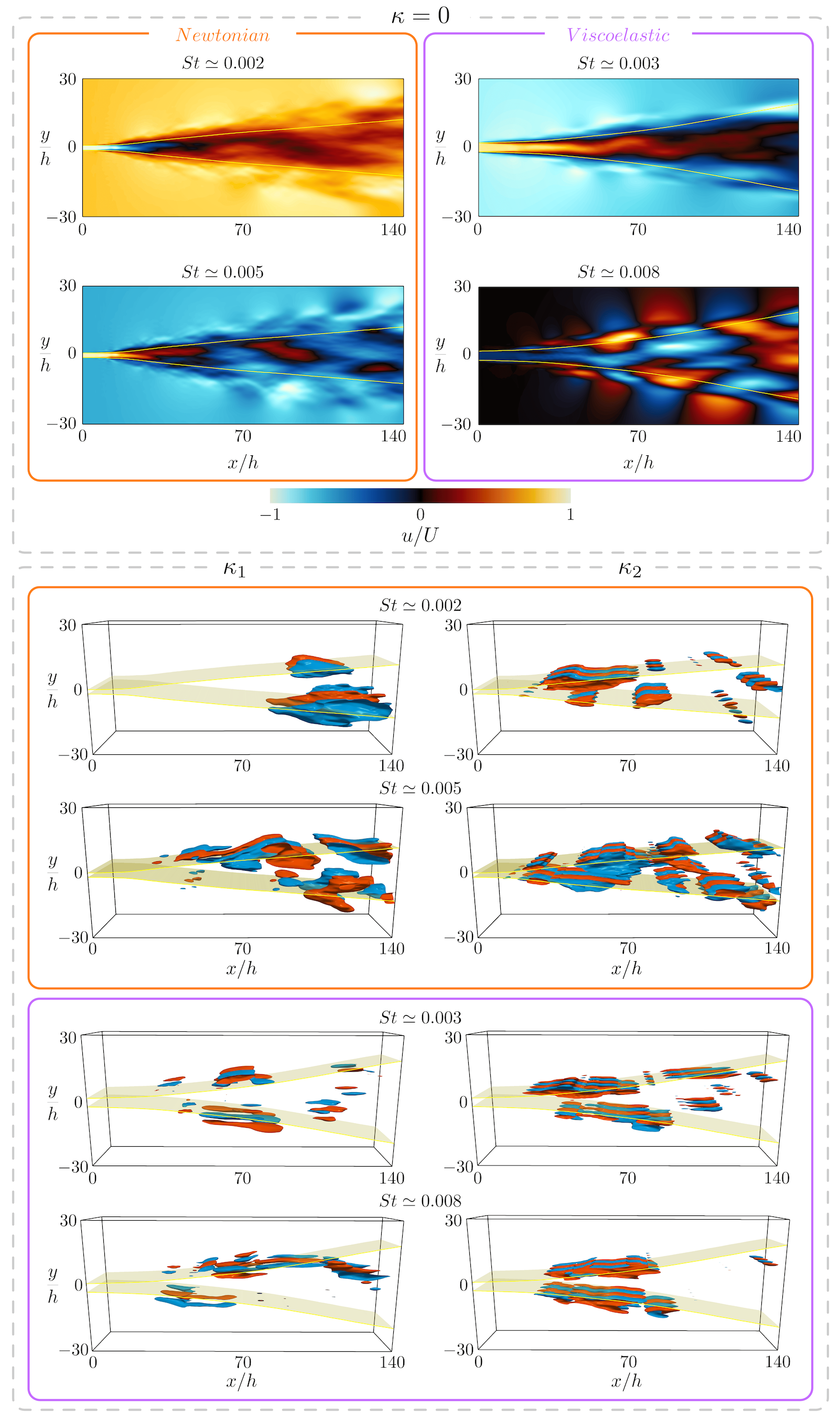}
    \caption{Spatial structure of two low-frequency spatio-temporal modes. The upper panel shows $xy$-planes at $z = 0$ for $\kappa = 0$, and the lower panel the three-dimensional iso-surfaces for $\kappa_1$ (left column) and $\kappa_2$ (right column). Both cases show the normalised streamwise velocity, with iso-surfaces indicating regions of magnitude $+0.5$ (red) and $-0.5$ (blue). The yellow lines and translucent surfaces mark the average jet thickness.}
    \label{fig:uvwstr-stkdL0L1L2}
\end{figure}
We complement the spatio-temporal spectra representing the spatial distribution of the flow structures. 
\revb{We first show in fig.~\ref{fig:uvwstr-stkdL0L1L2} the reconstruction of the leading spatio-temporal modes for $\kappa = 0$, and $\kappa_1$ and $\kappa_2$ for two low-frequency modes}. The $\kappa = 0$ mode is spanwise-homogeneous and therefore shown as a 2D slice, while non-zero wavenumbers are represented using 3D iso-surfaces. \revb{Note that the spatial modes are multiplied by $e^{i \omega t} = {\rm cos}\left(\omega t\right) + i{\rm sin}\left(\omega t\right)$, i.e., they are complex, so they change their phase with time $t$. Therefore, each subpanel shows an instantaneous realization of each mode, that for instance would be opposite if the phase of the mode is adjusted by $\pi$, i.e., positive iso-surfaces would become negative, and viceversa.} At first, the modes with $\kappa = 0$ highlight the dynamics of the fluid column in the Newtonian jet (orange panel). The magnitude of the modes is the largest at the potential core, where the flow is essentially two-dimensional. Downstream the potential core, where the flow is turbulent, the magnitude of the mode rapidly decays and becomes nearly zero elsewhere. At the same location, the shape of the mode with $St \simeq 0.003$ in the viscoelastic jet (purple panel) is reminiscent of flapping oscillations, that are driven by large-scale coherent structures, that are represented at $St \simeq 0.008$, indicated by the negative correlation between both halves of the jet from \reva{$x/h \approx 50$} \citep{Goldschimdt1973FlappingJet, Gortari1981JFluidsEngFlapping}. In Newtonian jets, the antisymmetric mode dominates the dynamics at the far-field from across Reynolds numbers \citep{Thomas1986DevelopJet, Antonia1983OrganizedJet}. \revc{The same mode is more dominant at low Reynolds numbers \citep{deo2008influence, suresh2008reynolds, Soligo2025JFMNewtonJet}, where the presence of larger structures causes a more vigorous mixing. Similar large-scale, coherent structures were also observed in high-Reynolds numbers viscoelastic jets, promoting a more effusive large-scale stirring \citep{Guimaraes2025PRFMixing}}. This is also in good agreement with the faster spreading rate and centreline velocity decay reported by \cite{soligo2023non} in low-Reynolds, viscoelastic jets when compared to Newtonian turbulent jets.

\revb{On the other hand, the modes with non-zero wavenumbers $\kappa_1$ and $\kappa_2$ show three-dimensional streamwise-elongated structures}, that are similar to the streaky structures reported by \cite{nogueira2019streaks} and  \cite{pickering2020KHOrrLU} in Newtonian turbulent round jets. In the Newtonian jet (orange panel), large-scale streamwise-elongated structures have a global presence: they arrange downstream the potential core along the jet edges, though they have great penetration within the jet core. Structures associated with $\kappa_2$ are located closer to the potential core compared to those for $\kappa_1$. The modes in the viscoelastic jet (purple panel) have a spatial distribution similar to those of the Newtonian jet---after the potential core and along the jet edges---but they are overall disappear after \reva{$x/h \approx 100$}, they are less massive, and they have little penetration within the jet core. This description correlates with the spatio-temporal spectra. The amplitude of the modes with $\kappa_1$ and $\kappa_2$ in the Newtonian jet is greater compared to those from the viscoelastic case, hence streamwise-elongated structures are more regular and persistent in the former. On the contrary, the streamwise-elongated structures in the viscoelastic jet are disrupted by the large-scale structures at the far-field, thus breaking them and homogenising the flow.

\begin{figure}
    \centering
    \includegraphics[scale=0.03, keepaspectratio]{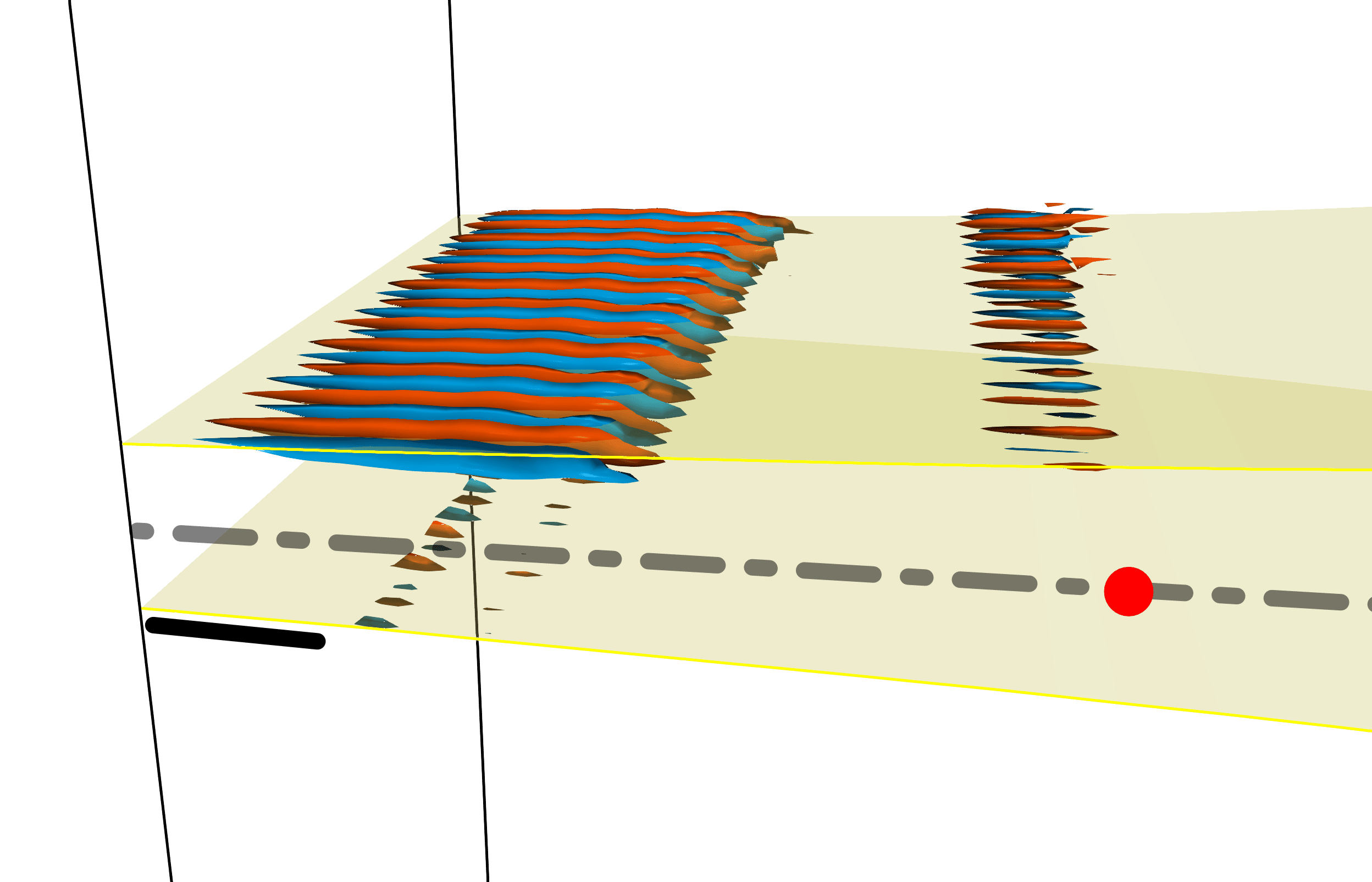}
    \put(-73,32){\tiny $6$}
    \put(-72,18){\tiny $0$}
    \put(-80,18){\small $\frac{y}{h}$}
    \put(-66,7){\tiny $2h$}
    \put(-74,4){\tiny $-6$}
    \put(-85,55){\revaa{\small $a)$}}
    \put(-55,55){$St \simeq 0.003$}
    \hspace{0.3cm}
    \includegraphics[scale=0.03, keepaspectratio]{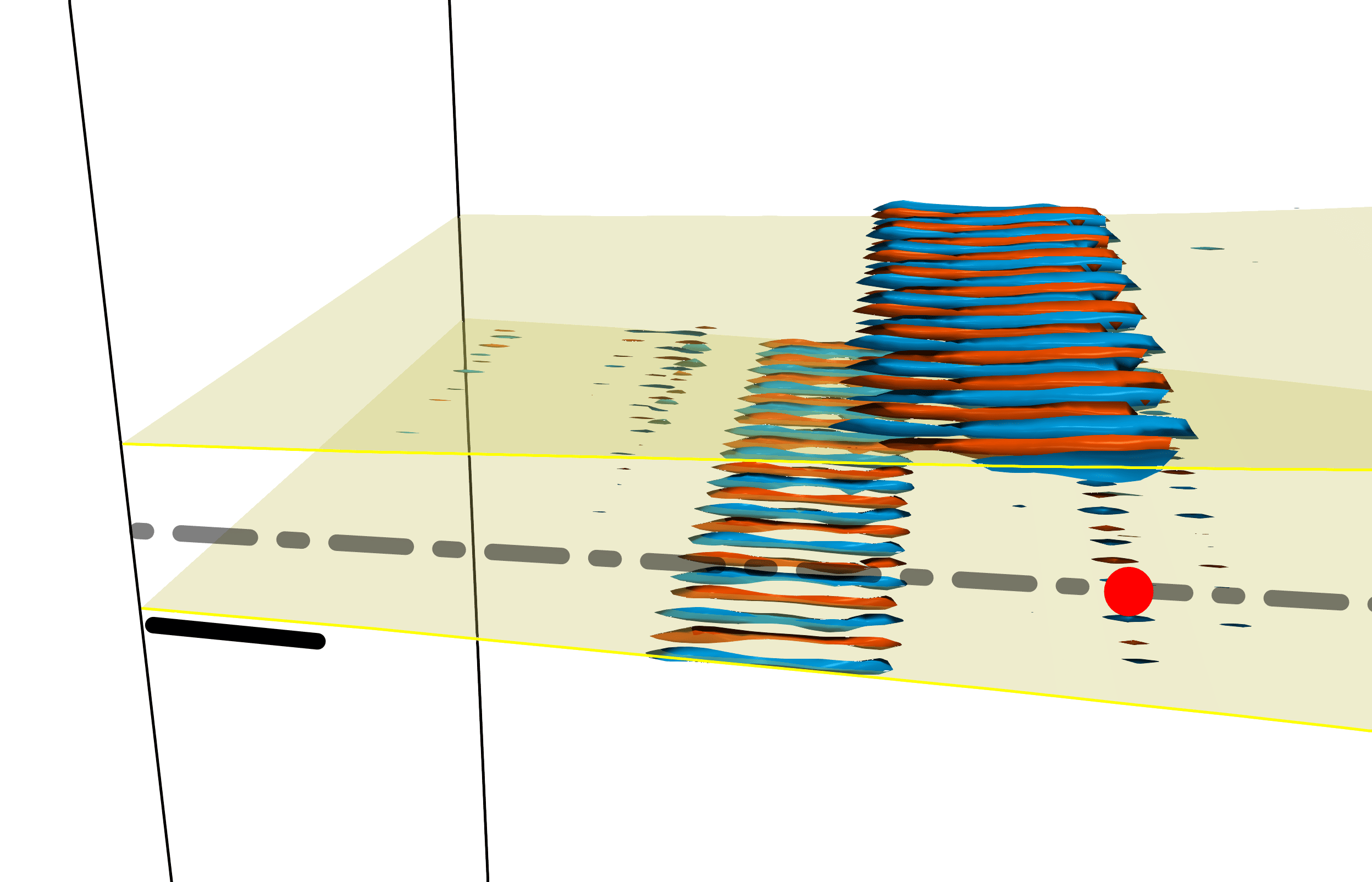}
    \put(-73,32){\tiny $6$}
    \put(-72,18){\tiny $0$}
    \put(-80,18){\small $\frac{y}{h}$}
    \put(-66,7){\tiny $2h$}
    \put(-74,4){\tiny $-6$}
    \put(-85,55){\revaa{\small $b)$}}
    \put(-55,55){$St \simeq 0.007$}
    \hspace{0.3cm}
    \includegraphics[scale=0.03, keepaspectratio]{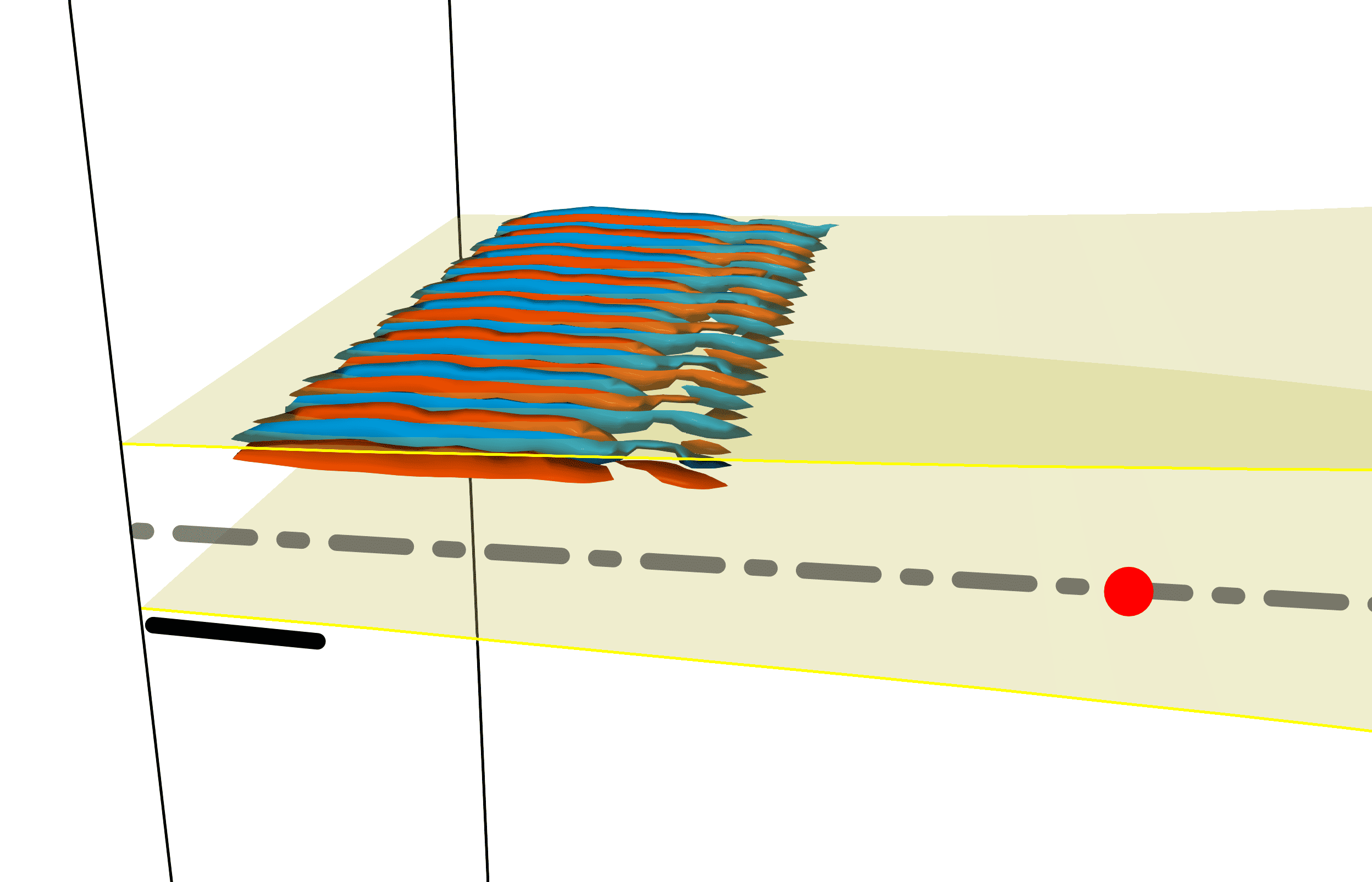}
    \put(-73,32){\tiny $6$}
    \put(-72,18){\tiny $0$}
    \put(-80,18){\small $\frac{y}{h}$}
    \put(-66,7){\tiny $2h$}
    \put(-74,4){\tiny $-6$}
    \put(-85,55){\revaa{\small $c)$}}
    \put(-55,55){$St \simeq 0.008$}
    \hspace{0.3cm}
    \includegraphics[scale=0.03, keepaspectratio]{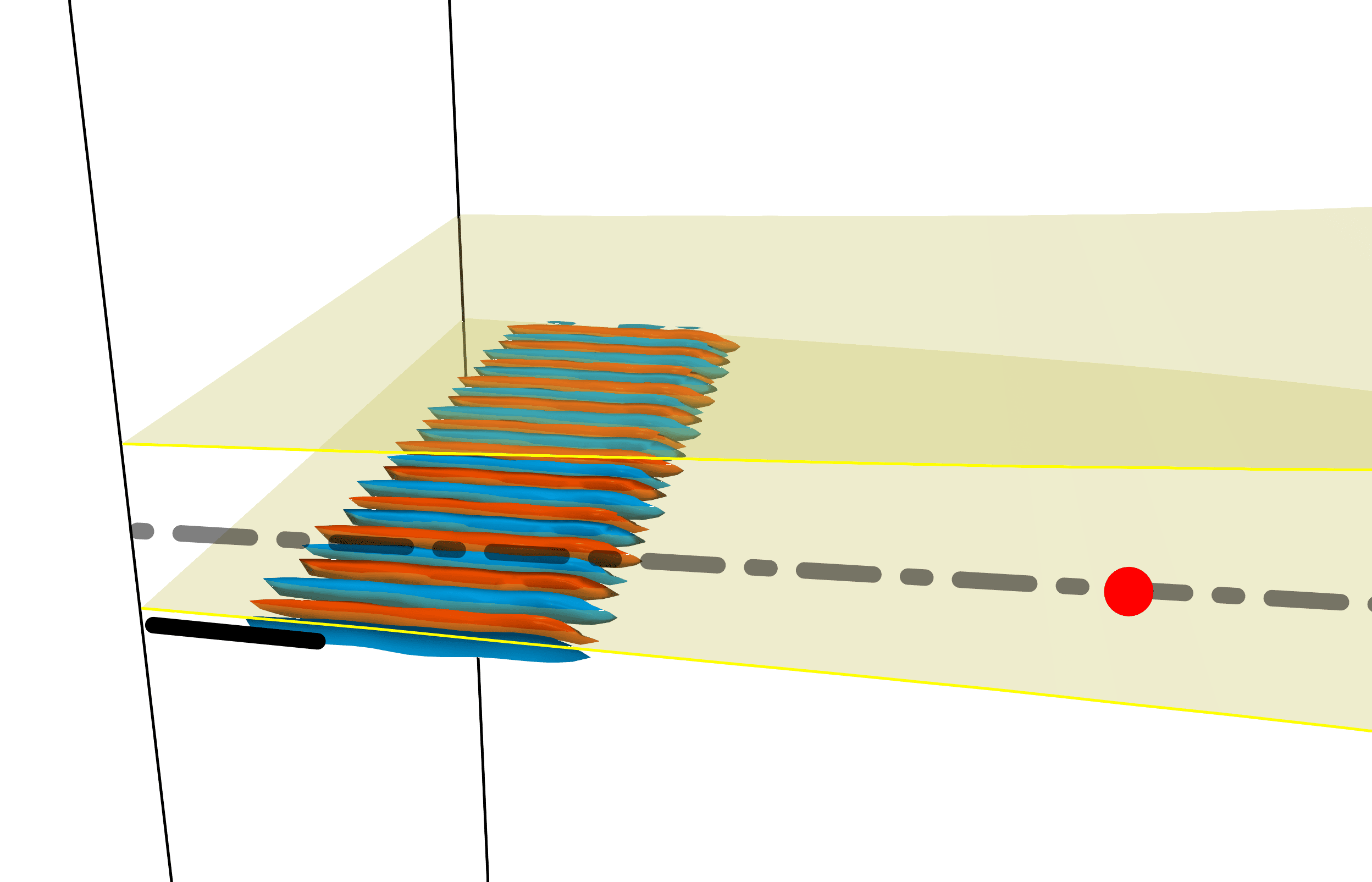}
    \put(-73,32){\tiny $6$}
    \put(-72,18){\tiny $0$}
    \put(-80,18){\small $\frac{y}{h}$}
    \put(-66,7){\tiny $2h$}
    \put(-74,4){\tiny $-6$}
    \put(-85,55){\revaa{\small $d)$}}
    \put(-55,55){$St \simeq 0.011$}
    \vspace{0.3cm}
    \includegraphics[scale=0.03, keepaspectratio]{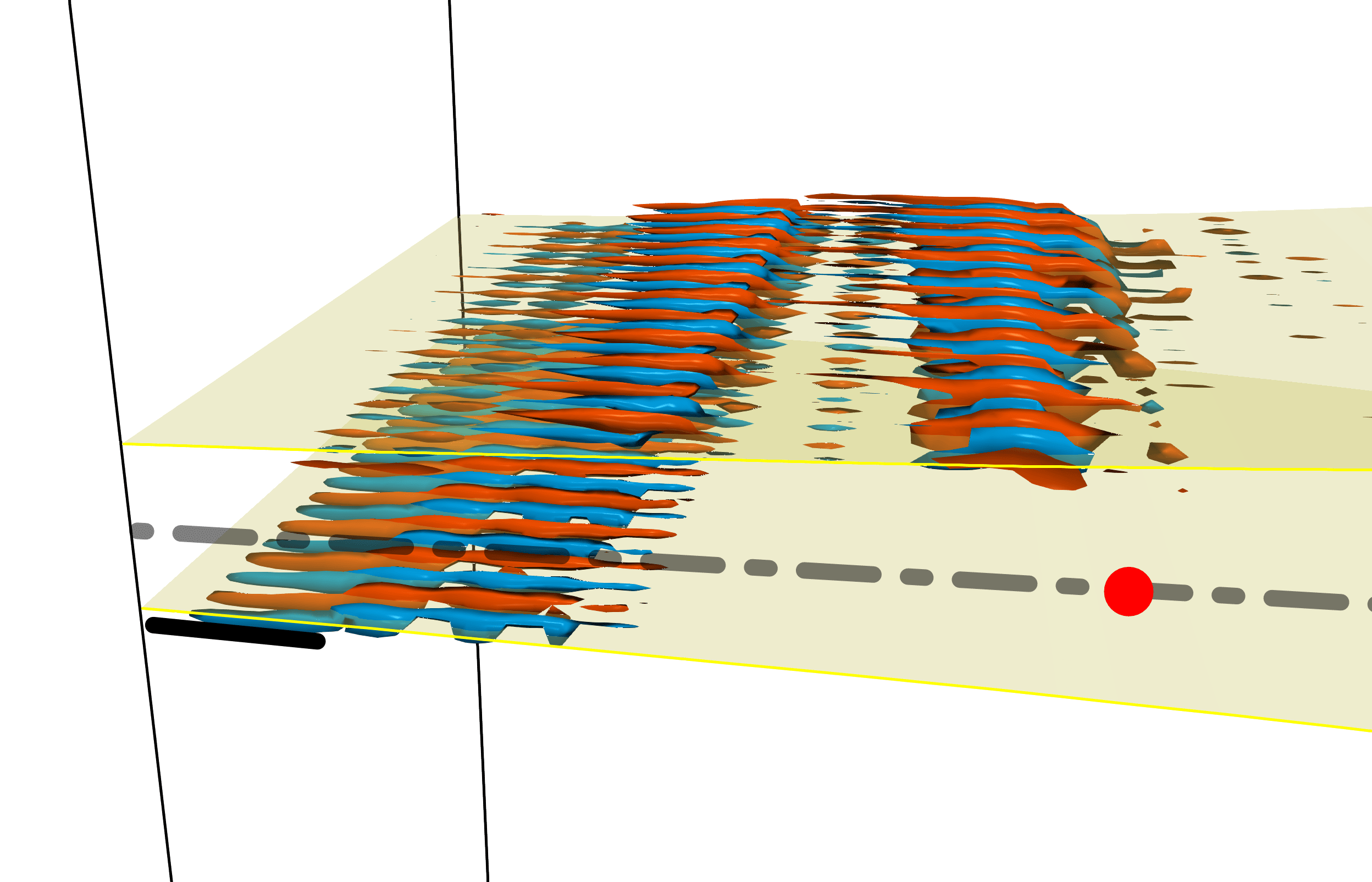}
    \put(-73,32){\tiny $6$}
    \put(-72,18){\tiny $0$}
    \put(-80,18){\small $\frac{y}{h}$}
    \put(-66,7){\tiny $2h$}
    \put(-74,4){\tiny $-6$}
    \put(-85,55){\revaa{\small $e)$}}
    \hspace{0.3cm}
    \includegraphics[scale=0.03, keepaspectratio]{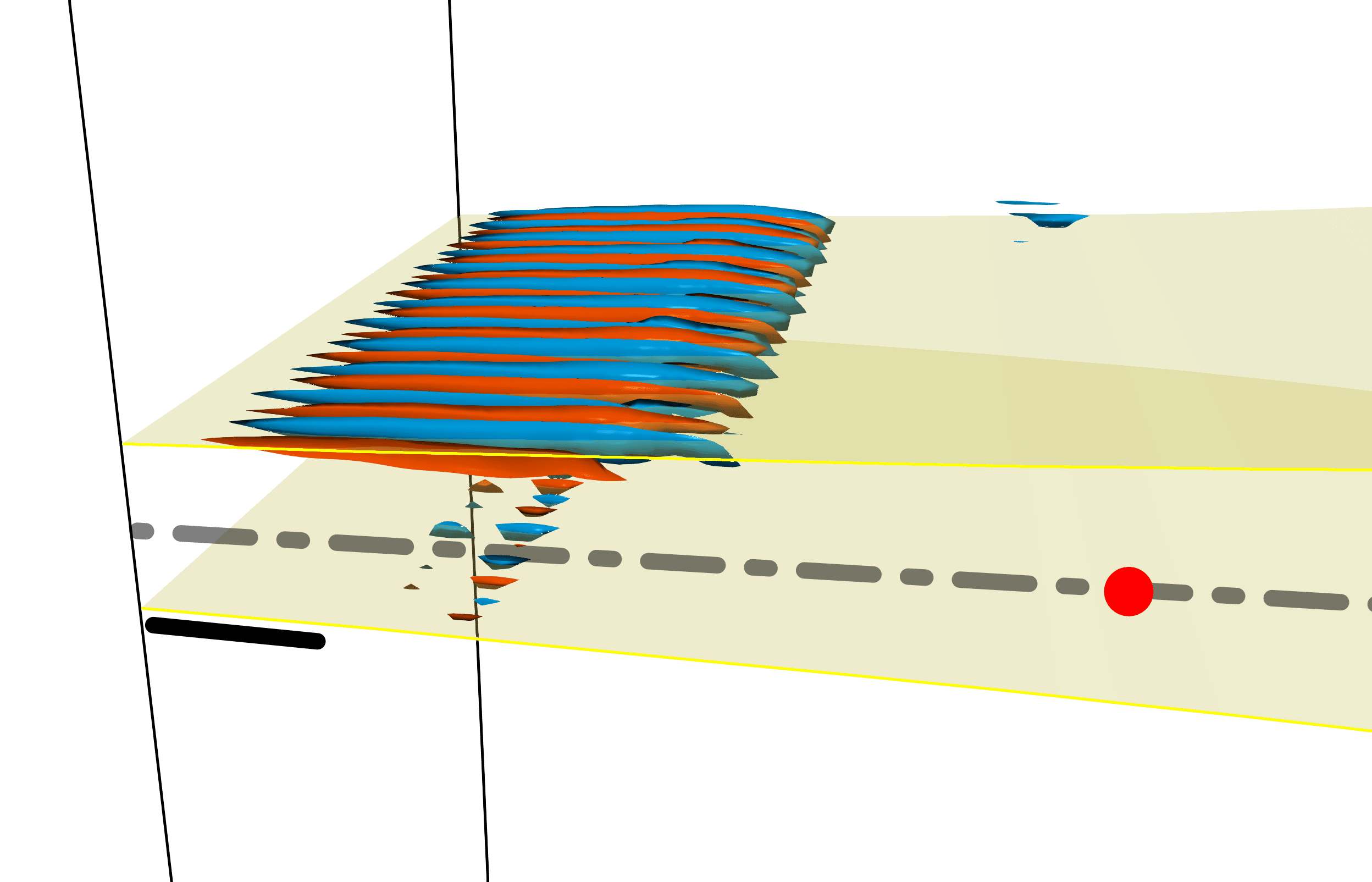}
    \put(-73,32){\tiny $6$}
    \put(-72,18){\tiny $0$}
    \put(-80,18){\small $\frac{y}{h}$}
    \put(-66,7){\tiny $2h$}
    \put(-74,4){\tiny $-6$}
    \put(-85,55){\revaa{\small $f)$}}
    \hspace{0.3cm}
    \includegraphics[scale=0.03, keepaspectratio]{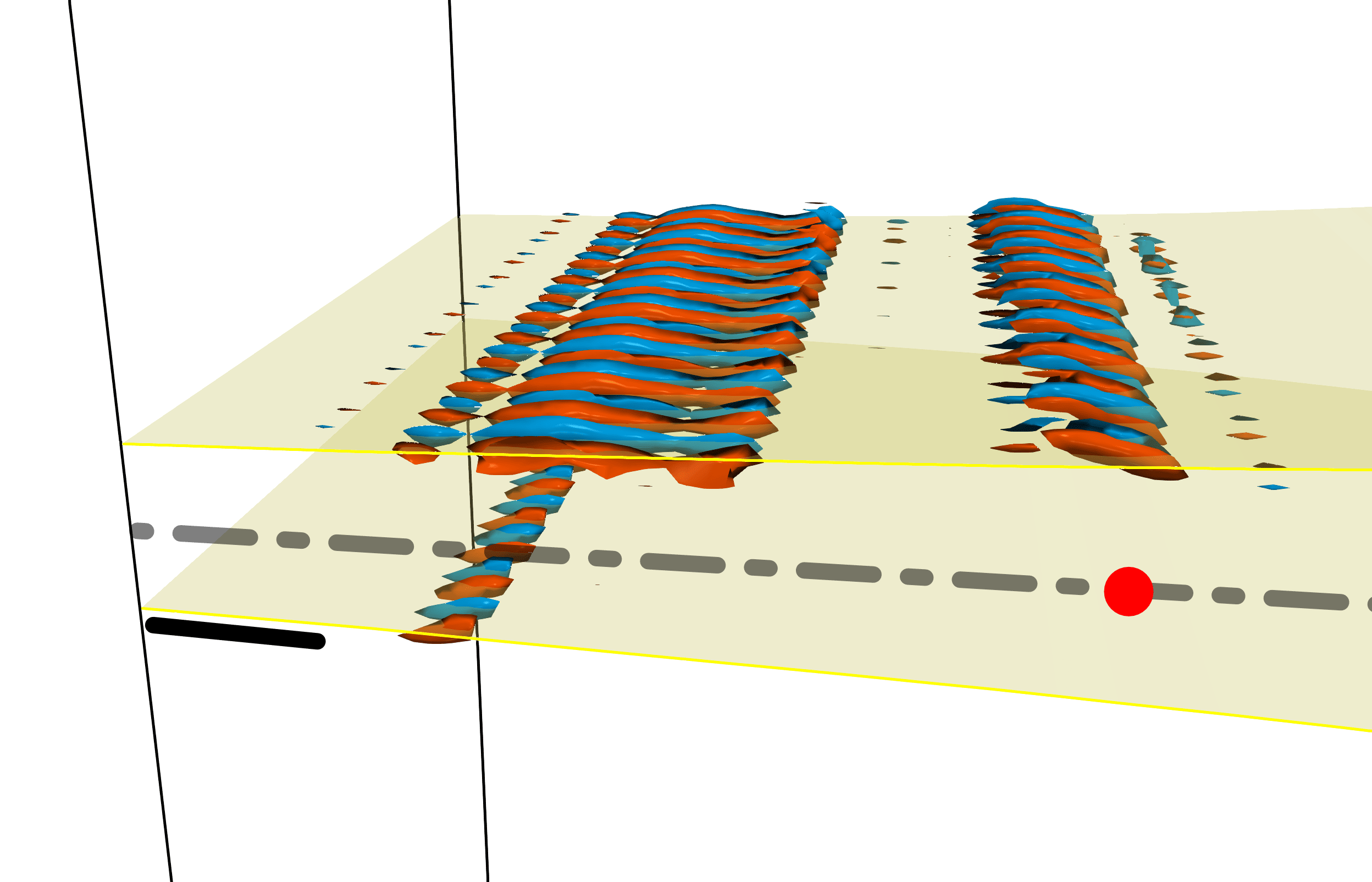}
    \put(-73,32){\tiny $6$}
    \put(-72,18){\tiny $0$}
    \put(-80,18){\small $\frac{y}{h}$}
    \put(-66,7){\tiny $2h$}
    \put(-74,4){\tiny $-6$}
    \put(-85,55){\revaa{\small $g)$}}
    \hspace{0.3cm}
    \includegraphics[scale=0.03, keepaspectratio]{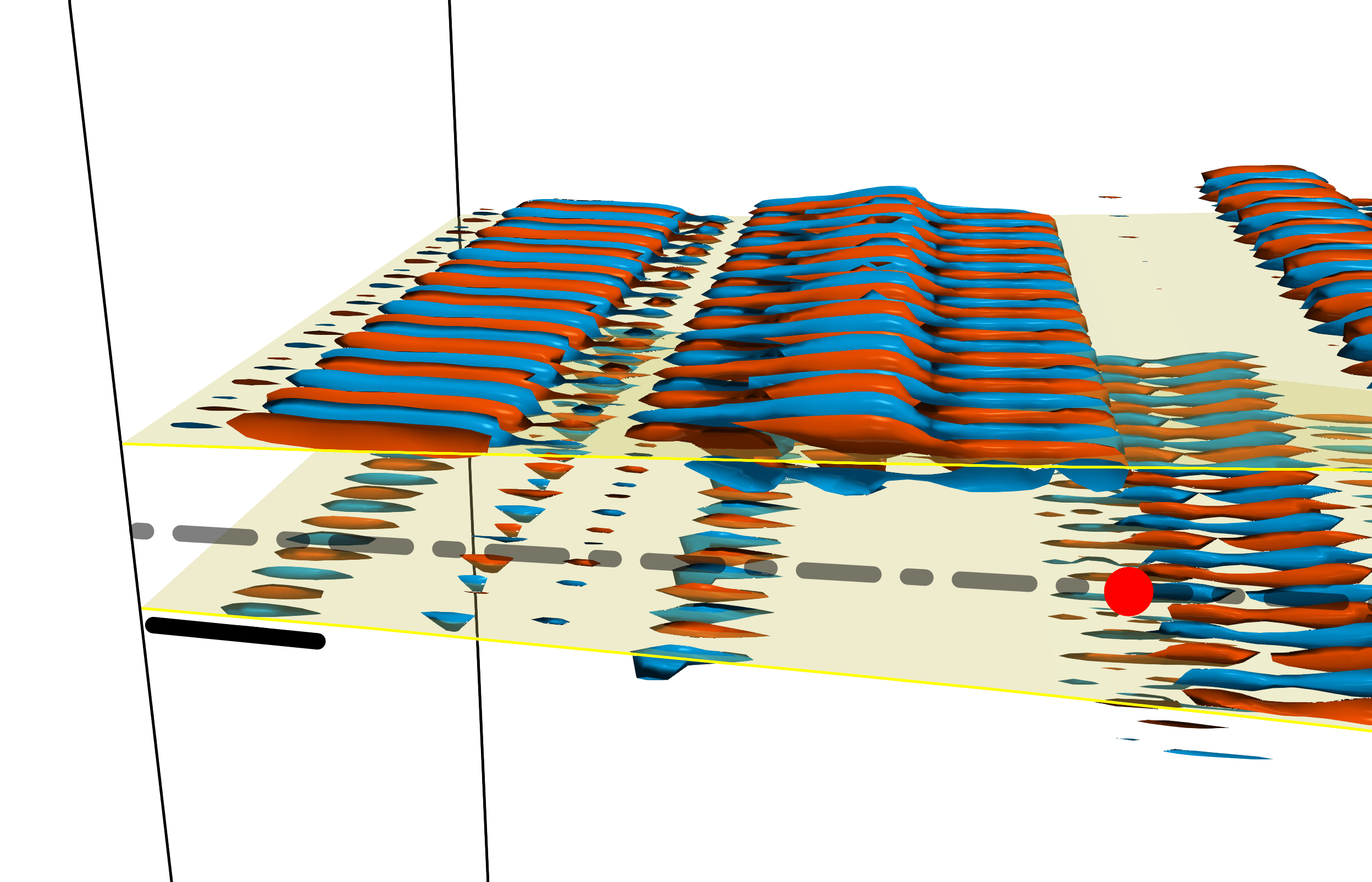}
    \put(-73,32){\tiny $6$}
    \put(-72,18){\tiny $0$}
    \put(-80,18){\small $\frac{y}{h}$}
    \put(-66,7){\tiny $2h$}
    \put(-74,4){\tiny $-6$}
    \put(-85,55){\revaa{\small $h)$}}
    \caption{Reconstruction of the near-field streaks in the viscoelastic jet. Real (panels \textit{a-d}) and imaginary (panels \textit{e-h}) components of the spatio-temporal modes with wavenumber $\kappa = 12 \kappa_{1}$. Three-dimensional iso-surfaces represent the normalised streamwise velocity of magnitude $+0.5$ (red) and $-0.5$ (blue). \revaa{The black translucent dashed-dotted line indicate the centreline of the jet, and the red circle the coordinate $x/h = 20$.}}
    \label{fig:uvwstr-stkdL_12}
\end{figure}
While higher wavenumbers indicate smaller structures, they are significantly different between jets. At low frequency, these relate exclusively to the near-field streaks in the viscoelastic case, represented in fig.~\ref{fig:uvwstr-stkdL_12}. The smallest wavenumber associated with them is $\kappa_{12} = 12 \kappa_{1}$ (wavenumbers greater than $\kappa_2$ but smaller than $\kappa_{12}$ correspond with streamwise-elongated structures located after the potential core), while its harmonics capture finer streaks in the same region. Across all low frequencies, this mode shows groups of streaks that are at most $6h$ long, and they extend up to \revaa{$x/h \approx 20$} from the inlet along the upper and lower shear layers. These structures are well-resolved (about three grid points thick), and they resemble the streaky pattern at the potential core in fig.~\ref{fig:new_vjet}\textit{g}, with the mode capturing twenty-four alternating high- and low- speed streaks organised along the spanwise axis, consistent with the flow visualisation. The real and imaginary parts of the mode provide insight into the temporal dynamics of the streaks. Because the DMD modes are complex conjugates, their real and imaginary components represent the same structure shifted by half a period. Here, the two are not identical, suggesting that the flow reorganises within half a period. The streaks in the potential core therefore do not behave as standing structures, but they travel downstream and are disrupted locally in the streamwise direction.

To summarise, the spatio-temporal analysis of the low-frequency modes reveals large-scale, streamwise-elongated structures in the far-field of both jets, and smaller-scale streaks in the near-field of the viscoelastic jet. While the description of the far-field is reasonably similar between cases, the presence of the near-field streaks influences the dynamics at the potential core in the viscoelastic jet. Here, high- and low- speed streaks induce regions of localised high-shear and high-elongation among approaching streaks, that stretch the polymer in the streamwise direction and ultimately induce the transition to turbulence due to elastic instabilities. This purely elastic pathway allows the transition to turbulence in viscoelastic planar jets at low Reynolds and high polymer elasticity, and it differs from the stabilising effect at higher Reynolds \citep{rallison1995instability, ray2015absolute, yamani2021spectral, yamani2023spatiotemporal}. 

\begin{figure}
    \centering
    \includegraphics[width=\textwidth, keepaspectratio]{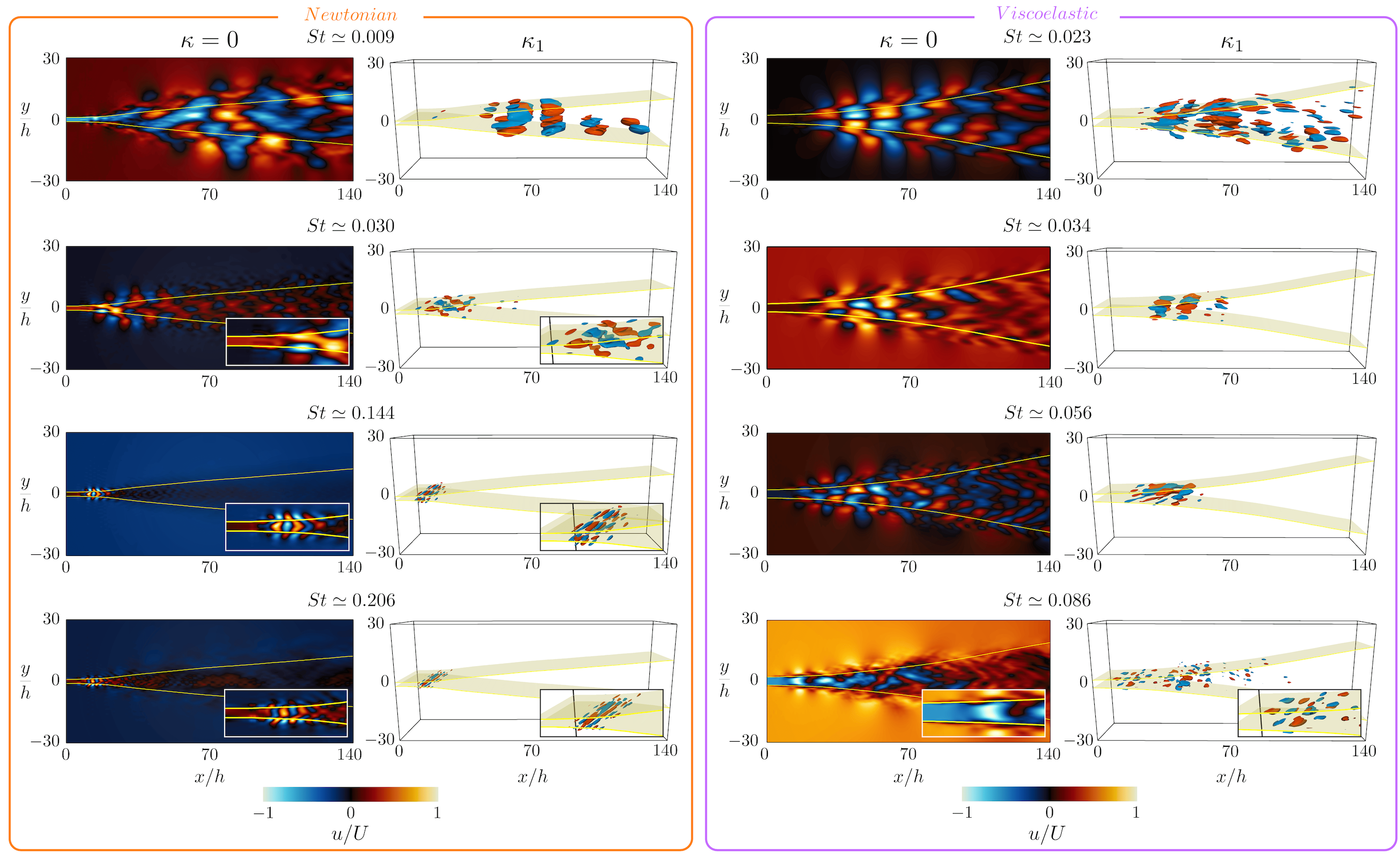}
    \caption{\revb{Spatial structure of four high-frequency spatio-temporal modes. For each jet, the left column shows $xy$-planes at $z = 0$ for $\kappa = 0$, and the right column the three-dimensional iso-surfaces for $\kappa_1$. Both cases show the normalised streamwise velocity, with iso-surfaces indicating regions of magnitude $+0.5$ (red) and $-0.5$ (blue). Insets provide a closer view at the near-field up to $x \approx 30h$. The yellow lines and translucent surfaces mark the average jet thickness.}}
    \label{fig:uvwwp-stkdL0L1}
\end{figure}
We conclude characterising in fig.~\ref{fig:uvwwp-stkdL0L1} the spatial shape of four high-frequency modes in the Newtonian and viscoelastic jets. In the Newtonian jet (orange panel), the streamwise velocity field takes the form of wave packets at $\kappa = 0$. However, their placement changes from $St \geq 0.144$: for lower $St$, wave packets are placed downstream the potential core at the upper and lower halves of the jet, while for higher $St$ they are at the potential core along the shear layers. Similarly, \cite{pickering2020KHOrrLU} reported that the dominant mechanism at zero-wavenumber in high-Reynolds, Newtonian round jets changes from Orr to Kelvin-Helmholtz at $St \approx 0.3$ (based on the diameter of the nozzle), that is independent of the Reynolds number; for non-zero wavenumbers, the dominant response of the flow is determined by Kelvin-Helmholtz. We do not find similitude with the Kelvin-Helmholtz wave packets for $\kappa_1$ at $St \simeq 0.009$ and $0.030$ but at $St \simeq 0.144$ and $0.206$---spanwise-homogeneous and aligned along the shear layer---that are consistent with the description of Kelvin-Helmholtz wave packets.

In the viscoelastic jet (purple panel), the modes with $\kappa = 0$ indicate Orr-type wave packets, exception made to $St \simeq 0.086$, that shows primarily the motion of the fluid column at the near-field. On the other hand, the shape of the modes with wavenumber $\kappa_1$ changes significantly with frequency: at $St \simeq 0.023$, structures are rather streamwise-coherent (\cite{pickering2020KHOrrLU} also related non-zero wavenumbers at low-frequency to modes of streaky structures) while the Orr wave packets are recovered at $St \simeq 0.034$ and $0.056$. Conversely, $St \simeq 0.086$ shows structures scattered throughout the domain, though those at the near-field, located at the shear layers, resemble the Kelvin-Helmholtz wave packets.

\section{Local analysis of the viscoelastic planar jet}\label{sec:inj}
In the previous section, we showed the existence of the near-field streaks, that potentially interact with the flow instability of the jet at the potential core. The previous analysis was global and thus it is difficult to fully assess the role of streaks in the near-field region. To address this, we now apply HODMD to the viscoelastic case at smaller three-dimensional boxes that enclose the flow near the inlet. Three boxes are considered with increasing streamwise length: $l_x / h = 20, 30,$ and $40$. All boxes are centred on the jet centreline, have a jet-normal length of $l_y / h = 20$, and span the full width of the domain. In particular, the smallest box covers the potential core, while longer ones extend partially into the turbulent region. 

\begin{figure}
     \centering
     \includegraphics[width=\textwidth, keepaspectratio]{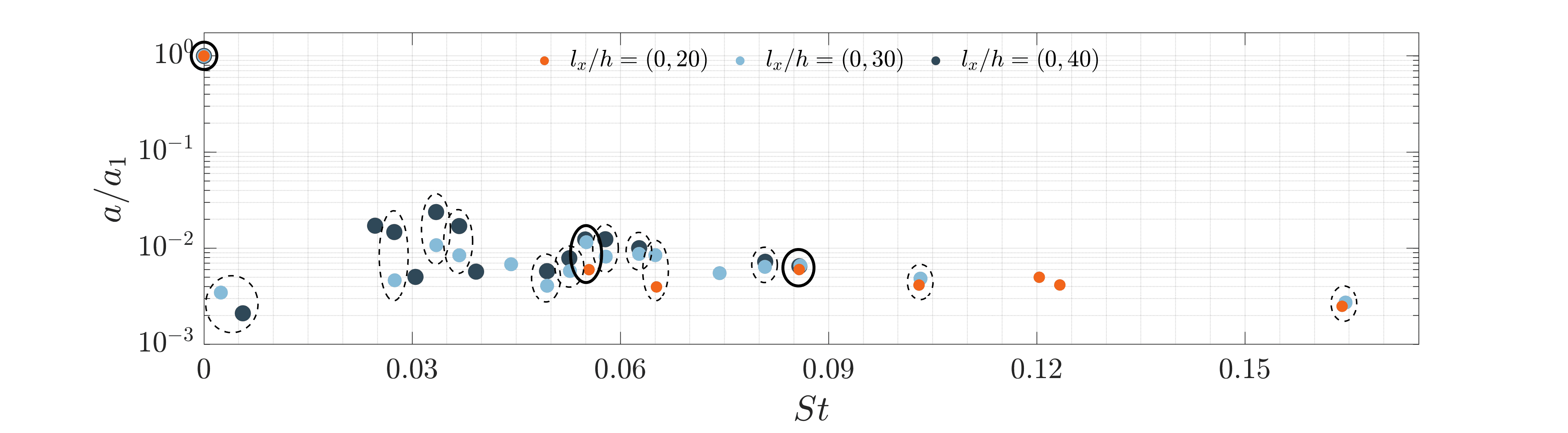}
     \caption{Local HODMD spectrum for robust frequencies. Non-dimensional frequency, $St$, compared to the normalised amplitude, $a / a_1$, with $a_1$ the largest amplitude in each series. Black solid circles mark frequencies detected in all three boxes, while black dashed circles indicate those present in two boxes.}
     \label{fig:hodmd-inj_spectrum}
\end{figure}
Figure~\ref{fig:hodmd-inj_spectrum} shows the spectrum of robust modes in the three boxes. The method was recalibrated for the local analysis: $d$ is varied from $60$ to $120$ (larger $d$ allows to capture more complex dynamics because of the higher dimension of the delay-embedding space), while the threshold $\varepsilon$ is kept the same as the global analysis, with the addition of $\varepsilon = 4 \cdot 10^{-4}$ to spot smaller amplitude modes. In the following, we show the results for $d = 70$ in the smallest box and $d = 80$ in the remaining two, while $\varepsilon_1$ and $\varepsilon_2$ are set to $6 \cdot 10^{-4}$ in all cases. 
\revb{The smallest box yields $7$ robust modes, compared to $16$ and $14$ for the longer ones.} The dominant mode also shifts to lower frequencies with distance---$St \simeq 0.086$ in the smallest box, $St \simeq 0.056$ in the middle-sized one, and $St \simeq 0.034$ in the longest box---consistent with the growing influence of slower, larger-scale dynamics farther from the potential core. Similarly, the cumulative amplitude of the high frequency modes is higher in the smallest box since scales are faster and smaller, so high frequencies are more relevant. We should point out that some of these modes likely persist in the longer boxes, but their amplitude may fall below the threshold $\varepsilon$ used in the analysis, thus not being detected by the method. Moreover, we identify a mode with very small frequency in the longer boxes, though it is not the case for the smallest one, where dynamics associated to lower frequencies (streaks) are hindered in the smallest box because of the dominance of smaller-scale, higher-frequency modes. Although some of the high-frequency modes, in particular those in the smallest box, might be related to the jet instability, none of the extracted frequencies show temporal growth---the jets are stable in time---, thus indicating that the instabilities are not absolute in nature. 

\begin{figure}
     \centering
     \includegraphics[scale=0.27, keepaspectratio]{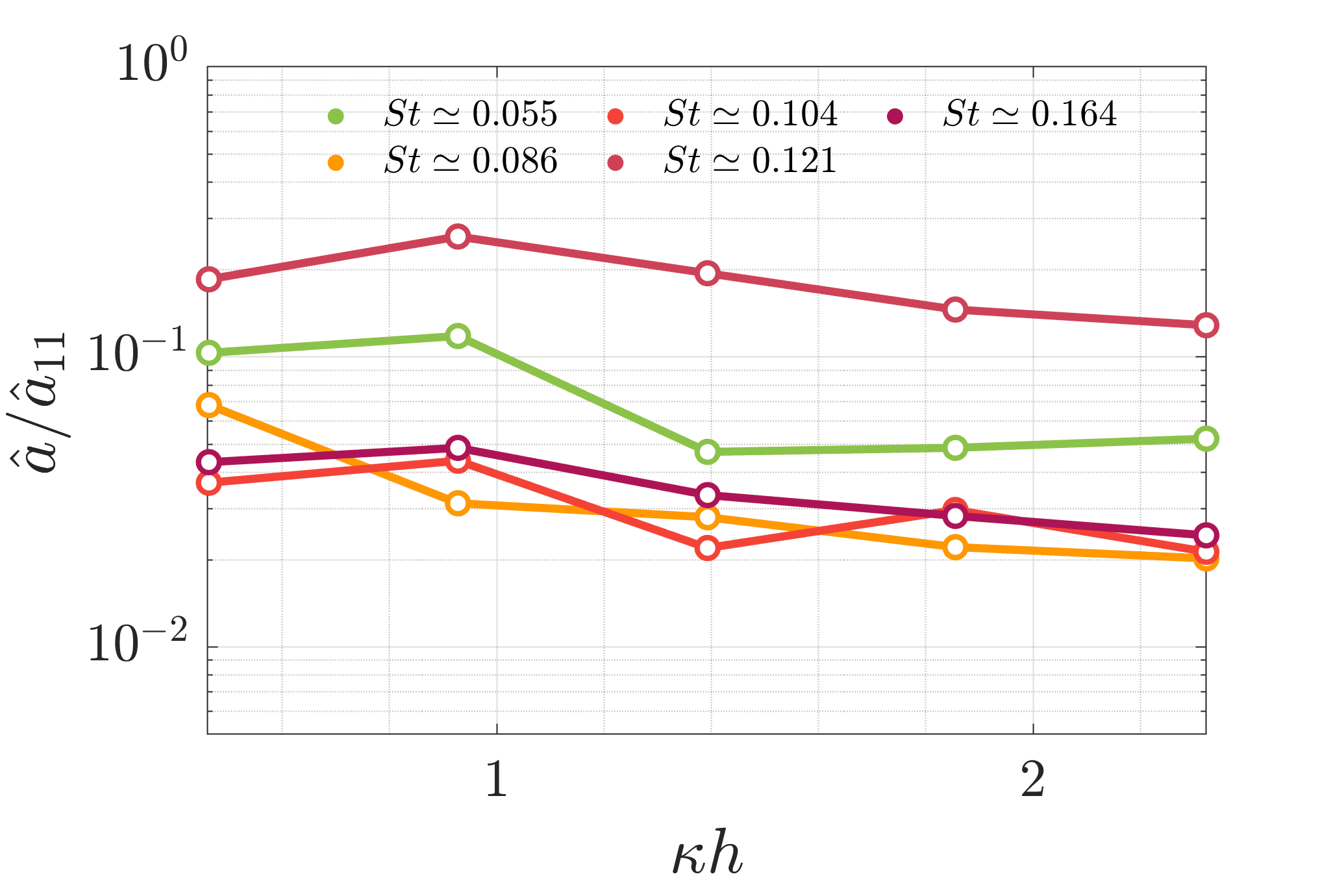}
     \put(-88,88){\small $l_x / h = \left( 0, 20 \right)$}
     \hspace{-0.3cm}
     \includegraphics[scale=0.27, keepaspectratio, trim=30 0 0 0, clip]{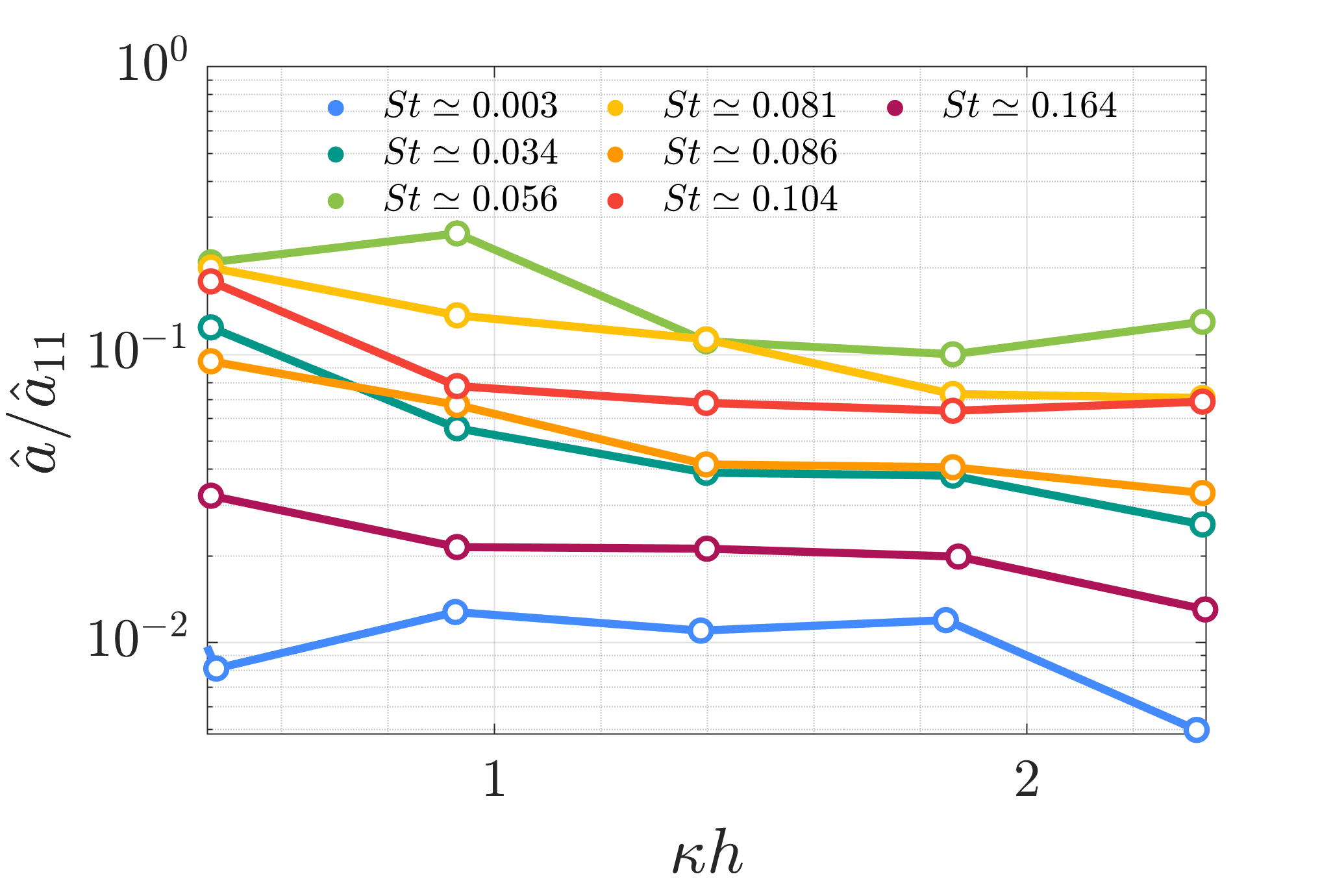}
     \put(-88,88){\small $l_x / h = \left( 0, 30 \right)$}
     \hspace{-0.3cm}
     \includegraphics[scale=0.27, keepaspectratio, trim=30 0 0 0, clip]{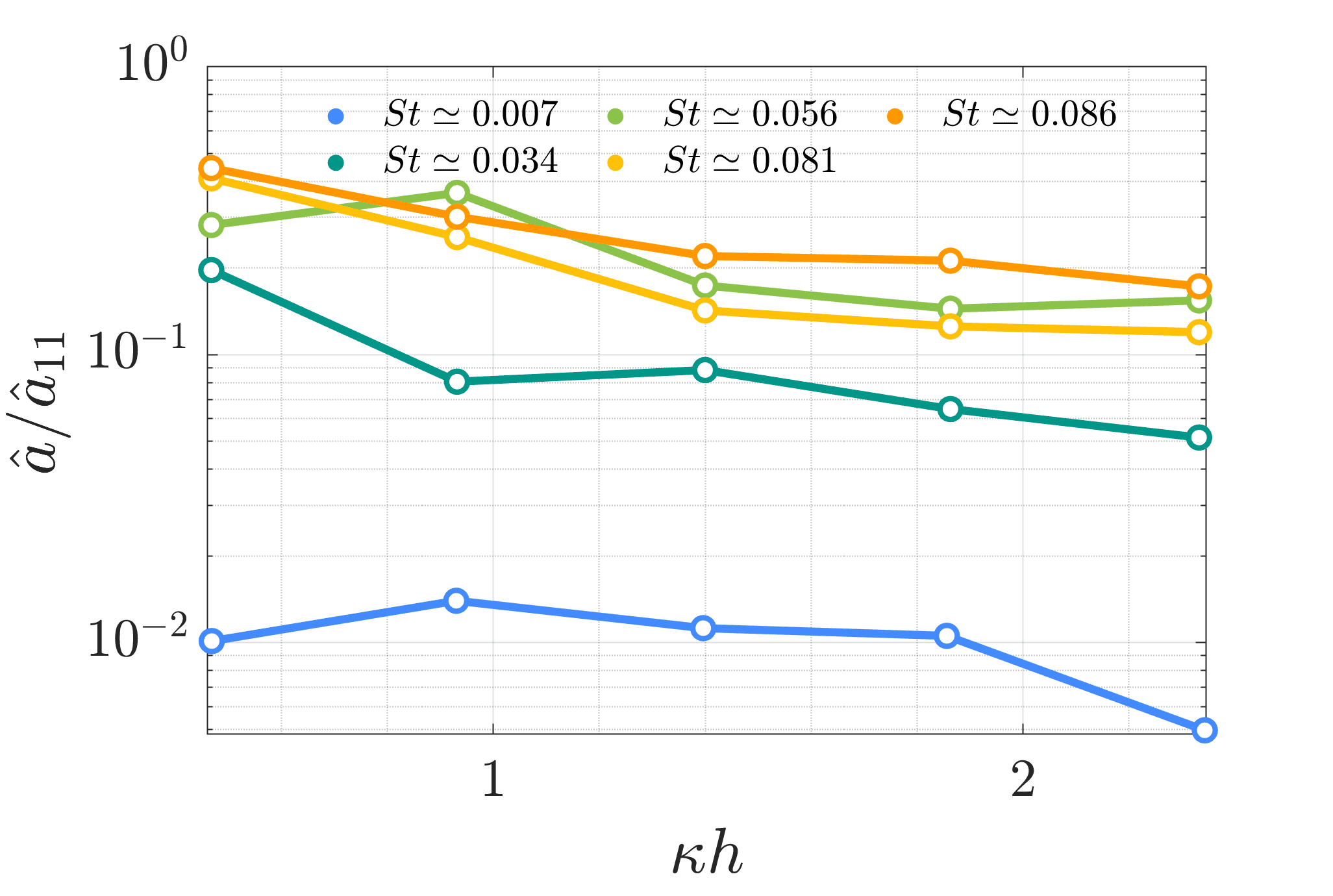}
     \put(-88,88){\small $l_x / h = \left( 0, 40 \right)$}
     \caption{Local spatio-temporal spectra. \revb{Normalised} spanwise wavenumber, $\kappa h$, compared to their normalised amplitude, $\hat{a} / \hat{a}_{11}$, from robust modes computed in all three boxes. Similar frequencies have same colours among plots.}
     \label{fig:stkd-inj_spectrum}
\end{figure}
We extend the temporal analysis by decomposing the modes in the spanwise direction. Figure~\ref{fig:stkd-inj_spectrum} shows the spatio-temporal spectra obtained in all three boxes. Overall, the amplitude of the non-zero wavenumbers increases with box length, highlighting the growing importance of three-dimensional structures in the turbulent region. The dominant wavenumber also shifts from $\kappa_2 = 2 \kappa_{min}$ in the smallest box to $\kappa_1 = \kappa_{min}$ in the longer ones, indicating that larger structures emerge downstream the potential core. In the two longer boxes, the lowest frequency mode has the smallest amplitude, so streamwise-elongated structures are sub-dominant in the near-field. Moreover, the dominant wavenumber is $\kappa_2$, in good agreement with the global analysis, which suggested that smaller streamwise-elongated structures, so higher wavenumbers, are located closer to the potential core. This trend is similar at higher frequency, with $\kappa_1$ modes increasing their amplitude the most with box length. In particular, $\kappa_1$ at $St \simeq 0.081$ and $0.086$ grow significantly, eventually surpassing $St \simeq 0.056$ in the longest box. Finally, the amplitude of $St \simeq 0.121$ and $0.164$ decreases with the box length ($St \simeq 0.121$ is not found outside the smallest box), where $\kappa_2$ for $St \simeq 0.121$ dominates the spectrum in the smallest box.

\begin{figure}
    \centering
    \includegraphics[scale=0.23, keepaspectratio]{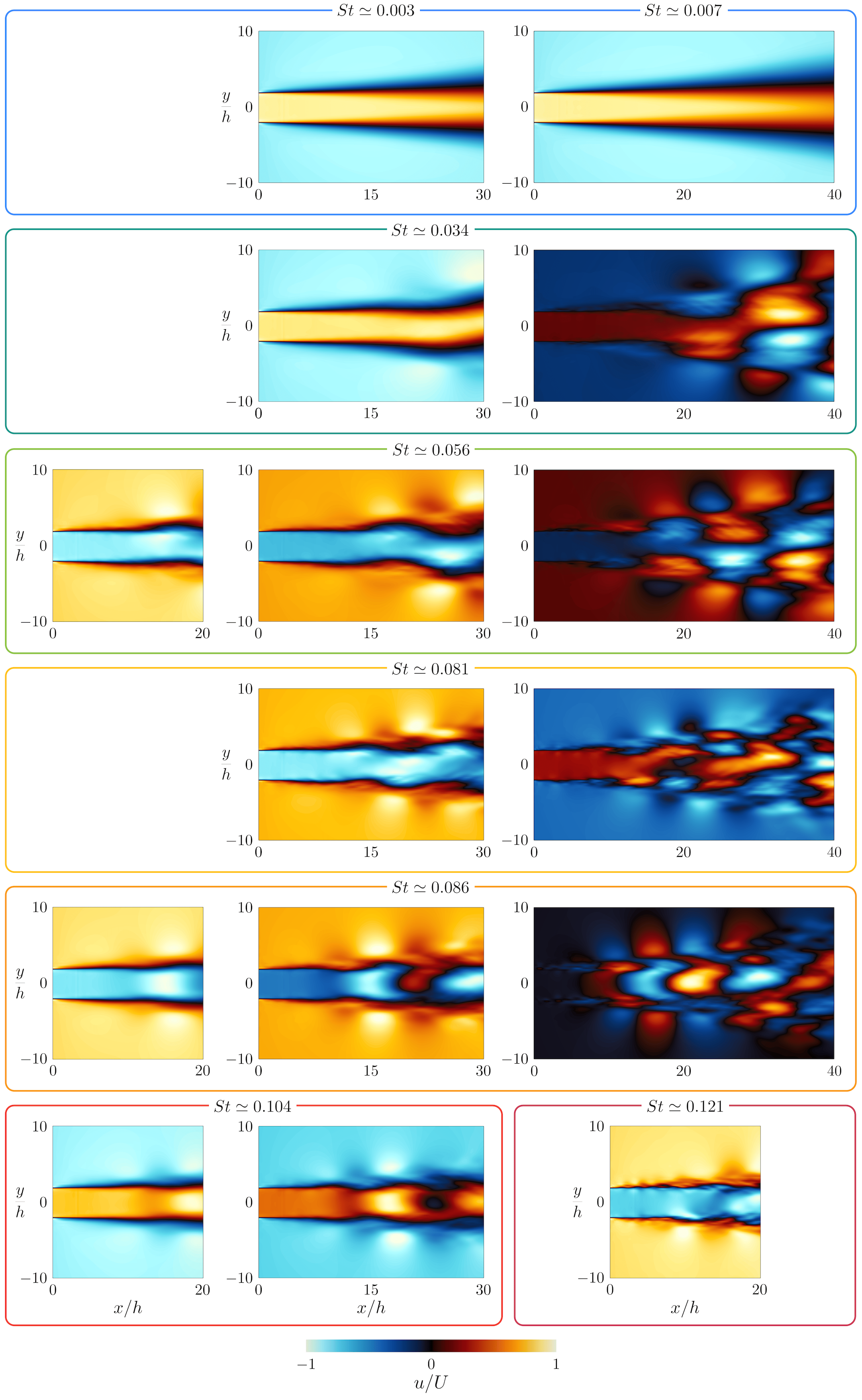}
    \caption{Reconstruction of the bulk flow at the near-field. Two-dimensional $xy$-planes at $z = 0$ of the normalised streamwise velocity component for the modes with $\kappa = 0$.} 
    \label{fig:stkd-inj_modesL0}
\end{figure}
Next, we visualise the spatio-temporal modes in all three boxes. We first show the modes with $\kappa = 0$ in fig.~\ref{fig:stkd-inj_modesL0}. In the global analysis, these modes were linked to two-dimensional wave packets. Here, we find qualitatively similar structures in the longest box, while the smaller boxes show the motion of the bulk flow. Starting with the longest box, $\kappa = 0$ at the lowest frequency $St \simeq 0.007$ matches the mean flow, and it shows the dissipation of the jet downstream the potential core as it spreads. At higher frequency, wave packet-like structures appear. They are located in the upper and lower halves of the jet for $St \simeq 0.034$ and $0.056$, while they align with the centreline for $St \simeq 0.086$, while the flow structures at $St \simeq 0.081$ resemble a combination of the mentioned three modes. Looking to the smaller boxes for the same frequencies, we can connect these patterns to the antisymmetric or sinuous mode ($St \simeq 0.034, 0.056$) and the symmetric or varicose mode ($St \simeq 0.086$) of the bulk flow. In the smallest boxes, the modes clearly show the sinuous mode for $St \simeq 0.056$ and the varicose one for $St \simeq 0.086$ and $0.104$, that are visible from \reva{$x \approx  10h$}. Even though both modes are obtained in the viscoelastic jet, the sinuous mode is amplified the most downstream, as indicated by the growth of the magnitude of the mode, that also correlates with the growth of its temporal amplitude in fig.~\ref{fig:hodmd-inj_spectrum}. Similarly, the mode at $St \simeq 0.086$ experiences a similar growth in space, though it is confined to \reva{$10h \lessapprox x  \lessapprox 30h$}. Interestingly, for $St \simeq 0.121$, we observe fluctuations mainly in the shear layer rather than in the jet column, that destabilise the flow downstream. \cite{yamani2023spatiotemporal} reported a similar transition induced by a shear layer instability in elasto-inertial turbulent planar jets. At high Reynolds, this instability arises from the interplay between inertia and elasticity, driven by the crowding and dilation of streamlines within the edge region of the jet \citep{rallison1995instability}. In elastic turbulence, inertia is negligible, and high-shear originates from the near-field streaks  instead, as shown in the previous section.

\begin{figure}
    \centering
    \includegraphics[scale=0.038, keepaspectratio]{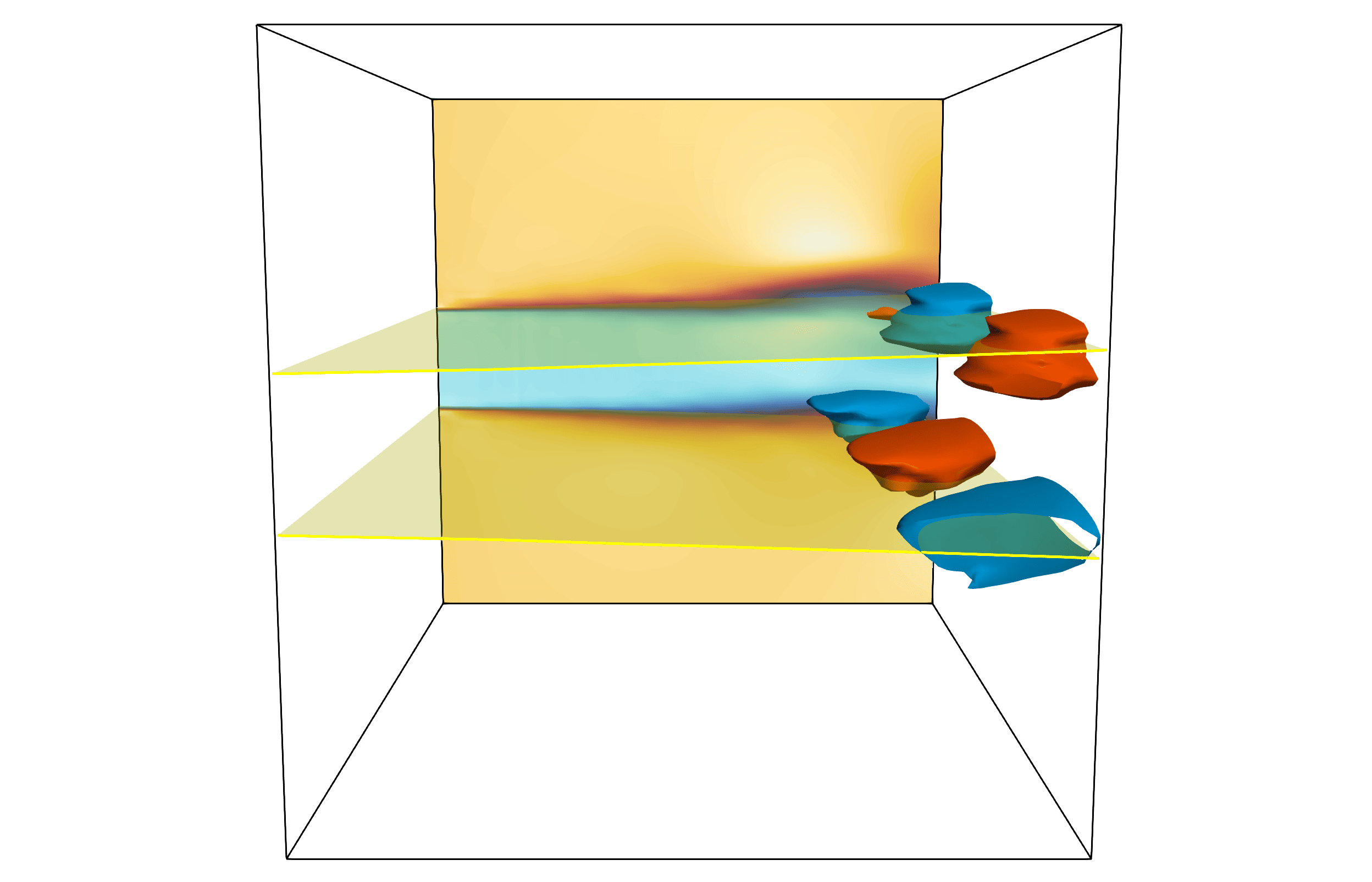}
    \put(-89,57){\small $10$}
    \put(-84,27){\small $0$}
    \put(-96,27){\large $\frac{y}{h}$}
    \put(-93,-1){\small $-10$}
    \put(-77,-8){\small $0$}
    \put(-24,-8){\small $20$}
    \put(-93,69){\small $a)$}
    \put(-68,69){$St \simeq 0.056$}
    \includegraphics[scale=0.038, keepaspectratio]{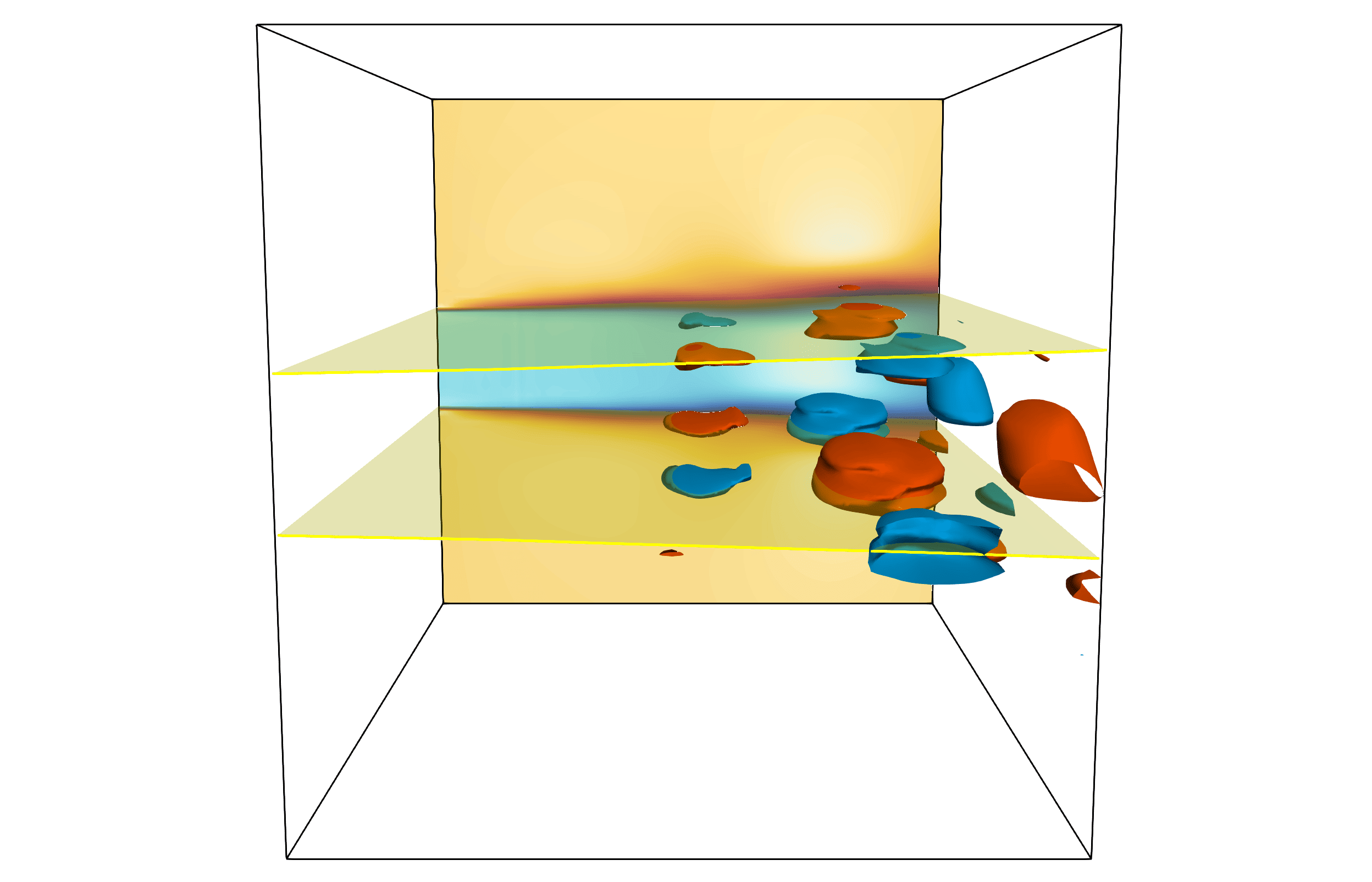}
    \put(-89,57){\small $10$}
    \put(-84,27){\small $0$}
    \put(-93,-1){\small $-10$}
    \put(-77,-8){\small $0$}
    \put(-24,-8){\small $20$}
    \put(-93,69){\small $b)$}
    \put(-68,69){$St \simeq 0.086$}
    \includegraphics[scale=0.038, keepaspectratio]{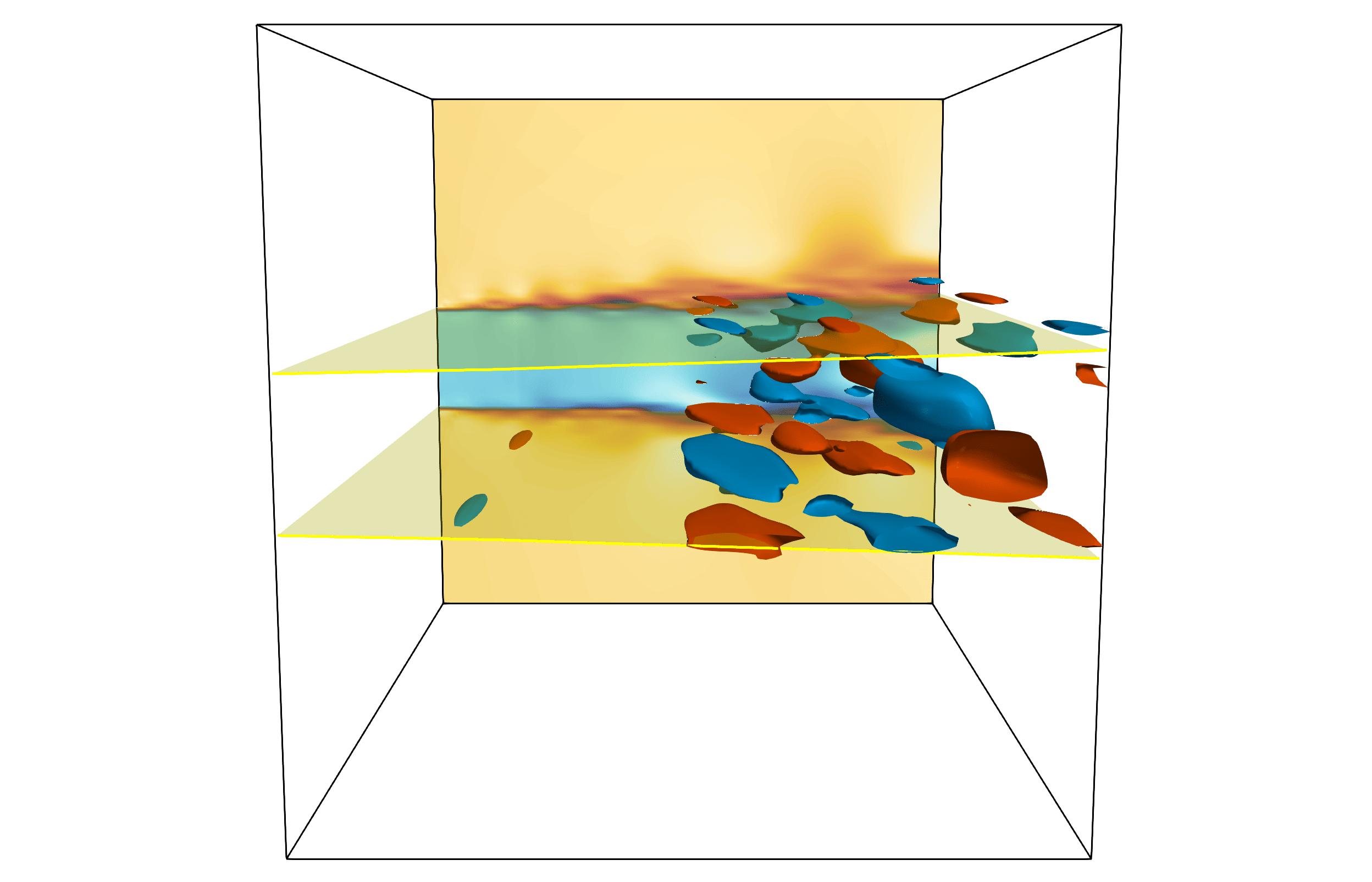}
    \put(-89,57){\small $10$}
    \put(-84,27){\small $0$}
    \put(-93,-1){\small $-10$}
    \put(-77,-8){\small $0$}
    \put(-24,-8){\small $20$}
    \put(-93,69){\small $c)$}
    \put(-68,69){$St \simeq 0.121$}
    
    \vspace{0.2cm}
    \includegraphics[scale=0.038, keepaspectratio]{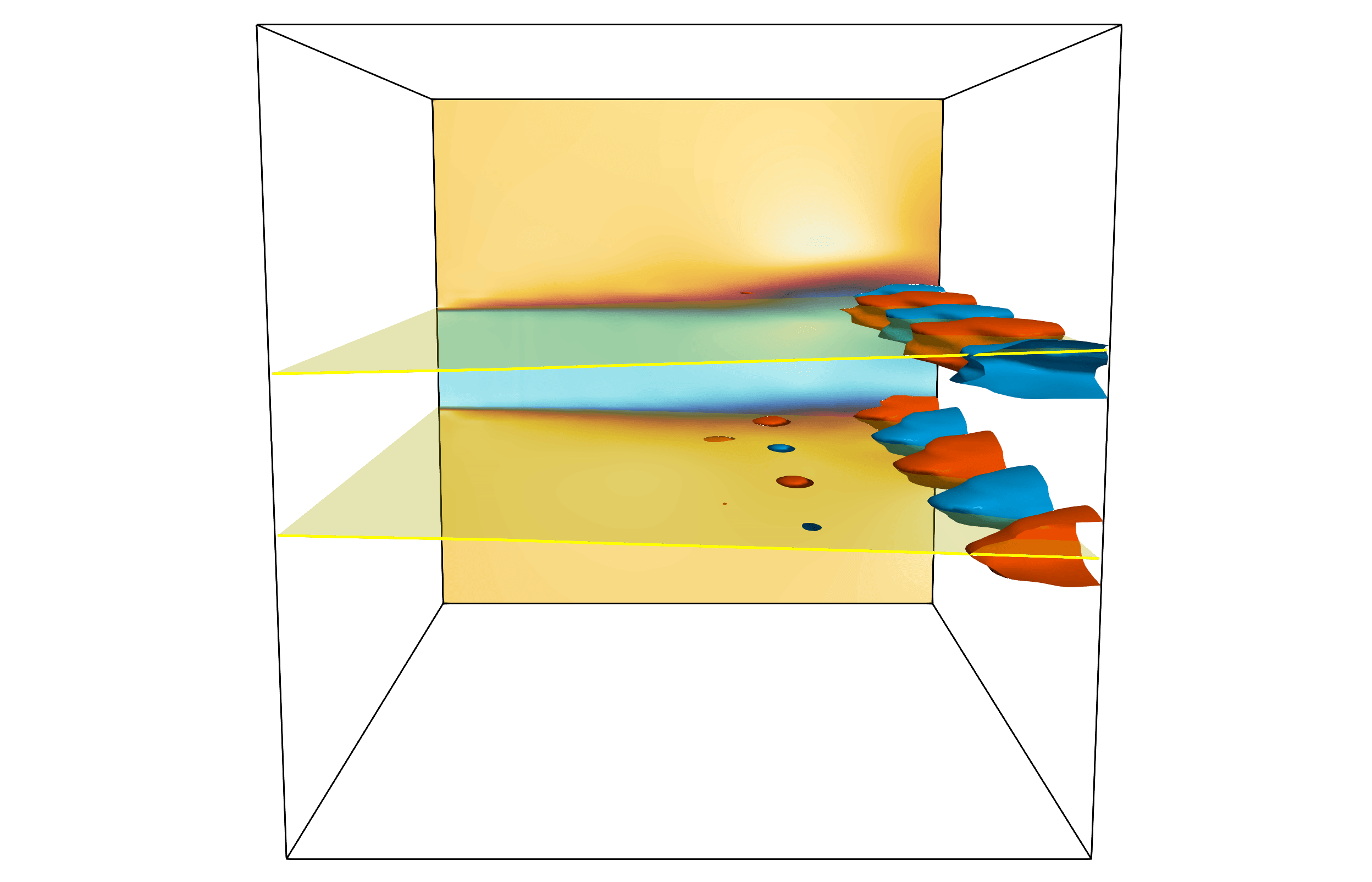}
    \put(-89,57){\small $10$}
    \put(-84,27){\small $0$}
    \put(-96,27){\large $\frac{y}{h}$}
    \put(-93,-1){\small $-10$}
    \put(-77,-8){\small $0$}
    \put(-24,-8){\small $20$}
    \put(-53,-16){$x/h$}
    \put(-93,69){\small $d)$}
    \includegraphics[scale=0.038, keepaspectratio]{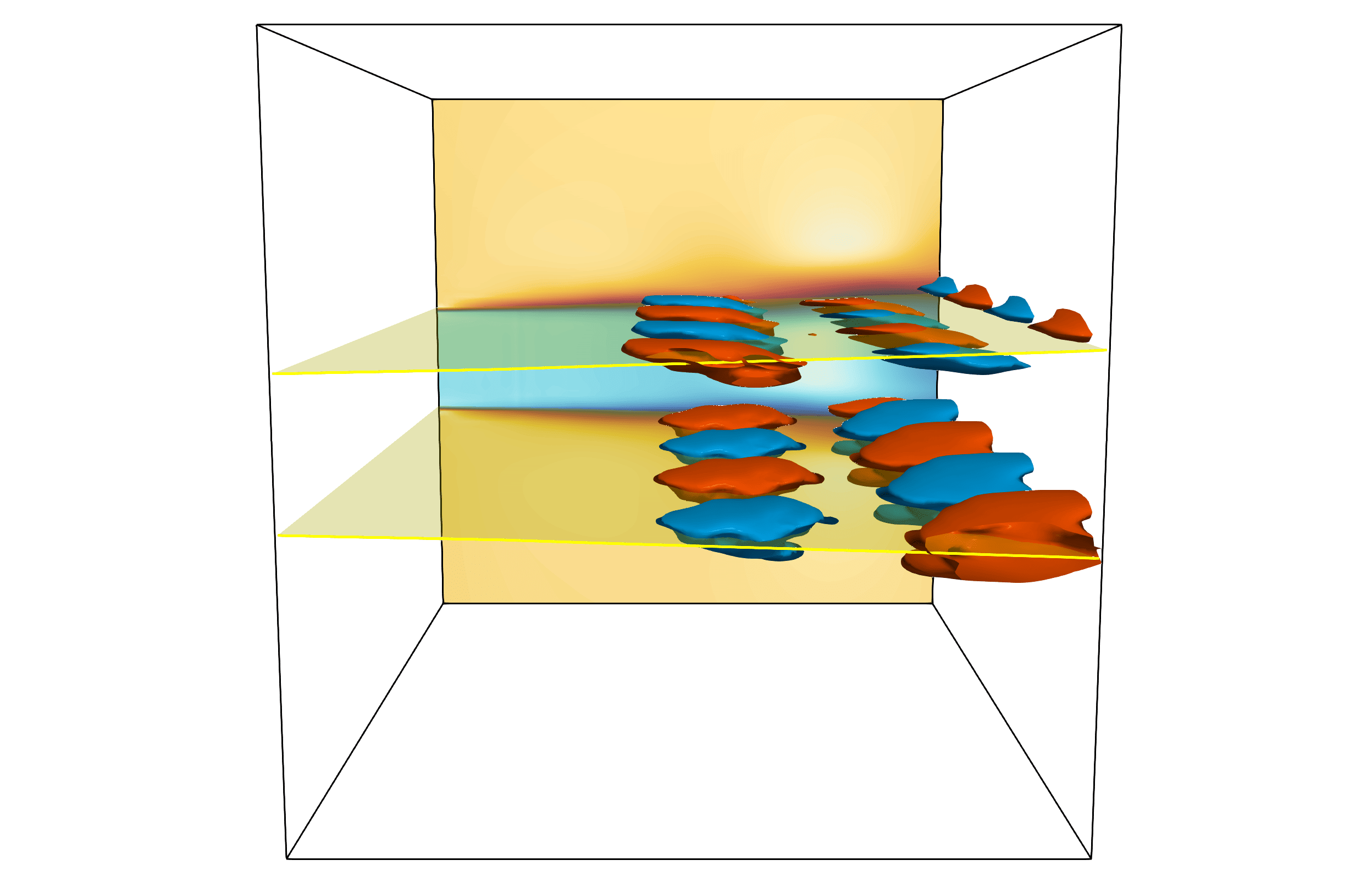}
    \put(-89,57){\small $10$}
    \put(-84,27){\small $0$}
    \put(-93,-1){\small $-10$}
    \put(-77,-8){\small $0$}
    \put(-24,-8){\small $20$}
    \put(-53,-16){$x/h$}
    \put(-93,69){\small $e)$}
    \includegraphics[scale=0.038, keepaspectratio]{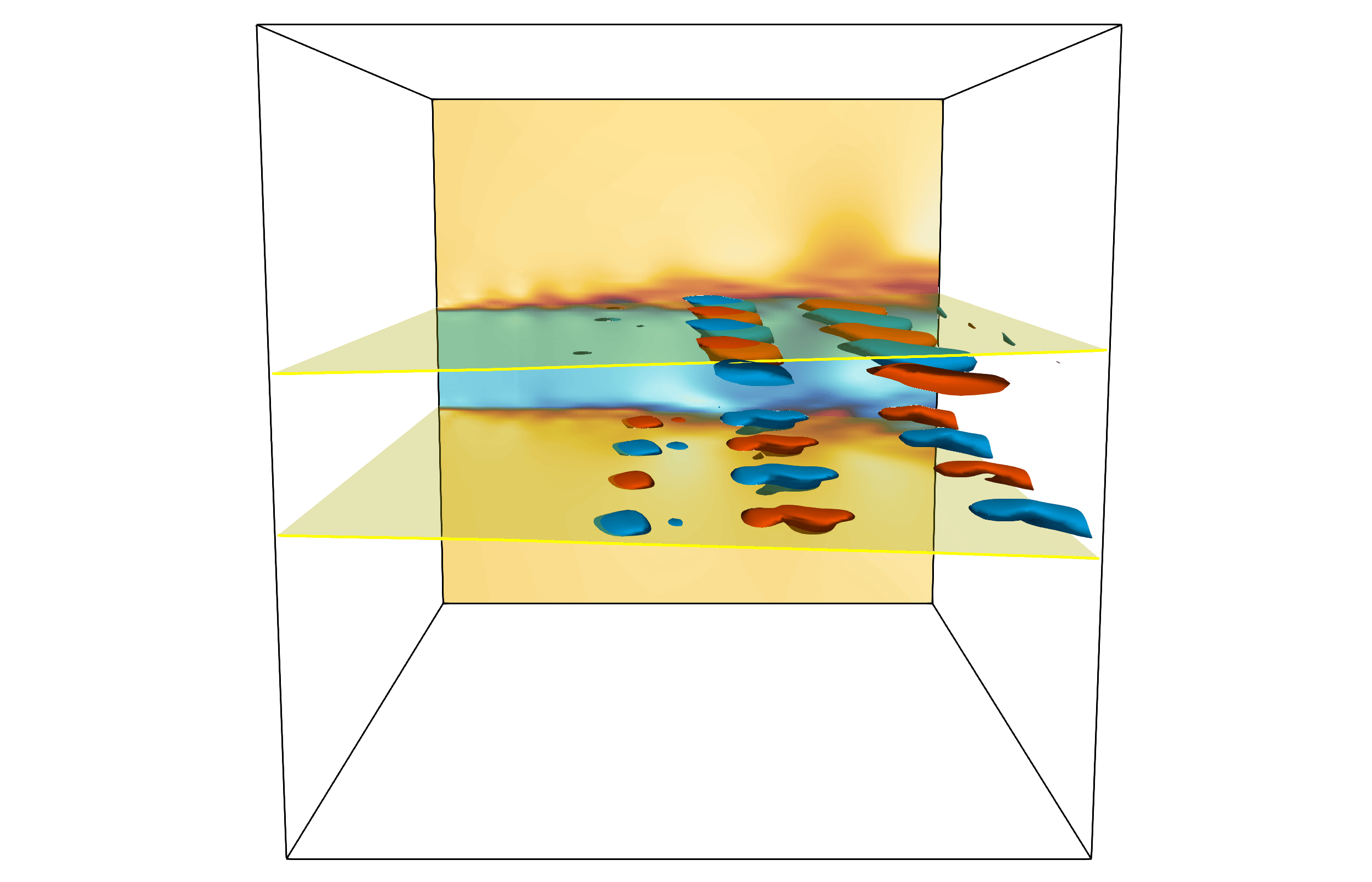}
    \put(-89,57){\small $10$}
    \put(-84,27){\small $0$}
    \put(-93,-1){\small $-10$}
    \put(-77,-8){\small $0$}
    \put(-24,-8){\small $20$}
    \put(-53,-16){$x/h$}
    \put(-93,69){\small $f)$}
    \caption{Reconstruction of the wave packet modes at the near-field. Three-dimensional iso-surfaces of the modes with $\kappa_1$ (\textit{a-c}) and $\kappa_2$ (\textit{d-f}) of the normalised streamwise velocity for values of $+0.5$ (red) and $-0.5$ (blue). The background corresponds to $xy$-planes at $z = 0$ of the mode with $\kappa = 0$ for the same frequency. The yellow translucent surfaces mark the average jet thickness.}
    \label{fig:uvw-stkdL12}
\end{figure}
The results for $\kappa = 0$ reveal that the near-field streaks not only modify the bulk flow, but they also interact with the coherent structures at the same region. This interaction ultimately drives their breakdown, as illustrated in fig.~\ref{fig:uvwstr-stkdL_12}. To study this in more detail, we show in fig.~\ref{fig:uvw-stkdL12} the three-dimensional modes with $\kappa_1$ and $\kappa_2$ for $St \simeq 0.056$, $0.086$ and $0.121$, that have the largest amplitudes. In all cases, the modes resemble wave packet structures. At $St \simeq 0.056$, these structures are at the shear layer, whereas at $St \simeq 0.086$ and $0.121$ they appear at both the shear layer and jet column for different wavenumbers. We therefore distinguish two types of modes: a shear layer mode at higher wavenumber and a dual mode (shear layer and jet column) mode at lower wavenumber. The spatio-temporal analysis separates their contribution and quantifies their importance based on the relative spatio-temporal amplitude: the dual mode is more relevant at $St \simeq 0.086$, while the shear layer mode dominates at $St \simeq 0.121$. Moreover, the shear layer mode develops earlier in the potential core than the dual mode---more specific the structures at the jet column---and its cumulative amplitude across frequencies exceeds that of the dual mode. Taken together, these findings indicate that shear layer dynamics play a more significant role at the potential core. This classification aligns with theoretical and experimental results. The shear layer mode generally appears at higher wavenumbers that the jet column mode---the dual mode in this case---that merge at sufficiently high elasticity \citep{rallison1995instability, yamani2023spatiotemporal}. Finally, both modes, especially the shear layer one, interact with the near-field streaks. The interaction is the most relevant at $St \simeq 0.086$ and $0.121$, owing to the earlier onset of the structures. The near-field streaks also interact with the mode at $St \simeq 0.056$, though such interaction occurs near the end of the potential core at \reva{$x/h \approx 20$} between the wave packets and the remnants of the near-field streaks that broke upstream. This interplay between streaks and wave packets constitutes the mechanism responsible for streak breakdown at the near-field, and it completes the effect of the streaks---modification of bulk flow and interaction with flow instability---at the potential core. This interplay is elastic in nature, where both near-field streaks and wave packets are driven by the elasticity of the flow.

\section{Polymer dynamics}\label{sec:trc}
We conclude our analysis by examining the trace of the conformation tensor to highlight similarities and differences between coherent structures in the velocity and polymer fields of the viscoelastic planar jet. The spatio-temporal analysis is applied globally to the three-dimensional trace of the conformation tensor. The calibration from STKD is analogous to that used for the velocity field and, in particular, we show the results for the set of parameters $d = 80$ and $\varepsilon_1 = \varepsilon_2 = 6 \cdot 10^{-4}$.

\begin{figure}
    \centering
    \includegraphics[scale=0.39, keepaspectratio]{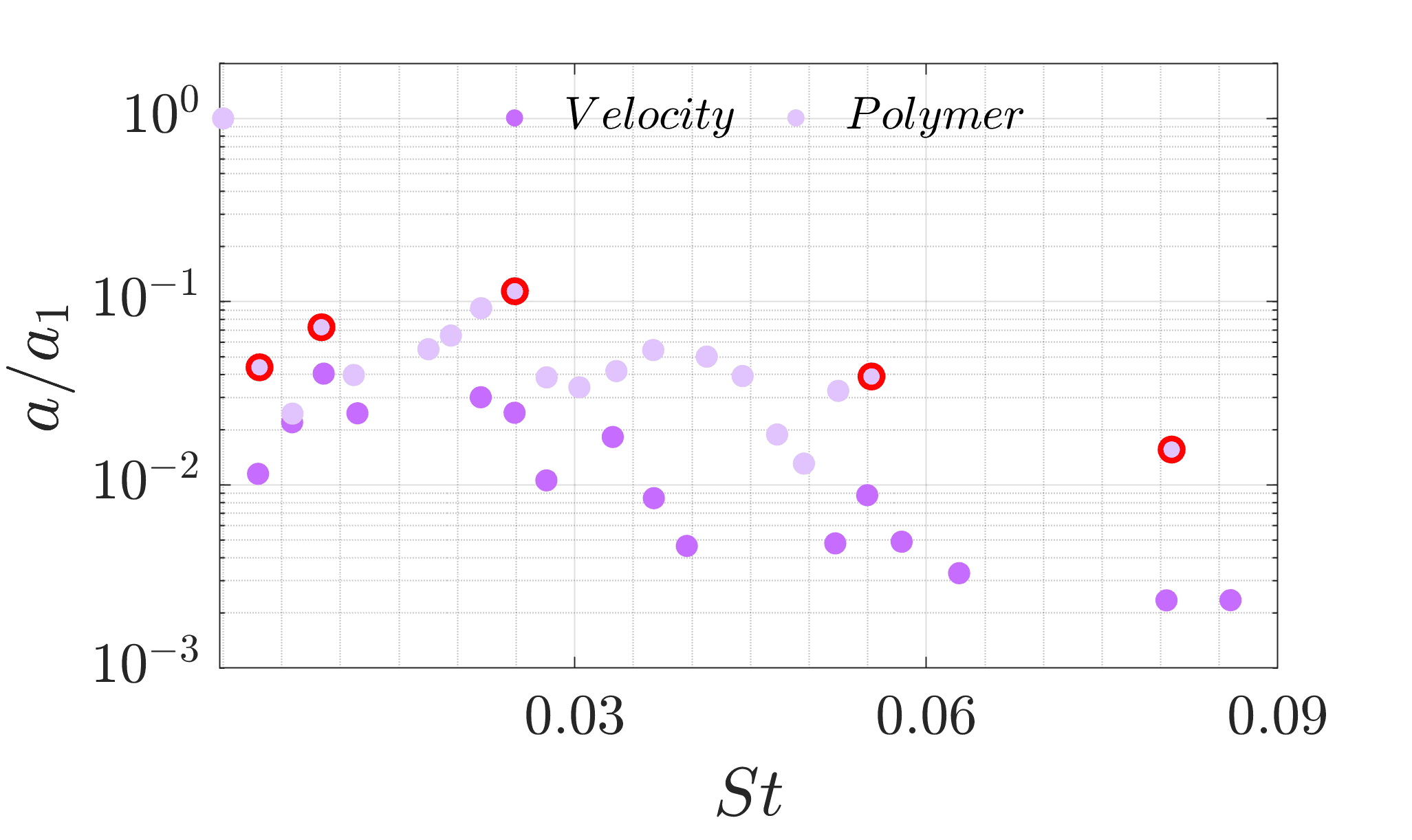}
    \put(-190,100){\small $a)$}
    \hspace{-0.4cm}
    \includegraphics[scale=0.39, keepaspectratio]{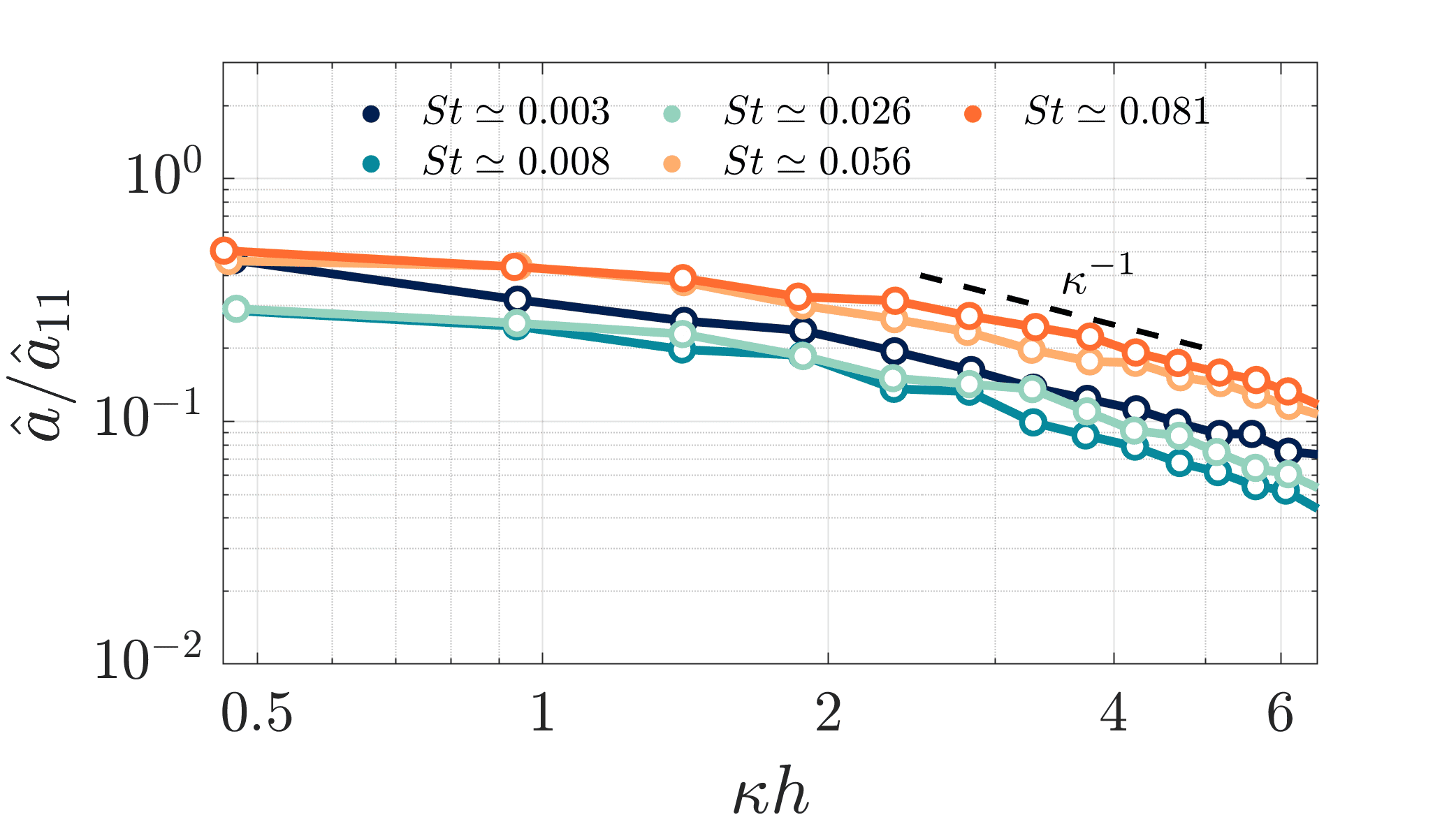}
    \put(-190,100){\small $b)$}
    \caption{Temporal (\textit{a}) and spatio-temporal \revc{(\textit{b}) spectra of the trace of conformation tensor}. Panel \textit{a} shows the non-dimensional temporal frequency, $St$, compared to the normalised amplitude, $a / a_1$, with $a_1$ the largest amplitude in each series of robust modes from the polymer field (light purple), that are compared to those from the velocity field (purple) reported in fig. \ref{fig:hodmd-spectra}. Red outlines indicate modes of the polymer that are promoted to the spatio-temporal analysis; panel \textit{b} shows the normalised spanwise wavenumber, $\kappa h$, compared to the normalised spatio-temporal amplitude, $\hat{a} / \hat{a}_{11}$, with $\hat{a}_{11}$ the largest amplitude for each robust mode. \revc{A power-law decay of the normalized amplitude is suggested for high wavenumbers.}}
    \label{fig:hodmd2trC-spectra}
\end{figure}
Figure \ref{fig:hodmd2trC-spectra} summarises the spatio-temporal spectra. Panel \textit{a} shows the temporal frequencies and amplitudes of the robust modes, that are compared to those from the velocity field. Panel \textit{b} presents the corresponding spanwise spectra for the subset of robust modes highlighted with red outlines in panel \textit{a}. The velocity and polymer fields share similar temporal dynamics: several robust frequencies identified in the analysis of the velocity field also appear in the corresponding for the polymer field. However, the polymer strongly amplifies these modes, indicated by their higher amplitude, that is roughly an order of magnitude larger. This amplification is particularly evident at higher frequencies, where the dominant frequency shifts to $St \simeq 0.026$, suggesting that faster dynamics are more relevant for reconstructing the polymer field. The spanwise spectrum also resembles that of the velocity field. In both cases, amplitudes decay monotonically with increasing wavenumber, indicating that smaller scales contribute progressively less to the reconstruction of the temporal dynamics. For the polymer, however, this decay is less pronounced, \revc{as indicated by smaller exponent of the power-law decay of the normalized amplitude ($-1$ compared to $-1.5$ in the velocity field)}, reflecting a more complex structure. Moreover, the spatio-temporal amplitudes of the non-zero wavenumbers at low frequency are higher, indicating that the streaky structure of the polymer field is more regular compared to the velocity field.

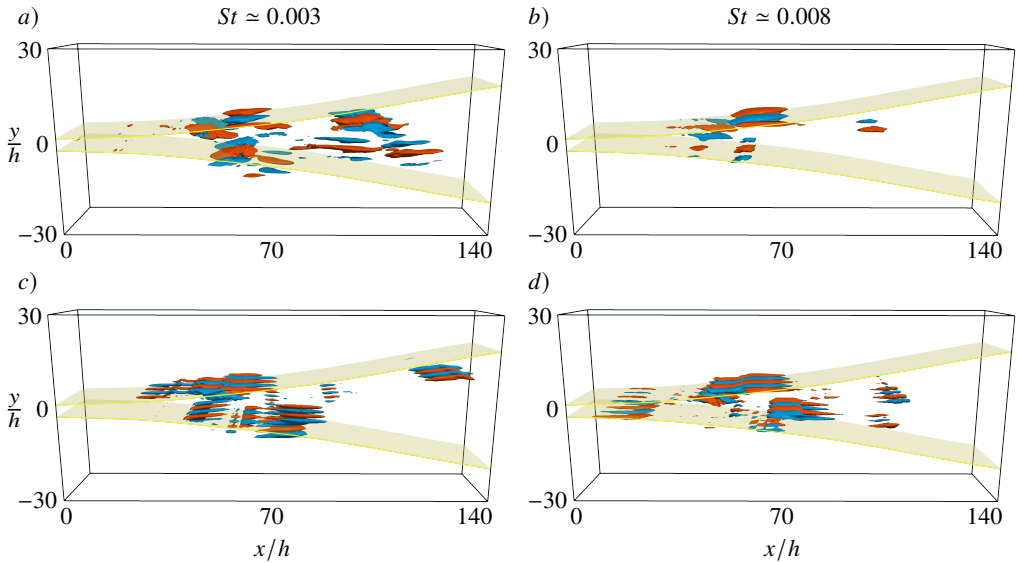
\begin{figure}
    \centering
    \includegraphics[scale=0.07, keepaspectratio]{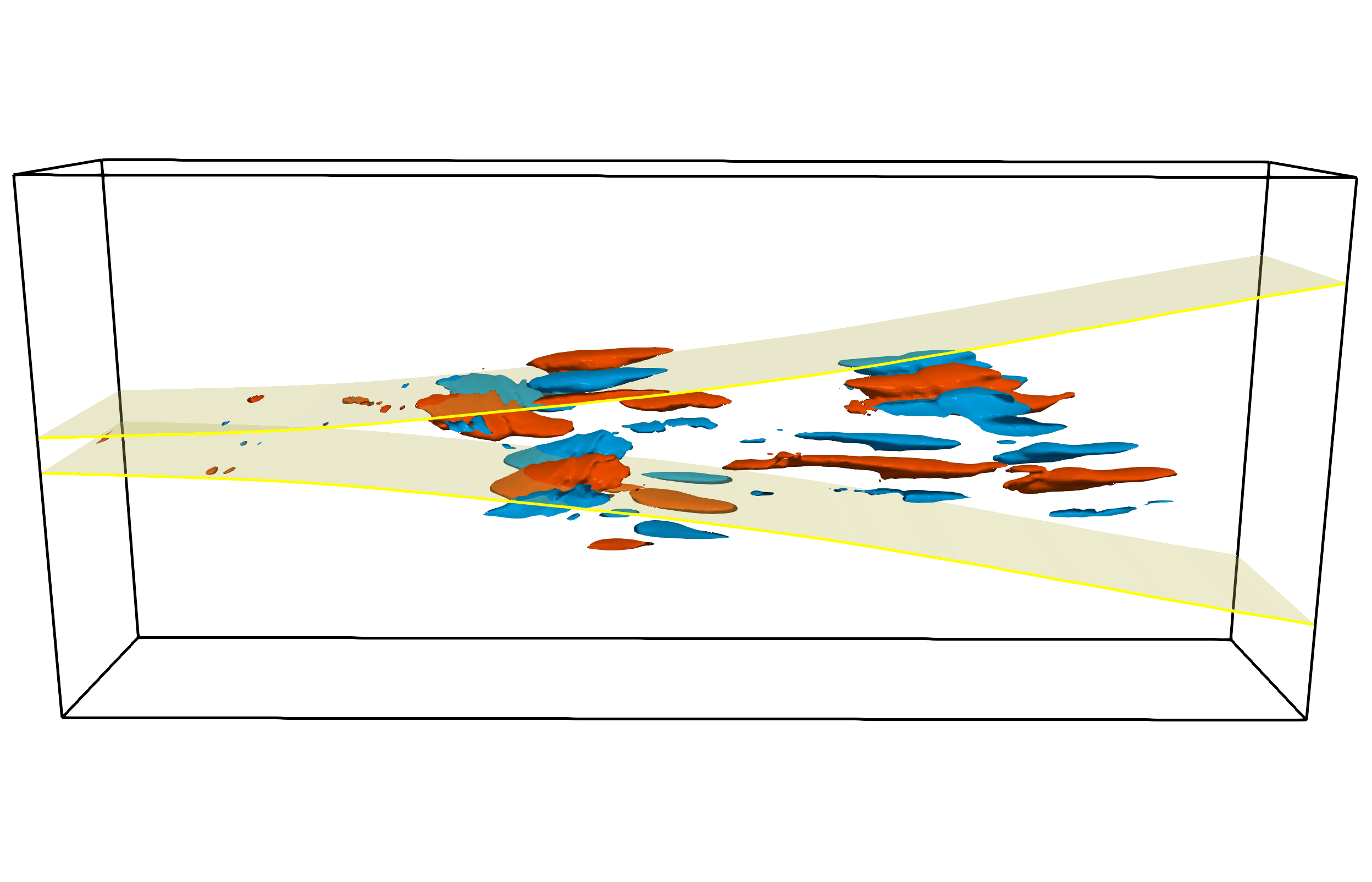}
    \put(-185,87){\small $30$}
    \put(-178,52){\small $0$}
    \put(-185,18){\small $-30$}
    \put(-190,52){\large $\frac{y}{h}$}
    \put(-169,12){\small $0$}
    \put(-94,12){\small $70$}
    \put(-20,12){\small $140$}
    \put(-185,100){\small $a)$}
    \put(-110,100){$St \simeq 0.003$}
    \hspace{0.5cm}
    \includegraphics[scale=0.07, keepaspectratio]{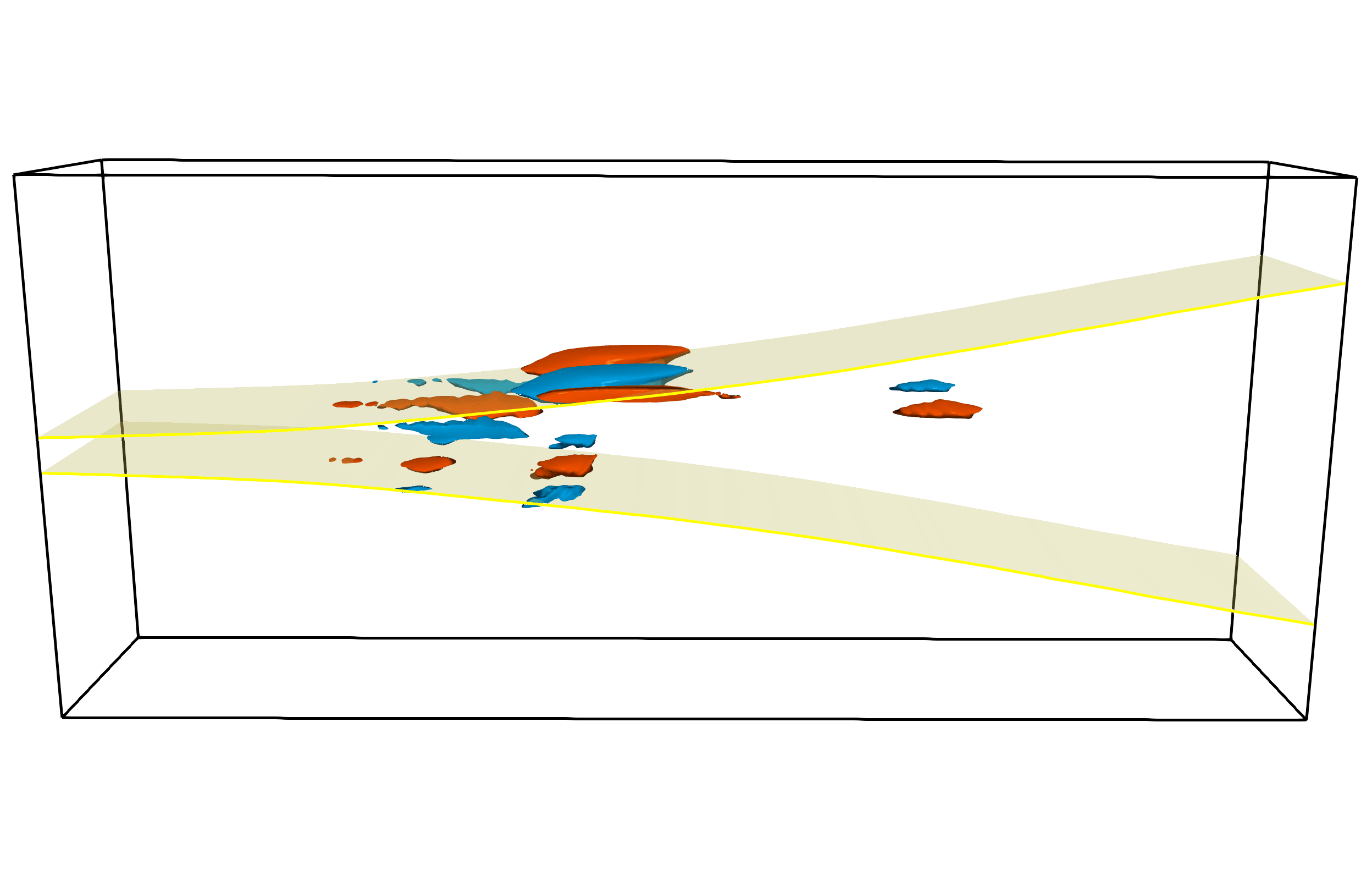}
    \put(-185,87){\small $30$}
    \put(-178,52){\small $0$}
    \put(-185,18){\small $-30$}
    \put(-169,12){\small $0$}
    \put(-94,12){\small $70$}
    \put(-20,12){\small $140$}
    \put(-185,100){\small $b)$}
    \put(-110,100){$St \simeq 0.008$}
    \vspace{-0.5cm}
    \includegraphics[scale=0.07, keepaspectratio]{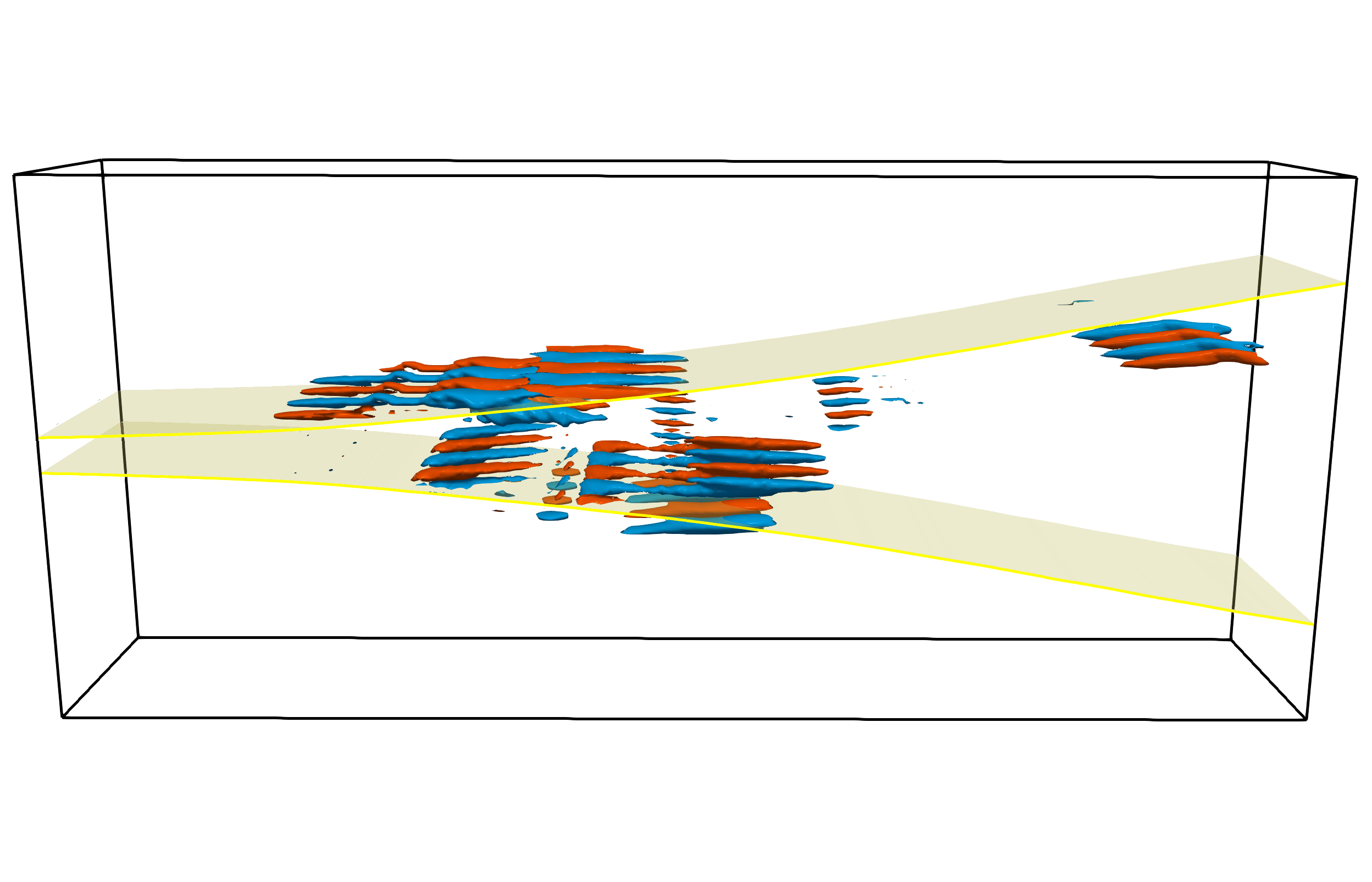}
    \put(-185,87){\small $30$}
    \put(-178,52){\small $0$}
    \put(-190,52){\large $\frac{y}{h}$}
    \put(-185,18){\small $-30$}
    \put(-169,12){\small $0$}
    \put(-94,12){\small $70$}
    \put(-20,12){\small $140$}
    \put(-96,0){$x/h$}
    \put(-185,100){\small $c)$}
    \hspace{0.5cm}    
    \includegraphics[scale=0.07, keepaspectratio]{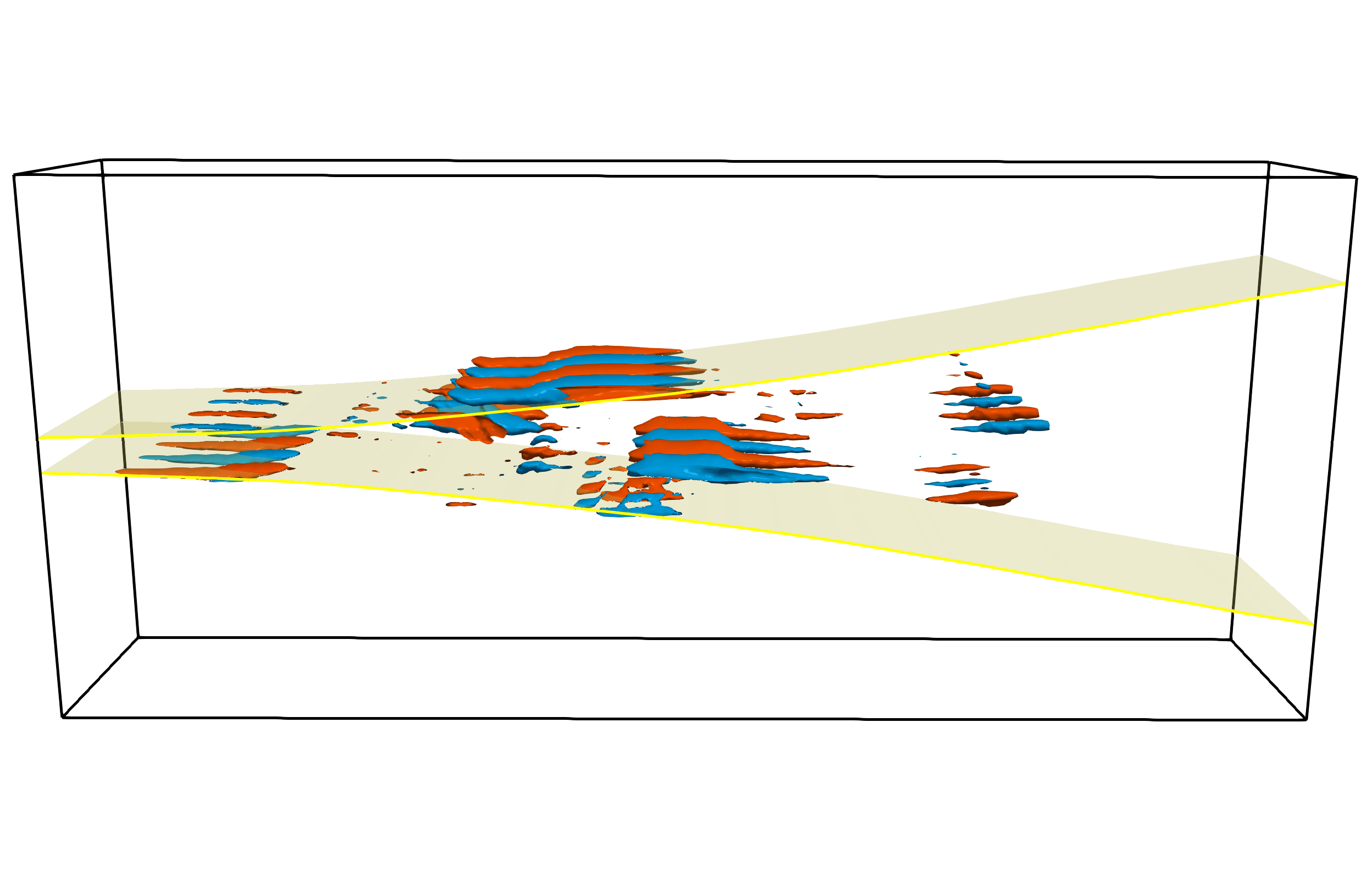}
    \put(-185,87){\small $30$}
    \put(-178,52){\small $0$}
    \put(-185,18){\small $-30$}
    \put(-169,12){\small $0$}
    \put(-94,12){\small $70$}
    \put(-20,12){\small $140$}
    \put(-96,0){$x/h$}
    \put(-185,100){\small $d)$} 
    \caption{Spatial structures of two low frequency spatio-temporal modes from the polymer field. Panels \textit{a} and \textit{b} show the modes with $\kappa_1$, and panels \textit{c} and \textit{d} with $\kappa_2$. Three-dimensional iso-surfaces are plotted for magnitudes $+0.5$ (red) and $-0.5$ (blue). The yellow translucent surfaces mark the average jet thickness.}
    \label{fig:stkd2trC-fil}
\end{figure} 
We now compare the spatio-temporal structures of the polymer with those from the velocity field. First, we show in fig.~\ref{fig:stkd2trC-fil} the two leading non-zero wavenumbers, $\kappa_1$ and $\kappa_2$, at low-frequency. The first distinctive structure of the polymer field are polymer filaments stretched in the streamwise direction. These filaments are connected to the same low frequencies of the velocity streaks, exhibit a qualitatively similar spatial organisation, and populate the same region of the flow. At higher wavenumber (panels \textit{c} and \textit{d}), smaller filaments are located more upstream, similar than the velocity streaks, though they populate a region closer to the inlet (panel \textit{d}). The size of these structures is larger compared to the velocity streaks at the same location. At the potential core, the strong shear generated from high- and low- speed streaks stretches the polymer; the only structure capable of stretching the polymer are indeed the near-field streaks. This confirms that polymer stretching occurs precisely in this region, supporting the view that elasticity plays a decisive role in triggering the transition to elastic turbulence.

\begin{figure}
    \centering
    \includegraphics[scale=0.045, keepaspectratio]{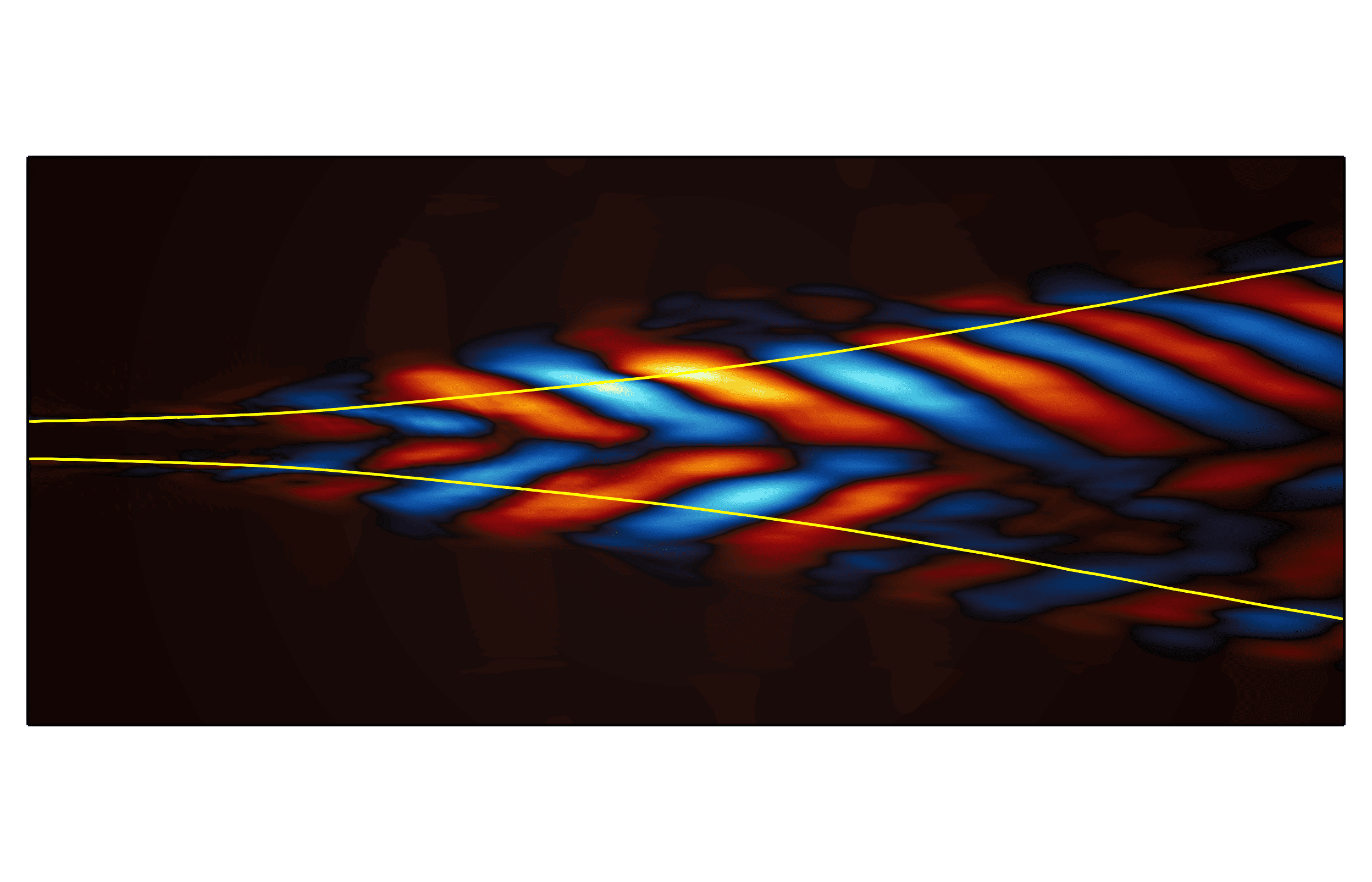}
    \put(-123,57){\small $30$}
    \put(-118,34){\small $0$}
    \put(-128,11){\small $-30$}
    \put(-130,34){\large $\frac{y}{h}$}
    \put(-113,3){\small $0$}
    \put(-60,3){\small $70$}
    \put(-10,3){\small $140$}
    \put(-62,-9){$x/h$}
    \put(-75,67){$St \simeq 0.026$}
    \hspace{0.5cm}
    \includegraphics[scale=0.045, keepaspectratio]{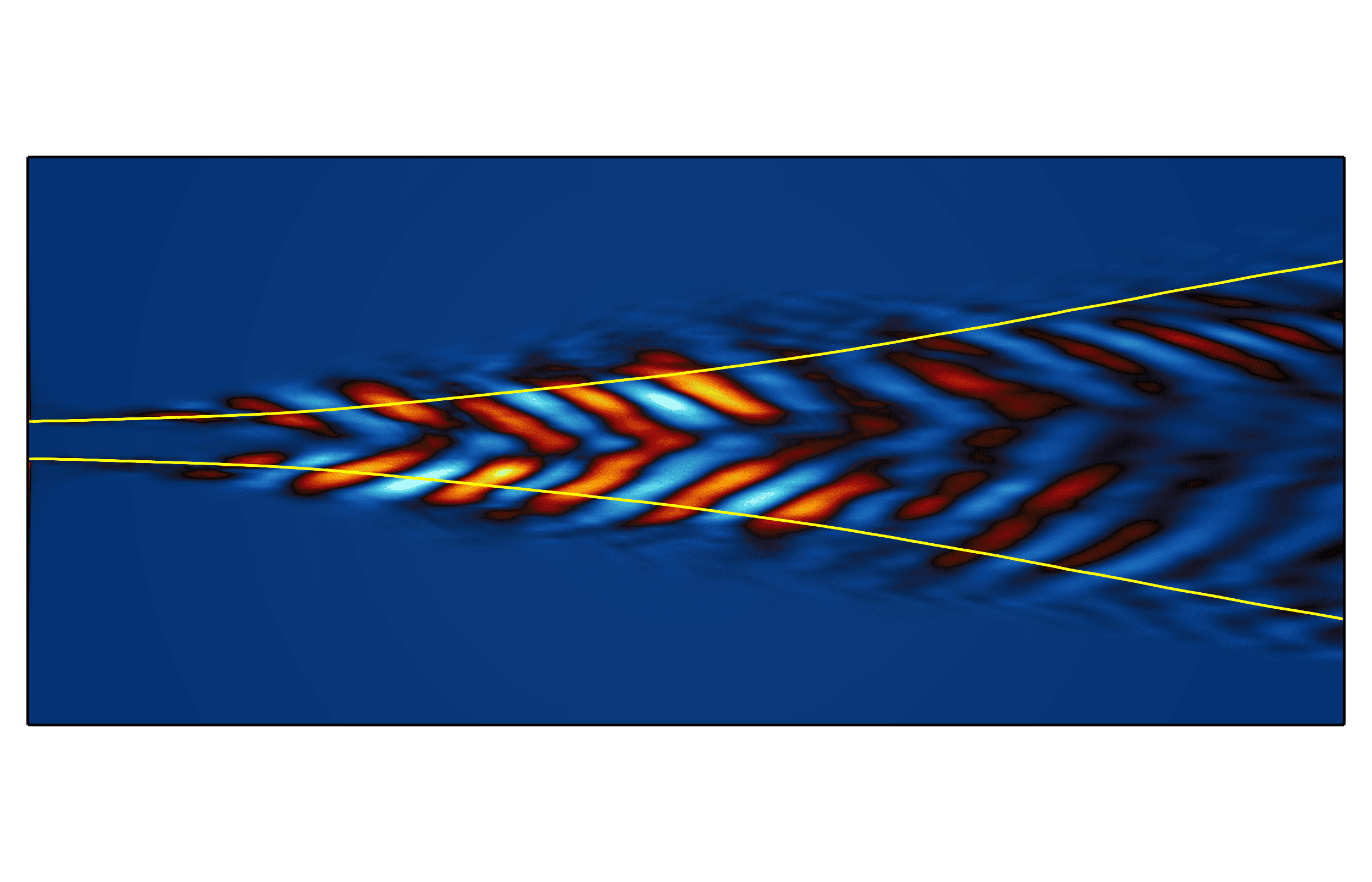}
    \put(-123,57){\small $30$}
    \put(-118,34){\small $0$}
    \put(-128,11){\small $-30$}
    \put(-113,3){\small $0$}
    \put(-60,3){\small $70$}
    \put(-10,3){\small $140$}
    \put(-62,-9){$x/h$}
    \put(-75,67){$St \simeq 0.056$}
    \hspace{0.5cm}
    \includegraphics[scale=0.045, keepaspectratio]{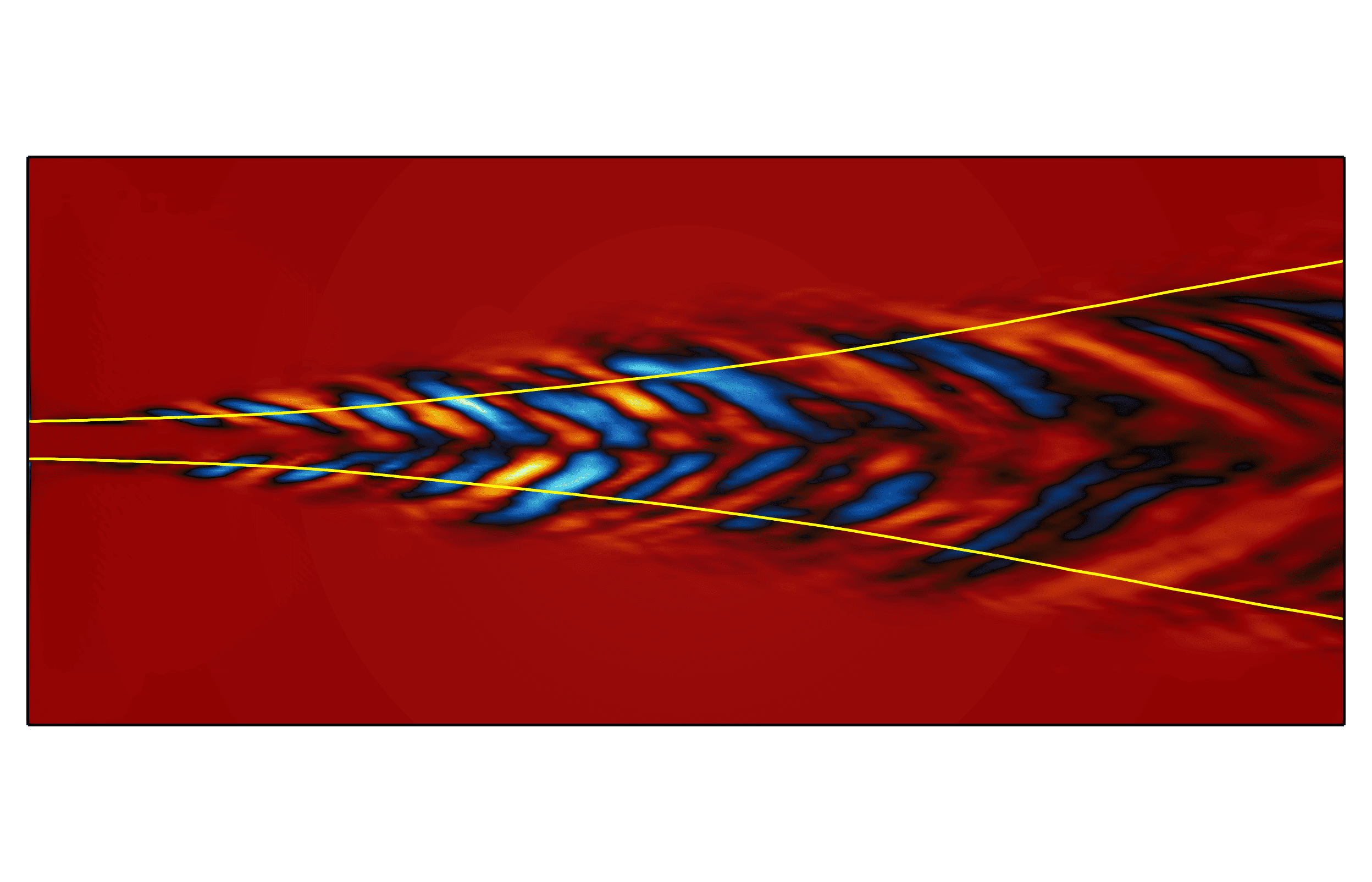}  
    \put(-123,57){\small $30$}
    \put(-118,34){\small $0$}
    \put(-128,11){\small $-30$}
    \put(-113,3){\small $0$}
    \put(-60,3){\small $70$}
    \put(-10,3){\small $140$}
    \put(-62,-9){$x/h$}
    \put(-75,67){$St \simeq 0.081$}
    
    \includegraphics[width=\textwidth, keepaspectratio, trim=0 80 0 2400, clip]{figures/figure11.png}
    \put(-255,-8){\small $-1$}
    \put(-195,-8){\small $0$}
    \put(-140,-8){\small $1$}
    \put(-200,-20){$u/U$}
    \caption{Spatial structures of three high frequency spatio-temporal modes from the polymer field. Two-dimensional $xy-$planes at $z = 0$ of the normalised streamwise velocity component for the modes with $\kappa = 0$. The yellow lines mark the average jet thickness.}
    \label{fig:stkd2trC-arrow}
\end{figure}
Lastly, the second distinctive structure of the polymer field are the centre-mode structures. These are shown in fig.~\ref{fig:stkd2trC-arrow}. In this case, polymer sheets first appear in both shear layers, are stretched by the faster flow in the jet core, and then merge at the centre further downstream. At $St \simeq 0.056$ and $0.081$, stretching occurs immediately after the potential core, and the resulting structures are smaller than those observed at $St \simeq 0.026$. The organisation also changes with frequency. While the structure is antisymmetric with respect to the centreline at $St \simeq 0.026$, it becomes symmetric from $St \simeq 0.056$. Similar highly extended sheets of polymer stress have been reported in two-dimensional elasto-inertial channel flows \citep{Dubief2023ARFMEIT}. These are connected to the centre-mode instability of viscoelastic pipe and channel flows at high Reynolds \citep{Garg2018PRLCentMode, Khalid2021JFMCentMode}, that persist at vanishing inertia \citep{Buza2022JFMCMLowRe}. While similar centre-mode structures have been observed in elastic turbulent channel flows \citep{Morozov2022PRLETArr, Rota2024PRFvChan}, they also appear in free shear flow configurations \citep{BertiJFM20082DETKF, Lewy2025JFMETKFCentMode}. Most of these observations were done in a two-dimensional domain; in fact, the centre-mode structure of the viscoelastic planar jet is two-dimensional, where the superposition of $\kappa = 0$ with the leading non-zero wavenumber and its harmonics modulates the two-dimensional structure in the three-dimensional space. While it is believed that elasto-inertial turbulence is essentially a two-dimensional phenomenon \citep{Sid2018PRFEIT2D, Dubief2023ARFMEIT}, elastic turbulence in planar jets is necessarily three-dimensional: the near-field streaks, that are inhomogeneous structures in the spanwise direction, play a relevant role in the transition to elastic turbulence. These will be unavoidably missed if the domain is fully two-dimensional, remarking the three-dimensional nature of elastic turbulence.

\section{Conclusions}\label{sec:concl}
We perform a data-driven modal analysis of Newtonian and viscoelastic turbulent planar jets using STKD, where we represent the flow dynamics as a superposition of modes related to the large-scale coherent structures. This sparse representation facilitates the interpretation of the complex dynamics in turbulent jet flows, allowing us to compare the most relevant structures between the high-Reynolds, Newtonian and the low-Reynolds, viscoelastic jets. \reva{We therefore compare a fully-developed Newtonian turbulent jet with a fully-developed elastic turbulent jet; in the later case, turbulence is sustained solely by elasticity, owing to the low Reynolds number and high Deborah number, with the equivalent low-Reynolds, Newtonian jet being laminar \citep{sato1960stability, sato1964experimental, soligo2023non}.}

The global analysis of the velocity field reveals the presence of slow streaky structures and fast wave packets. While their overall characterisation is similar, there are subtle differences. Low-frequency streaky structures have a global presence in the Newtonian jet, particularly in the far-field, where they also extend deep into the jet core. In contrast, they are less dominant in the viscoelastic jet, where the strong flapping motion at the wake disrupts the far-field streaks, rendering the flow more uniform and homogeneous. High-frequency modes show closer agreement between the two cases, with structures consistent with the description of Orr and Kelvin-Helmholtz wave packets. Moreover, the global analysis captures the characteristic instability in the Newtonian jet, where symmetries in the high-frequency modes align with the symmetric, varicose instability, known to dominate at high Reynolds numbers \citep{Antonia1983OrganizedJet, Thomas1986DevelopJet, suresh2008reynolds, deo2008influence}. In the viscoelastic jet, the wave packets more closely resemble the Orr wave packets, while evidence of Kelvin-Helmholtz wave packets, particularly at intermediate frequencies and non-zero wavenumbers, is more inconclusive.

The most striking difference between the two jets is the presence of streaks in the near-field of the viscoelastic jet, that are absent in the Newtonian case and arise from flow elasticity. The near-field streaks are not standing structures: they travel and get disrupted by the flow instability of the jet. Local analysis of the potential core shows that the streaks interact with the high-frequency modes in the potential core. In particular, we identified two modes: a low-wavenumber dual mode with wave packets located in both the jet column and the shear layer, and a high-wavenumber shear layer mode. Similar modes have been reported by \cite{yamani2023spatiotemporal} in elasto-inertial turbulent planar jets using dynamic mode decomposition. These modes are elastic in nature---inertial effects are negligible in the viscoelastic jet---where their interaction with the streaks induces the breakdown of the later. At the same time, the streaks modify the bulk flow motion at the potential core. We observe both symmetric and antisymmetric dynamics: the symmetric motion dominates in the potential core, while the antisymmetric motion get amplified further downstream. The bulk flow is also perturbed at the shear layer, where disturbances emerge immediately after the inlet, grow downstream, and ultimately destabilise the jet. These observations are consistent with the transition reported along the jet edges in viscoelastic jets at higher Reynolds numbers \citep{rallison1995instability, yamani2021spectral, yamani2023spatiotemporal}. \revb{Therefore, we link the near-field streaks to the onset of elastic turbulence in viscoelastic planar jets.} We propose that high- and low- speed streaks generate regions of localised shear between adjacent streaks in the near-field of the jet. The high shear stretches the polymer in the streamwise direction, the resulting elastic stresses are amplified downstream, and they ultimately cause the transition to turbulence. Our findings thus point to the near-field streaks as a potential pathway to elastic turbulence in low-Reynolds, viscoelastic planar jets.

Finally, we applied the same analysis to the polymer field. The temporal analysis showed that high-frequency modes are more important for reconstructing the polymer dynamics, which mainly reflect the amplified response to the velocity field. As a result, the polymer field exhibits similar structures: streamwise-oriented filaments of stretched polymer at low frequency and centre-mode structures at high frequency. The polymer filaments correspond to the same low frequencies as the velocity streaks, share a similar spatial organisation, and occupy the same region of the flow. Notably, they also appear closer to the inlet at much smaller wavenumbers than the velocity streaks. This behaviour suggests that strong shear caused by high- and low- speed streaks stretches the polymer, reinforcing the idea that polymer dynamics contribute actively to the transition. Together, velocity streaks and polymer filaments highlight that elastic turbulence is inherently three-dimensional in viscoelastic planar jets, \revb{similarly to what observed recently in planar wall-bounded elastic turbulent flows \citep{Lellep2024PNASvChanET, Rota2024PRFvChan}}. In the planar jet, neglecting the inhomogeneous streaks and filaments would overlook key mechanisms driving the transition. On the other hand, we also identified high-frequency, centre-mode structures that resemble the arrowheads or narwhals reported in elasto-inertial and elastic turbulence in wall-bounded and free-shear flows \citep{BertiJFM20082DETKF, Morozov2022PRLETArr, Dubief2022PRFEITCohStr, Rota2024PRFvChan, Lewy2025JFMETKFCentMode}. Their presence confirms that the viscoelastic planar jet attains the elastic turbulent regime.

\backsection[Funding]{The research was supported by the Okinawa Institute of Science and Technology Graduate University (OIST) with subsidy funding to M.E.R. from the Cabinet Office, Government of Japan. M.E.R. also acknowledges funding from the Japan Society for the Promotion of Science (JSPS), grant 24K17210 and 24K00810. The authors acknowledge the computer time provided by the Scientific Computing \& Data Analysis section of the Core Facilities at OIST, and by HPCI, under the Research Project grants hp220099, hp230018, and hp250035. S.L.C. acknowledges the grant  PID2023-147790OB-I00 funded by MCIU/AEI/10.13039/501100011033/FEDER, UE. Part of this work was done during the 2023 Madrid Turbulence Workshop, organized by Prof. J. Jim{é}nez and made possible by ERC, Caust Grant ERC-AdG-101018287}

\backsection[Declaration of interests]{The authors report no conflict of interest.}

\backsection[Author ORCIDs]{\\
Christian Amor, \url{https://orcid.org/0000-0002-9710-7917} \\
Adri{á}n Corrochano, \url{https://orcid.org/0000-0002-1396-3119} \\
Giovanni Soligo, \url{https://orcid.org/0000-0002-0203-6934} \\
Soledad Le Clainche, \url{https://orcid.org/0000-0003-3605-7351} \\
Marco Edoardo Rosti, \url{https://orcid.org/0000-0002-9004-2292}}

\bibliographystyle{jfm}
\bibliography{totalbib}

\end{document}